\documentclass[a4paper,11pt]{article}
\usepackage{jheppub} 
\usepackage{lineno}
\usepackage[normalem]{ulem}
\usepackage{xcolor}
\usepackage{mathtools}
\usepackage{comment}

\usepackage{footnote}
\usepackage{mathrsfs}
\usepackage{physics}
\usepackage{amssymb,bm}
\usepackage{amsmath}
\usepackage{slashed}
\usepackage{graphics}
\usepackage{graphicx}
\usepackage{tikz}
\usepackage{comment}
\usepackage{dsfont}

\definecolor{lcolour}{rgb}{0,0.34,0.25}
\definecolor{citcolour}{rgb}{0.73,0.31,0.28}

\hypersetup{
  linkcolor=lcolour,
  citecolor=citcolour,
  urlcolor=blue
}

\newcommand{\eqn}[1]{Eq.\,\eqref{#1}}
\newcommand{\beq}{\begin{equation}}
\newcommand{\eeq}{\end{equation}}

\newcommand{\der}{\mathrm{d}}
\newcommand{\eq}{Eq.\,}
\newcommand{\eqs}{Eqs.\,}
\newcommand{\fig}{Fig.\,}
\newcommand{\po}{\mathbb{P}}
\newcommand{\Nc}{N_\mathrm{c}}
\newcommand{\Cf}{C_\mathrm{F}}
\newcommand{\Qs}{Q_\mathrm{s}}
\newcommand{\bj}{\mathrm{Bj}}
\newcommand{\Qbar}{\Bar{Q}}
\newcommand{\Qtil}{\tilde{Q}}
\newcommand{\Qhat}{\hat{Q}}

\newcommand{\xbj}{x_\mathrm{Bj}}
\newcommand{\xpo}{x_\mathbb{P}}
\newcommand{\xqq}{x_{q\bar{q}}}
\newcommand{\xbarqg}{x_{\bar{q}g}}

\newcommand{\Mqqg}{M_{q\bar{q}g}}
\newcommand{\Mqq}{M_{q\bar{q}}}
\newcommand{\qqg}{q\bar{q}g}
\newcommand{\qbarq}{q\bar{q}}
\newcommand{\barqg}{\bar{q}g}
\newcommand{\as}{\alpha_\mathrm{s}}
\newcommand{\aem}{\alpha_{\mathrm{e.m.}}}

\newcommand{\xb}{\boldsymbol{\mathrm{x}}_1}
\newcommand{\yb}{\boldsymbol{\mathrm{x}}_2}
\newcommand{\zb}{\boldsymbol{\mathrm{x}}_3}
\newcommand{\bb}{\boldsymbol{\mathrm{b}}}
\newcommand{\rb}{\boldsymbol{\mathrm{r}}}
\newcommand{\Rb}{\boldsymbol{\mathrm{R}}}
\newcommand{\rt}{r_\perp}
\newcommand{\Rt}{R_\perp}
\newcommand{\kq}{k_1}
\newcommand{\kaq}{k_2}
\newcommand{\kg}{k_3}
\newcommand{\kbq}{\boldsymbol{\mathrm{k}}_1}
\newcommand{\kbaq}{\boldsymbol{\mathrm{k}}_2}
\newcommand{\kbg}{\boldsymbol{\mathrm{k}}_3}
\newcommand{\kto}{\boldsymbol{\Delta}}  %=k1+k2+k3=0
\newcommand{\Pb}{\boldsymbol{\mathrm{P}}} 
\newcommand{\Pt}{P_\perp} 
\newcommand{\Kb}{\boldsymbol{\mathrm{K}}}  
\newcommand{\qb}{\boldsymbol{\mathrm{q}}}  
\newcommand{\Kt}{K_\perp}  
\newcommand{\bKb}{\boldsymbol{\bar{\mathrm{K}}}}  
\newcommand{\Deltab}{\boldsymbol{\mathrm{\Delta}}}  
\newcommand{\ktq}{k_{1\perp}}
\newcommand{\ktaq}{k_{2\perp}}
\newcommand{\ktg}{k_{3\perp}}
\newcommand{\PN}{P^-_N}

\newcommand{\zq}{z_1}
\newcommand{\zaq}{z_2}
\newcommand{\zg}{z_3}

\newcommand{\polg}{\varepsilon_\sigma}
\newcommand{\polp}{\varepsilon_\lambda}
\newcommand{\hq}{h_1}
\newcommand{\haq}{h_2}
\newcommand{\Sm}{\mathcal{S}}
\newcommand{\EDqqg}{\text{ED}_{\gamma\rightarrow q\bar{q}g}}

\newcommand{\EDqqp}{\text{ED}_{\gamma \rightarrow q\bar{q}^\prime}}
\newcommand{\EDqpq}{\text{ED}_{\gamma \rightarrow q^\prime\bar{q}}}
\newcommand{\EDaqg}{\text{ED}_{\bar{q}'\rightarrow\bar{q}g}}
\newcommand{\EDqg}{\text{ED}_{q'\rightarrow qg}}

\title{\boldmath TMD factorization in diffractive heavy-quark production in photon-nucleus collisions}

\author[a]{Paul Caucal,}
\author[b,c]{Patricia Gimeno-Estivill,}
\author[d]{Edmond Iancu,}
\author[b,c]{Tuomas Lappi}
\author[e,f,g]{and Farid Salazar}
 \affiliation[a]{SUBATECH UMR 6457 (IMT Atlantique, Université de Nantes, IN2P3/CNRS), 4 rue Alfred Kastler, 44307 Nantes, France}
\affiliation[b]{Department of Physics, University of Jyväskylä,
P.O. Box 35, 40014 University of Jyväskylä, Finland}
\affiliation[c]{Helsinki Institute of Physics,
P.O. Box 64, 00014 University of Helsinki, Finland}
\affiliation[d]{Universit\'{e} Paris-Saclay, CNRS, CEA, Institut de physique th\'{e}orique, F-91191, Gif-sur-Yvette, France}
\affiliation[e]{Department of Physics, Temple University, Philadelphia, Pennsylvania 19122, USA}
 \affiliation[f]{RIKEN-BNL Research Center, Brookhaven National Laboratory, Upton, New York 11973, USA}
 \affiliation[g]{Physics Department, Brookhaven National Laboratory, Upton, New York 11973, USA}

\emailAdd{caucal@subatech.in2p3.fr}
\emailAdd{patricia.p.gimenoestivill@jyu.fi}
\emailAdd{edmond.iancu@ipht.fr}
\emailAdd{tuomas.v.v.lappi@jyu.fi}
\emailAdd{farid.salazar@temple.edu}

\abstract{Using the Colour Glass Condensate effective theory, we study the diffractive production of a massive quark-antiquark pair accompanied by a gluon in coherent photon-nucleus collisions at high energy. This partonic configuration provides the leading twist contribution to the cross section in the correlation limit where two of the partons are hard and nearly back to back in the transverse plane, while the third one is semi-hard, with a transverse momentum of the order of the nuclear saturation momentum. We consider two scenarios: \texttt{(i)} a hard quark-antiquark pair together with a semi-hard gluon; in this case we demonstrate  transverse momentum dependent (TMD) factorization with a mass-dependent ``hard'' factor and the standard expression for the gluon diffractive TMD, and  \texttt{(ii)} a hard antiquark-gluon pair and a semi-hard quark; in this case we find TMD factorization with a mass-independent ``hard'' factor and a mass-dependent quark diffractive TMD, which represents a new result. We show that increasing the quark mass reduces (or even washes out) the effects of gluon saturation on the quark diffractive TMD. In particular, it leads to the suppression of the Cronin peak that we observe in the massless limit. Our results are the basis for future phenomenological studies of quarkonium and open charm production in the saturation regime in ultraperipheral collisions at the Large Hadron Collider, and in deep inelastic scattering at the Electron-Ion Collider. 
}

\begin{document}
\maketitle
\flushbottom
\section{Introduction}
At high collision energies, QCD scattering processes become sensitive to the physics of gluon saturation. In this regime, nonlinear interactions between gluons lead to the emergence of a semi-hard transverse momentum scale, the saturation scale $\Qs$ (in the ballpark of 1~GeV), which grows with energy, i.e. with decreasing $x$. The natural framework to study gluon saturation is provided by the Colour Glass Condensate (CGC) effective theory~\cite{Gelis:2010nm}. In order to be able to study saturation physics in the weak coupling regime, we should look for situations where $\Qs$ is large, i.e. go to heavy nuclei and as small $x$ as possible. On the other hand, to observe saturation physics, the kinematics of the measured process should in some way be sensitive to momentum scales of the order of $\Qs$. The physics of gluon saturation is also more visible in exclusive or diffractive observables, where the cross section is quadratically proportional to the gluon density in the target. These very generic considerations point to diffractive scattering off large nuclei  at high energies as particularly sensitive to the physics of saturation. 

Many precision measurements of the gluon saturation regime will be performed in the future Electron-Ion Collider EIC~\cite{Accardi:2012qut,Morreale:2021pnn}, but even before that much higher collision energies are reached at the LHC. In particular, ultraperipheral collisions (UPCs) feature a dilute photon probe as in DIS, but, since the photon is quasi-real, one needs another hard scale (among the kinematical variable of the process) to allow for weak coupling calculations.
An attractive choice from a theorist perspective is to study ``hard'' dijet production. Indeed much attention has in recent years been devoted to both inclusive \cite{Caucal:2021ent,Taels:2022tza,Iancu:2022gpw,Caucal:2023fsf,Caucal:2025xxh,Caucal:2026ymj,Caucal:2025zkl} and diffractive dijet production in UPCs and DIS~\cite{Boussarie:2014lxa,Boussarie:2016ogo,Iancu:2021rup,Iancu:2022lcw,Hatta:2022lzj,Fucilla:2022wcg,Iancu:2023lel,Hauksson:2024bvv} in the CGC formalism. Starting from \cite{Marquet:2009ca,Dominguez:2011wm}, one of the essential results of this body of work has been to establish that in the appropriate limit where the ``hard'' and ``semi-hard'' momentum scales are strongly separated, the CGC framework leads to  {\it TMD factorization}: the cross-section can be written as a product between a ``hard factor'' describing the hard scattering and transverse momentum dependent parton distributions (TMDs) for partons from the nuclear target. More recently, the NLO corrections to TMD factorization at small $x$ (and notably the emergence of the  Balitsky-Kovchegov (BK)/Jalilian-Marian–Iancu–McLerran–Weigert–Leonidov–Kovner (JIMWLK)~\cite{Balitsky:1995ub,Kovchegov:1999yj,JalilianMarian:1997jx,JalilianMarian:1997gr,Kovner:2000pt,Weigert:2000gi,Iancu:2000hn,Iancu:2001ad,Ferreiro:2001qy}, Dokshitzer– Gribov–Lipatov–Altarelli–Parisi (DGLAP)~\cite{Gribov:1972ri,Altarelli:1977zs,Dokshitzer:1977sg}, and Collins-Soper-Sterman (CSS)~\cite{Collins:1981uk,Collins:1981uw,Collins:1984kg,Collins:2011zzd} evolution equations) have also been investigated in the CGC effective theory~\cite{Taels:2022tza,Hauksson:2024bvv,Mueller:2012uf,Mueller:2013wwa,Caucal:2022ulg,Caucal:2024bae,Caucal:2024vbv,Altinoluk:2024vgg,Caucal:2025mth,Iancu:2025jsu}.

In addition to a high jet momentum, the hard scale can also be provided by a heavy quark mass. While experimentally a proper extraction of a  jet cross section can only happen at several tens of GeV,  charm and beauty quarks can be identified to quite low momenta, while nevertheless being perturbatively calculable. Thus it is often heavy quark observables that give the best kinematical reach into the saturation regime at the LHC.  Accordingly, the $J/\Psi$ is a much studied final state in exclusive processes~\cite{Klein:2019qfb}. Recently the UPC experimental program at the LHC has expanded to inclusive measurements of heavy flavor photoproduction~\cite{CMS:2025jjx,Nese:2025ohz}, which can be addressed both in the CGC~\cite{Gimeno-Estivill:2025rbw} and collinear factorization pictures~\cite{Cacciari:2025tgr}. A natural next step in this experimental program, giving access to the physics of gluon saturation,  would be measurements of heavy flavor in diffractive UPCs. Motivated by this prospect, it is our purpose in this paper to see if there is a diffractive TMD (DTMD) factorization also for massive quarks. This calculation at the parton level would, when complemented by an appropriate description of hadronization, provide a starting point for phenomenological studies of both open charm ($D,B$-meson) and quarkonium ($J/\Psi, \Upsilon$) diffractive photoproduction.

We will in this paper compute the diffractive production of a heavy quark pair with gluon emission in photon–nucleus collisions at high center-of-mass energy in the correlation limit. In this limit,   two partons (either a quark pair or a quark-gluon pair) are hard which, to leading power accuracy, demands that the third emitted parton is semi-hard and soft. (The additional gluon emission suppresses the cross section by an additional factor of $\alpha_s$  relative the leading order diffractive heavy-quark pair production; however, it is leading power in the hard scale as demonstrated in Ref.~\cite{Iancu:2021rup}.) Here we refer to a parton as hard when its mass and/or its transverse momentum either define the hard scale of the process (as is the case for photoproduction $Q^2=0$), or are  of the order of the hard scale given by the photon virtuality $Q^2$ in the case of DIS. A semi-hard parton, on the other hand,  has a transverse momentum of the order of the nuclear saturation scale, $\Qs(Y_\po,A)\equiv\Qs$. 
Besides, a parton is referred to as soft when it carries only a small  fraction of the longitudinal momentum of the probe, $z\ll 1$.
As shown in Ref.~\cite{Iancu:2021rup,Iancu:2022lcw,Hauksson:2024bvv}, the leading power contribution to the cross section arises from the emission of an unobserved soft parton. 
In the limit of large invariant mass of the diffractive system (small $\beta $) is sufficient to consider a purely eikonal emission. On the other hand, for smaller diffractive mass (large $\beta $), the longitudinal momentum fraction $z$ cannot be neglected when it is comparable to the square of the ratio of semi hard scale to hard scale.

In this paper, we will start the calculation in the dipole picture and use the Colour Glass Condensate to describe the saturation regime at small $x$ in the target nucleus. In the CGC formalism the calculation is factorized into a light cone wave function (LCWF) describing the partonic content of the probe, and a Wilson line operator describing the colour field of the target. The former, which we compute using Light Cone Perturbation Theory (LCPT), describes the perturbative splitting of a projectile photon into the quark-antiquark-gluon ($\qqg$) system. %\pc{
This CGC ``factorization" is \textit{a priori} valid for any final-state kinematics for the $q\bar q g$ system, up to power corrections suppressed by the center-of-mass energy (i.e.~subeikonal corrections). This contrasts with TMD factorization, which requires a hierarchy between two transverse scales: a hard scale and a semi-hard one. To derive a factorized form from the CGC cross section, one must carefully perform a power expansion and retain only the leading term in the ratio of the semi-hard to the hard scale. A common feature of this leading-power extraction~\cite{Marquet:2009ca,Iancu:2022lcw,Hauksson:2024bvv,Caucal:2025xxh,Caucal:2025mth} --- when a final state parton must be integrated out --- is that a soft ($z\ll 1$) final state parton in the CGC picture is reinterpreted as a degree of freedom of the target. In practice, a convenient mathematical procedure to ``transfer" this soft CGC parton into the target-like picture is to perform a change of variables from the longitudinal momentum fraction defined with respect to the projectile to one defined with respect to the target.

The TMD factorization of the cross section for diffractive 2+1 jet production has been demonstrated for massless partons in Refs.~\cite{Iancu:2021rup,Iancu:2022lcw, Hauksson:2024bvv}. In this manuscript, we consider massive quark production, which can hadronize into quarkonium or single/double open charm. By setting the virtuality $Q^2$ of the photon to zero and appropriate variables for diffraction, our results can be compared with future experimental data in ultraperipheral collisions (UPCs) at the LHC. Such a comparison is left for future publications.

This paper is organized as follows. In Section~\ref{Sec:kinem}
we present the  variables and kinematics used in diffractive scattering. 
In Section~\ref{Sec: soft gluon} we focus on the diffractive production of a hard quark pair with a soft gluon emission, and factorize the differential cross section into a hard factor and the gluon diffractive TMD distribution, in the limit of small and generic $\beta$.  In Section~\ref{Sec: soft quark} we study the diffractive production of a hard antiquark-gluon pair together with a soft quark. In this case, the TMD factorized cross section contains the quark diffractive TMD distribution. We conclude with a summary in Section~\ref{Sec: summary}. 
Appendix~\ref{sec: LCPT} 
introduces the necessary formulae in Light Cone Perturbation Theory to compute a general differential cross section, with more details on the instantaneous contribution  in Appendix
\ref{sec: instanteous contribution}. Appendices \ref{sec:DTMD-ms}
and \ref{sec:gDTMD ms} present a momentum space version of the calculation, which generalises the results obtained in the main text by allowing for a non-zero transfer of transverse momentum from the target to the projectile. In this framework, we also consider (in Appendix \ref{sec:DTMD-ms}) another interesting process for probing the quark diffractive TMD: the ``semi-inclusive'' scattering in which one measures a single (heavy quark) jet or hadron in the final state, with a transverse momentum much smaller than the photon virtuality. We expect this process to be measured too at the EIC.

\section{Diffractive \texorpdfstring{$\qqg$}{qqg} production in the correlation limit}
\label{Sec:kinem}

We use Light Cone Perturbation Theory~\cite{Brodsky:1997de} to calculate the parton-level differential cross section for the diffractive production of a heavy quark pair accompanied by a gluon in the dipole picture. In this picture, the splitting of a virtual photon into a quark-antiquark dipole, followed by a gluon emission
is described by the photon  light cone wavefunction (LCWF)~\cite{Kovchegov:2012mbw}. We evaluate the LCWF  at leading order in the strong coupling constant $\as$. At high energy the coherence time $t_\gamma$ of the photon fluctuation in the longitudinal direction is larger than the width $L$ of the target nucleus with radius $R_A$: $t_\gamma \simeq 2q^+/Q^2 \gg L=R_A/\gamma$, where $\gamma \gg 1$ is the Lorentz factor for the boosted nucleus. This implies that splitting of the photon  into a colourless dipole occurs long before the scattering with the target nucleus, which is described as a classical gluon field at small $x$ in the Colour Glass Condensate (CGC) framework \cite{Iancu:2002xk,Iancu:2003xm,Gelis:2010nm}. The characteristic feature of a diffractive process is that the interaction of the diffractive $\qqg$ system and the target nucleus with the target happens without transfering colour, through the exchange of what is known as the Pomeron ($\po$).  
  
In light cone coordinates \cite{Dirac:1949cp}, the photon with virtuality $q^2=-Q^2$ moves along the $x^+$-axis with momentum $q^\mu =(q^+,-Q^2/2q^+,\boldsymbol{0})$ and the target nucleus  along the $x^-$-axis with momentum $P_N^\mu=(0,\PN,\boldsymbol{0})$ per nucleon. The partons produced in the diffractive scattering have momenta $k^\mu_i = (k^+_i,k^-_i,\boldsymbol{\mathrm{k}}_i)$, where the label $i=1,2,3$ stands for respectively the quark, antiquark and gluon. The transverse components in bold are two dimensional, with $k_\perp=|\boldsymbol{\mathrm{k}}|$. 

In LCPT, a particle in momentum space is specified by its longitudinal and transverse momentum components: $\vec{k}=(z, \boldsymbol{\mathrm{k}})$, where it is convenient to express the longitudinal momentum as a fraction of the total momentum of the incoming particle $k^+=zq^+$. 
The minus light cone energy is determined by the mass-shell relation,  
\begin{equation}
\label{eq: intro: minus momentum}
\begin{aligned}
    k^-&=\frac{k_\perp^2+m^2}{2k^+}\,,  
\end{aligned}
\end{equation}
where $m$ denotes the quark mass\footnote{In LCPT, all intermediate particles are on-shell but the energy is not conserved at the interaction vertices. This is in contrast with covariant theory, where intermediate particles are off-shell but the energy is conserved \cite{Kovchegov:2012mbw}.}. However, the scattering off the classical gluon field at small-$x$, i.e. in the high energy limit, is more naturally  expressed in transverse coordinates. Therefore, a Fourier transform from transverse momentum space to transverse coordinate space is performed when describing the interaction with the target. In mixed space, a particle is specified by its longitudinal momentum and transverse coordinates: $(z,\boldsymbol{\mathrm{x}})$. The Fourier transform integrals used in this manuscript from momentum space $(z,\boldsymbol{\mathrm{k}})$ to mixed space $(z,\boldsymbol{\mathrm{x}})$ can be found, e.g., in the Appendix of Ref.\,\cite{Beuf:2022ndu}.  

We calculate diffractive $\qqg$ production in a kinematic regime where a hard scale $M_h$ is present, provided either by the large relative transverse momentum between two of the three produced particles, the virtuality of the photon, or the quark mass. More precisely, we shall define $M_h$ as
\begin{align}
    M_h^2&=Q^2+M_{ij}^2\,,
\end{align}
where $M_{ij}^2=(k_i+k_j)^2$ is invariant mass of the measured dijet pair in the final state. At this order of interest, the dijet is made of two partons and therefore $M_{ij}^2$ depends on the mass of these partons. For instance, for a dijet pair made of a $q\bar q$ pair, we have $M_{q\bar q}^2\equiv M_{12}^2\simeq (P_\perp^2+m^2)/(z_1z_2)$ where $\Pb=z_2\boldsymbol{\mathrm{k}}_1-z_1\boldsymbol{\mathrm{k}}_2$ is the relative transverse momentum of the $q\bar q$ pair.
This scale must be hard relative to the semi-hard saturation scale of the nucleus, $M_h\gg Q_s$. As we shall argue, the corresponding typical kinematic configuration is that of a hard dijet system with a transverse momentum imbalance $K_\perp$ much smaller than $M_h$, yet still controlled by $\Qs$, making it strongly sensitive to saturation effects in the nucleus.

The partons forming the measured hard dijet constitute a small-size system in transverse coordinate space and thus effectively behave as a single coloured particle from the point of view of scattering off the target. Furthermore, in this regime, the bulk of the cross section is controlled by configurations in which the third parton carries a small longitudinal momentum fraction $z$, of order $z\sim \Qs^2/M_h^2\ll 1$, and a transverse momentum of order the saturation scale. Indeed, for a coherent process, the sum of the three transverse momenta is very small, of the order of the inverse size of the target nucleus $1/R_A$, so it effectively vanishes. As such, the transverse momentum of the third parton is opposite to the dijet imbalance $K_\perp$, which, as noted above, is controlled by $\Qs$. As a consequence, this third, unmeasured, parton will be far from the two hard partons in transverse coordinate space, forming a large size partonic system that interacts strongly with the dense target.

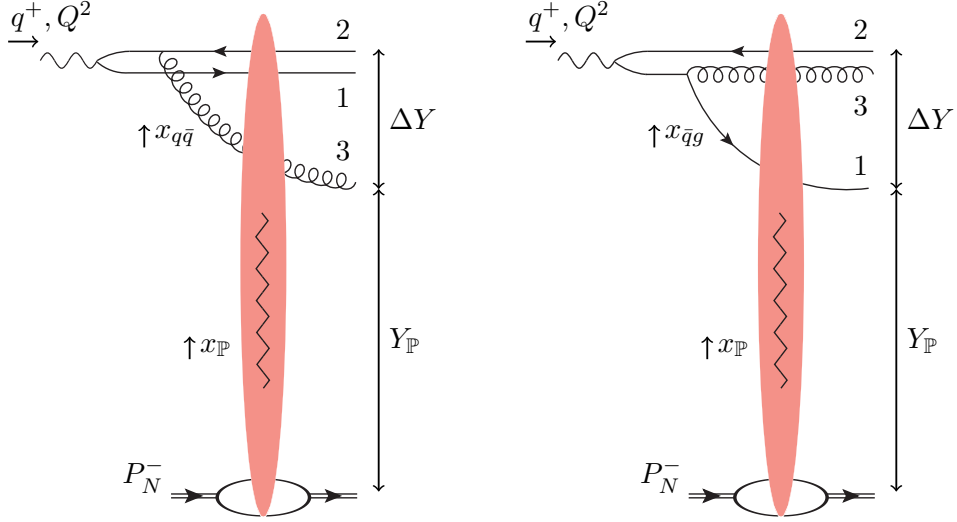
\begin{figure*}[tbp!]
\centerline{\includegraphics[width=0.3\textwidth]{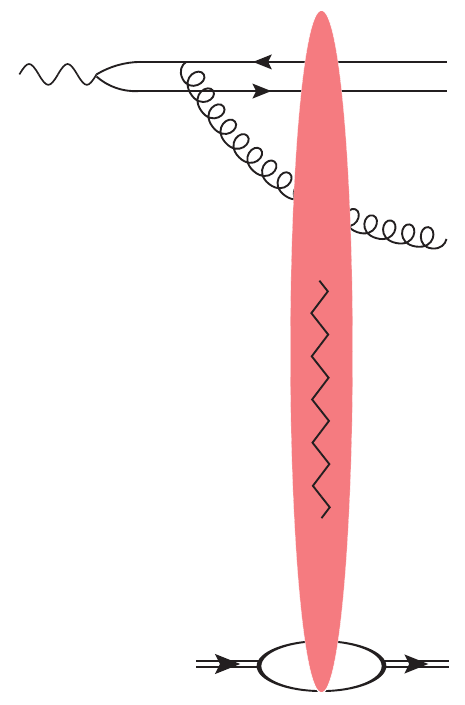}
\begin{tikzpicture}[overlay]
    \draw[line width=0.7pt, to-to](0.0cm,0.45cm) -- (0.0cm,4.45cm) node[midway, right] {$Y_\po$};
    \draw[line width=0.7pt, to-to](0.0cm,4.45cm) -- (0.0cm,6.3cm) node[midway, right] {$\Delta Y$};
    \draw[ line width=0.7pt, -to](-4.9cm,6.4cm) -- (-4.5cm,6.4cm);    
         \node[anchor=south west] at (-0.7cm,4.7cm) {$3$};
         \node[anchor=south west] at (-0.7cm,6.3cm) {$2$};
         \node[anchor=south west] at (-0.7cm,5.4cm) {$1$};
         \node[anchor=south west] at (-5cm,6.4cm) {$q^+,Q^2$};
         \node[anchor=south east] at (-2.7cm,0.3cm) {$\PN$};
    \draw[line width=0.7pt, ->] (-2.5cm,2.2cm) -- (-2.5cm,2.5cm) node[midway, right] {$\xpo$} ;
    \draw[line width=0.7pt, ->](-3.1cm,5cm) -- (-3.1cm,5.3cm) node[midway, right] {$\xqq$}; 
\end{tikzpicture}
\rule{5em}{0pt}
\includegraphics[width=0.3\textwidth]{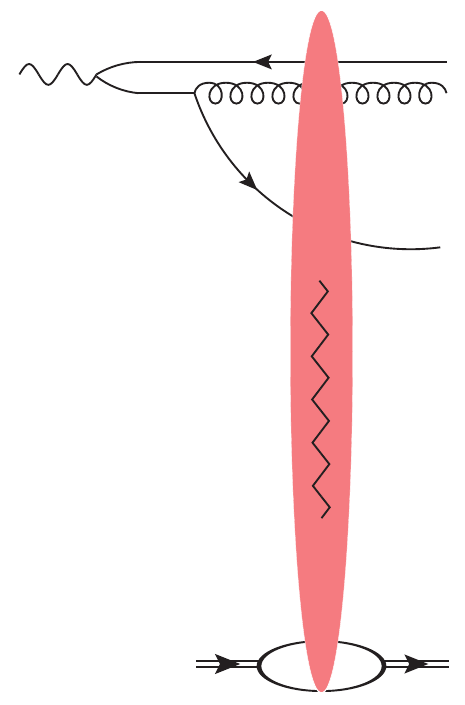}
\begin{tikzpicture}[overlay]
    \draw[ line width=0.7pt, to-to](0.0cm,0.45cm) -- (0.0cm,4.45cm) node[midway, right] {$Y_\po$};
    \draw[ line width=0.7pt, to-to](0.0cm,4.45cm) -- (0.0cm,6.3cm) node[midway, right] {$\Delta Y$};
    \draw[ line width=0.7pt, -to](-4.9cm,6.4cm) -- (-4.5cm,6.4cm);    
    \node[anchor=south west] at (-0.7cm,4.5cm) {$1$};
    \node[anchor=south west] at (-0.7cm,6.3cm) {$2$};
    \node[anchor=south west] at (-0.7cm,5.3cm) {$3$};
    \node[anchor=south west] at (-5cm,6.4cm) {$q^+,Q^2$};
    \node[anchor=south east] at (-2.7cm,0.3cm) {$\PN$};
    \draw[line width=0.7pt, ->] (-2.5cm,2.2cm) -- (-2.5cm,2.5cm) node[midway, right] {$\xpo$} ;
    \draw[line width=0.7pt, ->](-3.2cm,5cm) -- (-3.2cm,5.3cm) node[midway, right] {$\xbarqg$}; 
\end{tikzpicture}
}
\caption{General coherent diffractive $\qqg$ production with a soft gluon (left) and  soft quark (right) emission in the dipole picture. The virtual photon has momentum $q^+$ and virtuality $Q^2$, the target nucleus has momentum $\PN$. The interaction is represented by a salmon blob, and the Pomeron exchange with a zig-zag line. The coordinates, colour and helicity of the produced quark, antiquark and gluon are labelled respectively by the shorthand notation $1,2,3$. The minus longitudinal momentum fractions $\xpo$, $\xqq$ and $\xbarqg$ transferred from the target to the diffractive system and the rapidity gaps $Y_\po$, $\Delta Y$ described in the text are specified.} 
\label{fig: general diffractive scattering}
\end{figure*}

In this correlation limit, we will first calculate the diffractive process shown in the left diagram of \fig\ref{fig: general diffractive scattering}, where the pair of massive quarks are the hard partons and the emitted gluon is semi-hard and soft. Using the dipole picture as the starting point, we will obtain the TMD factorization of the differential cross section between a ``hard'' factor, which will depend on
the quark mass, the relative transverse momentum of $q\bar q$ pair and the virtuality of the photon (the possible hard scales of the process), and the massless diffractive gluon TMD distribution. The first describes the formation of the hard partons and the latter represents the unintegrated (or transverse momentum dependent) gluon distribution of the Pomeron.      

Subsequently, we will calculate the differential cross section for the diagram on the right in \fig\ref{fig: general diffractive scattering}, where the antiquark and gluon are the hard partons, with large relative transverse momentum, and the emitted quark is semi-hard and soft. In this correlation limit, the quark mass cannot be considered as a hard scale with respect to the transverse momentum imbalance of the pair or the saturation scale, and the differential cross section will factorize into a massless hard factor and the mass-dependent diffractive quark TMD distribution. Naturally the contribution where the antiquark is soft gives an equal contribution to the cross section, but can be obtained by symmetry from the soft quark contribution.

In both processes, we consider the target nucleus to remain intact in coherent diffraction. For homogeneous targets in the transverse plane, the transverse momentum transferred from the target to the diffractive $\qqg$ system can be estimated as $\Delta_{\perp}\sim 1/R_A$, with $R_A\approx A^{1/3}R_N$, and $R_N\simeq 1.1$~fm the nucleon-nucleon distance. This implies that for large nucleus, the total transverse momentum can be neglected\footnote{We will consider the effect of non-zero transverse momentum transfer of the target in the Appendices.}: $\kto=\kbq+\kbaq+\kbg\simeq 0$. Therefore, for the first diffractive process where a soft gluon is emitted, the total transverse momentum of the hard quark pair $\Kb\equiv \kbq + \kbaq$  is fixed by the respective momentum of the soft gluon $\Kb \simeq -\kbg$. Similarly, for the second diffractive process where a soft quark is emitted, the total transverse momentum of  the hard partons is $\Kb\equiv \kbaq + \kbg \simeq -\kbq$. 

Let us now introduce the standard variables of diffractive scattering. As it is shown in \fig\ref{fig: general diffractive scattering}, a signature of diffraction is the large rapidity gap $Y_\po=\ln(1/\xpo)$ between the target and the diffractive $\qqg$ system due to the colour neutral exchange.  Here $\xpo$ is the fraction of the target minus longitudinal momentum $\PN$ transferred to the diffractive system:
\begin{equation}
    \xpo\PN=\frac{1}{2q^+}\Bigg(\frac{\ktq^2+m^2}{\zq}+\frac{\ktaq^2+m^2}{\zaq}+\frac{\ktg^2}{\zg}+Q^2 \Bigg)\,,
\end{equation}
which implies,
\begin{equation}
\label{eq: intro: x Pomeron}
    \xpo=\frac{Q^2+\Mqqg^2}{2q\cdot P_N} \,.
\end{equation}

Specializing now to the soft gluon limit
$z_3\ll 1$ in the left diagram of \fig\ref{fig: general diffractive scattering}, the diffractive mass $\Mqqg^2$ of the three parton final state is
\begin{align}
    \label{eq: intro: diffractive mass qqg}
    \Mqqg^2&=(k_1+k_2+k_3)^2= \Bigg(\frac{\ktq^2}{\zq}+\frac{\ktaq^2}{\zaq}+\frac{\ktg^2}{\zg}+\frac{m^2}{\zq\zaq}\Bigg)=\Mqq^2+\Kt^2+\frac{\kg^2}{\zg}\,,
\end{align}
where the conditions $\kto \simeq 0$ and $\zg \ll 1$ and hence $\zq+\zaq\simeq 1$ have been used. For the hard $q\bar{q}$ system alone, the invariant  mass $\Mqq^2$ is
\begin{align}
\label{eq: intro: diffractive mass qq}
    \Mqq^2&=(k_1+k_2)^2 = \Bigg(\frac{\ktq^2}{\zq}+\frac{\ktaq^2}{\zaq}+\frac{m^2}{\zq\zaq}-\Kt^2\Bigg)\,.
\end{align}

The rapidity interval $\ln(1/\beta)$ represents the longitudinal phase space for developing the partonic structure of the projectile. Combined  with the rapidity gap $ Y_\po= \ln 1/\xpo$ between the diffractive system and the target it makes up the total rapidity interval available for the process, given by the Bjorken-$x$ ($\xbj$) variable $Y_\bj \equiv \ln (1/\xbj)$, as   $ Y_\bj =  \ln(1/\beta) + Y_\po$.
 In terms of the target momentum fractions the $\beta$ variable 
\begin{equation}
\label{eq: intro: beta}
    \beta\equiv\frac{Q^2}{Q^2+\Mqqg^2}\,,
\end{equation}
is interpreted as the momentum fraction of the Pomeron carried by the struck parton, which is related to Bjorken-$x$  as $\xbj = \xpo \beta$.
In the left diagram of \fig\ref{fig: general diffractive scattering}, the rapidity difference $\Delta Y=\ln(\xpo/\xqq)$ between the Pomeron and the hard $\qbarq$ pair is calculated with the minus longitudinal momentum fraction transferred to the hard $q\bar{q}$ system:
\begin{equation}
    \label{eq: intro: variable xqq}
    \xqq=\frac{1}{2q^+\PN}\Bigg(\frac{\ktq^2}{\zq}+\frac{\ktaq^2}{\zaq}+Q^2 +\frac{m^2}{\zq\zaq}\Bigg)=\frac{Q^2+\Mqq^2+\Kt^2}{2P_N\cdot q}\,.
\end{equation}
In order to obtain the diffractive gluon and quark TMD distributions, we will need to express the differential cross section calculated in the dipole picture as a function of the minus longitudinal momentum fractions $\xpo$, $\xqq$, $\xbarqg$ in the target frame. The ratio 
\begin{equation}
\label{eq: intro: variable x}
    x \equiv \frac{\xqq}{\xpo}\simeq\frac{Q^2+\Mqq^2+\Kt^2}{Q^2+\Mqqg^2}=\frac{Q^2+\Mqq^2+\Kt^2}{Q^2+\Mqq^2+\Kt^2+\ktg^2/\zg}\,,
\end{equation}
represents the momentum fraction of the gluon with respect to the Pomeron. In this interpretation, the soft gluon is now seen as emitted by the Pomeron instead of the $\qbarq$ pair~\cite{Hebecker:1997gp,Buchmuller:1998jv,Iancu:2021rup,Beuf:2022kyp,Iancu:2022lcw}. 

Similarly, for the soft quark emission represented by the right diagram in \fig\ref{fig: general diffractive scattering}, the momentum $x \xpo\PN$ carried by the t-channel virtual quark in the target frame gets transferred to the hard part of the amplitude and puts the final $\barqg$ on-shell. This argument implies $x\xpo=\xbarqg$, with 
\begin{equation}
\begin{aligned}
    \xbarqg&=\frac{1}{2q^+\PN} \Big(\frac{\ktaq^2+m^2}{\zaq}+\frac{\ktg^2}{\zg}+Q^2 \Big)\,.
\end{aligned}
\end{equation}

The change of variables from \textit{plus} longitudinal momentum fractions $z$ to \textit{minus} longitudinal momentum fractions $x$ represents a change of frame from the dipole picture to the target frame. In the first, the soft parton belonging to the photon wavefunction is emitted by the hard system. In the latter, the soft parton is seen as emitted by the Pomeron within the target wavefunction. 

\section{Gluon diffractive TMD}
\label{Sec: soft gluon}

In this section we calculate the {\it leading-twist} contribution to the cross section for the diffractive production of a {\it hard} quark-antiquark ($q\bar q$) pair --- the ``hard dijet''. 

As explained in the Introduction, by ``hard'' we mean that at least one of the three following transverse momentum scales is much larger than the nuclear saturation momentum $Q_s(\xpo, A)$ evaluated for a rapidity evolution which covers the rapidity gap  $Y_\po=\ln(1/\xpo)$:  \texttt{(i)} the relative transverse momentum 
$P_\perp$ of the produced $q\bar q$ pair,  \texttt{(ii)} the photon virtuality $Q\equiv \sqrt{Q^2}$, and  \texttt{(iii)} the (anti)quark mass $m$: $\textrm{max}(P_\perp^2,\, Q^2,\, m^2)\gg \Qs^2(\xpo, A)$. It turns out that the following combination of (potentially) hard scales
\begin{equation}\label{Mh}
M_h^2\,\equiv\,Q^2+M_{q\bar q}^2\simeq Q^2+\frac{P_\perp^2 +m^2}{z_1 z_2}\,,
\end{equation}
naturally enters the kinematics and the calculations, so we shall succinctly define our criterion for {\it hardness} as the condition $M_h^2\gg \Qs^2(\xpo, A)$.  When this condition is satisfied, we will find that the dijet cross section favours values $P_\perp$ which are hard as well, $P_\perp^2\gg \Qs^2$, so this is the most interesting case in practice.

\begin{figure*}[tbp!]
\centering{
\includegraphics[width=0.4\textwidth]{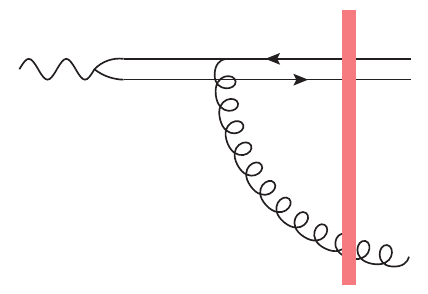}
\begin{tikzpicture}[overlay]
         \node[anchor=south east] at (-4.1cm,3.4cm) {$2'$};
         \node[anchor=south east] at (-1.9cm,3.4cm) {$2''$};
         \node[anchor=south east] at (-4.1cm,2.5cm) {$1'$};
         \node[anchor=south east] at (-0.5cm,3.4cm) {$2$};
         \node[anchor=south east] at (-0.5cm,2.5cm) {$1$};
         \node[anchor=south east] at (-0.5cm,0.9cm) {$a$};
         \node[anchor=south east] at (-1.9cm,1.6cm) {$b$};
\end{tikzpicture}
\rule{2.5em}{0pt}
\includegraphics[width=0.4\textwidth]{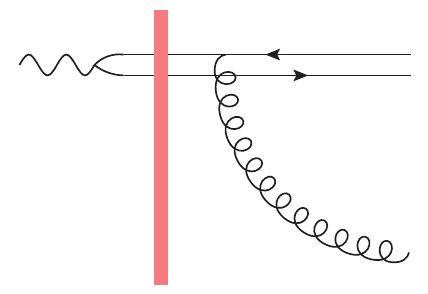}
\begin{tikzpicture}[overlay]
         \node[anchor=south east] at (-4.1cm,3.4cm) {$2'$};
         \node[anchor=south east] at (-3.1cm,3.4cm) {$2''$};
         \node[anchor=south east] at (-4.1cm,2.5cm) {$1'$};
         \node[anchor=south east] at (-0.5cm,3.4cm) {$2$};
         \node[anchor=south east] at (-0.5cm,2.5cm) {$1$};
         \node[anchor=south east] at (-0.5cm,0.9cm) {$a$};
\end{tikzpicture}
}
\caption{Amplitude for soft gluon emission from the antiquark before (left) and after (right) the interaction with the target, represented as an orange band. The labels $i=1,2$ denote the quark colour indices $\alpha_i$, with primes for intermediate quark states. The indices $a$ and $b$ denote the gluon colour after and before the interaction, respectively.
}
\label{fig: scattering soft gluon}
\end{figure*}
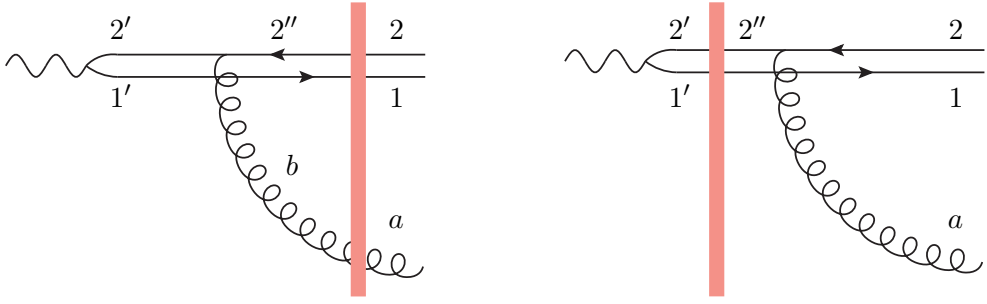

By ``leading twist'' (LT) we mean the contribution to the dijet cross section which decreases like $1/M_h^4$ when $M_h\to\infty$ (in practice, when  $M_h\gg Q_s$). As explained in Sect.~\ref{Sec:kinem}, to leading order in $\alpha_s$, the LT contribution is generated via ``2+1 jet'' configurations in which the $q\bar q$ pair is accompanied by a ``semi-hard'' gluon --- that is, a gluon with transverse momentum $\ktg\sim Q_s$ and small longitudinal momentum fraction $\zg\ll 1$. (Our subsequent calculations will show that, typically, $z_3 \lesssim \Qs^2/M_h^2\ll 1$.) Accordingly, the transverse momentum imbalance $K_\perp$ of the $q\bar q$ dijet is semi-hard as well: $K_\perp\simeq \ktg \sim Q_s$. Hence, in the particularly interesting case where $P_\perp\gg Q_s$, we also have $P_\perp\gg K_\perp$ and then the $q\bar q$ pair propagates nearly back-to-back in the transverse plane --- a configuration known as  ``the correlation limit''~\cite{Dominguez:2011wm}.

Our general strategy for the calculation follows the formalism laid out in more detail in in Appendix~\ref{sec: LCPT}: we will first compute the light cone wave function $\Psi^{\gamma^*\rightarrow \qqg}$ representing the quark-antiquark-gluon ($q\bar q g$) Fock-space component of the incoming photon (cf. the Fock state expansion in \eq\eqref{eq: LCPT: Fock state photon}), for a similar calculation see \cite{Beuf:2021qqa,Beuf:2022ndu,Kang:2023doo}. Then, we will calculate the scattering matrix element in \eq\eqref{eq: LCPT: NLO amplitude} and include the Wilson Lines \eqref{eq: LCPT: Definition Fund. WL} and \eqref{eq: LCPT: Definition Adj. WL} that represent the scattering with the target gluon field. We will then implement approximations based on the separation of scales
$M_h^2\gg \Qs^2$, with the purpose of retaining only the LT contribution to the cross section. Finally, via an appropriate change of longitudinal variables, we shall demonstrate that this LT contribution exhibits TMD factorization in terms of a {\it hard factor} --- the same as for {\it inclusive} quark-antiquark production with massive quarks~\cite{Dominguez:2011wm} --- and the {\it gluon diffractive TMD} --- the same as originally obtained in Refs.~\cite{Iancu:2022lcw,Hatta:2022lzj}.

\subsection{The photon light cone wavefunction}

In light cone perturbation theory (LCPT), there are {\it a priori} two possible time orderings for the amplitude describing quark-antiquark-gluon  ($q\bar q g$) production in photon-nucleus ($\gamma^* A$) DIS: the gluon can be emitted (by the quark, or by the antiquark) either {\it before}, or {\it after}, the interaction with the nuclear target --- a shockwave localised at LC time $x^+=0$. These two time-orderings are illustrated in Fig.~\ref{fig: scattering soft gluon}. For {\it diffractive} production in the kinematics of interest --- i.e. for 2+1 jets where the $q\bar q$ pair is hard and accompanied by a semi-hard gluon ---, the  leading twist contribution to the cross section comes exclusively from the first diagram, where the gluon is emitted before crossing the shockwave and hence  can interact with it. Indeed, only in this case the transverse extent of the colliding system is large enough (namely, of order $R\sim 1/\Kt\sim 1/Q_s$) to allow for strong scattering and thus avoid the suppression due to colour transparency (which is particularly severe for elastic scattering). Accordingly, in what follows we shall concentrate on the left graph  in Fig.~\ref{fig: scattering soft gluon} {\it alone}.

To start with, we use LCPT to  compute the quark-antiquark-gluon ($q\bar q g$) Fock-space component of the light cone wavefunction (LCWF) of the virtual photon at LC time $x^+=0$ but in the absence of the collision. The effects of the scattering with the nuclear target will be added in a subsequent subsection.

\subsubsection{The general structure}

The LCWF $\Psi^{\gamma^*\rightarrow \qqg}$ describes the splitting $\gamma^*\rightarrow q\bar q$ of the virtual photon followed by a gluon emission by any of the two fermions (see Fig.~\ref{fig: soft gluon emission}). As explained in Appendix~\ref{sec: LCPT}, it includes both regular and instantaneous contributions, corresponding to the non-local (in LC time) and the instantaneous piece of the intermediate fermion propagator, respectively:
\begin{equation}
    \Psi^{\gamma^* \rightarrow \qqg} = \Psi ^{\gamma^* \rightarrow \qqg}_{\text{reg}}+ \Psi ^{\gamma^* \rightarrow \qqg}_{\text{inst}} \,.
\end{equation}
In the regular piece, the gluon can be emitted by either the quark or the antiquark,  as 
shown in \fig\ref{fig: soft gluon emission}, so we have two contributions: $\Psi_{\text{reg}}^{\gamma^*\rightarrow \qqg} = \Psi^{\hyperlink{diag:a}{(a)}} + \Psi^{\hyperlink{diag:b}{(b)}}$, with   

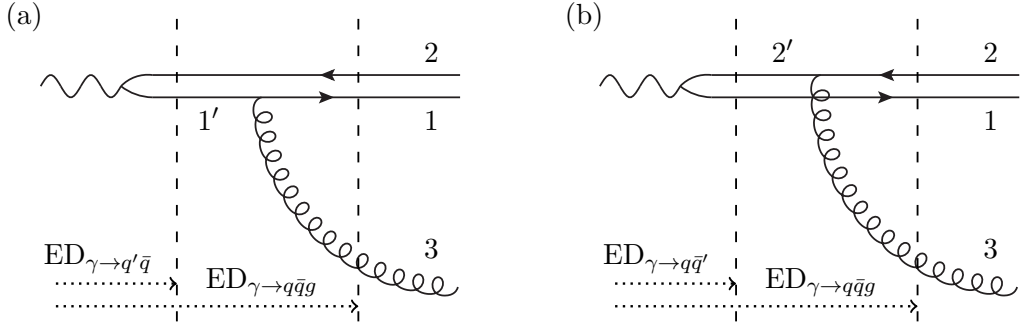
\begin{figure*}[t]
\centerline{
\includegraphics[width=0.4\textwidth]{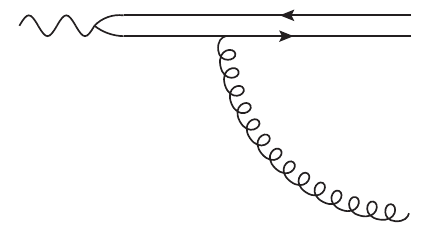}
\begin{tikzpicture}[overlay]
\node[anchor=south west] at (-6.5cm,3.5cm) {\hypertarget{diag:a}{(a)}};
    \draw[dash pattern=on 4pt off 6pt,line width=0.7pt, -](-4.1cm,3.8cm) -- (-4.1cm,-0.2cm);
    \draw[dash pattern=on 4pt off 6pt,line width=0.7pt, -](-1.7cm,3.8cm) -- (-1.7cm,-0.2cm);
    \draw[dotted, line width=0.9pt, -to](-5.7cm,0.3cm)-- (-4.1cm,0.3cm);
    \draw[dotted, line width=0.9pt, -to](-5.7cm,0cm) -- (-1.7cm,0cm);
         \node[anchor=south east] at (-3.4cm,2.2cm) {$1'$};
         \node[anchor=south east] at (-0.5cm,3.1cm) {$2$};
         \node[anchor=south east] at (-0.5cm,2.2cm) {$1$};
         \node[anchor=south east] at (-0.5cm,0.5cm) {$3$};
         \node[anchor=south east] at (-4.3cm,0.3cm) 
         {ED$_{\gamma \rightarrow q'\bar{q}}$};
        \node[anchor=south east] at (-2.1cm,0cm) {ED$_{\gamma \rightarrow\qqg}$};
\end{tikzpicture}
\rule{2.5em}{0pt}
\includegraphics[width=0.4\textwidth]{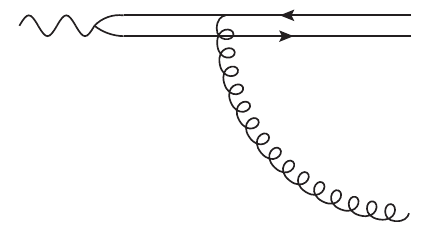}
\begin{tikzpicture}[overlay]
\node[anchor=south west] at (-6.5cm,3.5cm) {\hypertarget{diag:b}{(b)}};
    \draw[dotted, line width=0.9pt, -to](-5.7cm,0.3cm)-- (-4.1cm,0.3cm);
    \draw[dotted, line width=0.9pt, -to](-5.7cm,0cm) -- (-1.7cm,0cm);
    \draw[dash pattern=on 4pt off 6pt,line width=0.7pt, -](-4.1cm,3.8cm) -- (-4.1cm,-0.2cm);
    \draw[dash pattern=on 4pt off 6pt,line width=0.7pt, -](-1.7cm,3.8cm) -- (-1.7cm,-0.2cm);
         \node[anchor=south east] at (-3.2cm,3.1cm) {$2'$};
         \node[anchor=south east] at (-0.5cm,3.1cm) {$2$};
         \node[anchor=south east] at (-0.5cm,2.2cm) {$1$};
         \node[anchor=south east] at (-0.5cm,0.5cm) {$3$};
         \node[anchor=south east] at (-4.3cm,0.3cm) 
         { ED$_{\gamma \rightarrow q\bar{q}'}$};
        \node[anchor=south east] at (-2.1cm,0cm) {$\EDqqg$};
\end{tikzpicture}
}
\caption{Soft gluon emission from (a) the quark and (b) the antiquark. Particle momenta are labeled using numbers as a shorthand notation, with a prime to denote intermediate-state particles. The dotted arrow indicates the time ordering of the energy denominator before the intermediate particles cross the shockwave.}
\label{fig: soft gluon emission}
\end{figure*}
\begin{align}
      \label{eq: soft gluon: emission LCWF A}
   \Psi^{\hyperlink{diag:a}{(a)}}&=\sum_{q'}\frac{ee_f\bar{u}(\vec{\kq'}) \slashed{\varepsilon}_\lambda(\vec{q})v(\vec{\kaq})}{\EDqpq} \frac{(-gt^a)\bar{u}(\vec{\kq})\slashed{\varepsilon}_\sigma^*(\vec{\kg})u(\vec{\kq'})}{\EDqqg} \, , \\ 
\label{eq: soft gluon: emission LCWF B}
   \Psi^{\hyperlink{diag:b}{(b)}}&= \sum_{\bar{q}'}\frac{ee_f\bar{u}(\vec{\kq}) \slashed{\varepsilon}_\lambda(\vec{q})v(\vec{\kaq'})}{\EDqqp}\frac{gt^a\bar{v}(\vec{\kaq'})\slashed{\varepsilon}_\sigma^*(\vec{\kg})v(\vec{\kaq})}{\EDqqg }  \,.
\end{align}
Here $\vec{k}=(k^+,\boldsymbol{k})$, and $\polp$ and $\polg$ denote the polarizations of the photon and gluon, with $\lambda,\sigma = \pm$ (see below for more details on the polarization vectors). The first factor in \eqs\eqref{eq: soft gluon: emission LCWF A} and \eqref{eq: soft gluon: emission LCWF B} represents the photon splitting and the  second factor the gluon emission. We explicitly write the relative minus between the emission from the quark and from the antiquark in the LCPT diagrammatical rules, and denote with a prime the variables of the intermediate quark and antiquark. For the phase-space integrals we use the following shorthand notations:
\begin{align}
    &\sum_{q'}\equiv\sum_{\hq'\alpha'_1}\int\frac{\der^3 \vec{\kq'}}{2k'^+_1(2\pi)^3}(2\pi)^3\delta^3(\vec{\kq'}-\vec{\kq}-\vec{\kg})=\sum_{\hq'\alpha'_1}\frac{1}{2(\zq+\zg)q^+}\simeq
    \sum_{\hq'\alpha'_1}\frac{1}{2q^+\zq} \, , \\
    &\sum_{\bar{q}'}\equiv\sum_{\haq'\alpha'_2}\int\frac{\der^3 \vec{\kaq'}}{2k'^+_2(2\pi)^3}(2\pi)^3\delta^3(\vec{\kaq'}-\vec{\kaq}-\vec{\kg})=\sum_{\haq'\alpha'_2}\frac{1}{2(\zaq+\zg)q^+}\simeq\sum_{\haq'\alpha'_2}\frac{1}{2q^+\zaq} \,,
\end{align}
where $h^\prime_{1,2}$ and $\alpha^\prime_{1,2}$ are helicity and respectively colour indices for the intermediate fermion.

In writing the approximate equalities above, we anticipated  the fact that the relevant partonic configurations are such that the gluon is soft, $\zg\ll 1$, which allowed us to neglect $\zg$ next to $\zq$ or $\zaq$. That said, $\zg$ is not {\it arbitrarily} small --- as we shall see, it takes a typical value  $z_3 \sim \Qs^2/M_h^2$, with $M_h$ the hard scale introduced in \eqn{Mh} ---, hence some care must be taken when performing approximations which rely on the smallness of $z_3$: such approximations must be correlated with the hierarchy of transverse momentum scales in the problem. That is, $\zg$ can be neglected only when its effects are not amplified by a large ratio of transverse scales. With this in mind, the energy denominators in \eqs\eqref{eq: soft gluon: emission LCWF A} and \eqref{eq: soft gluon: emission LCWF B} can be estimated as follows
\begin{subequations}
\begin{align}   
    \EDqpq&=q^- - {k'_1}^{-} -k_2^{-} 
    =-\frac{1}{2q^+\zq\zaq}\big(\ktaq^2+\Qbar^2 +m^2 \big) \,,
    \\*[0.2cm] 
    \EDqqp&=q^- - k_1^{-} -{k'_2}^{-}  
    =-\frac{1}{2q^+\zq\zaq}\big(\ktq^2+\Qbar^2 +m^2 \big) \,,
   \\*[0.2cm] 
     \EDqqg&=q^- - k_1^{-} -k_2^{-}  -k_3^{-} 
    =-\frac{1}{2q^+\zg} \big(\ktg^2 + \Omega^2\big) \,,  \label{eq: soft gluon: energy denominator qqg}
     \end{align}
\end{subequations}
where we repeatedly used $z_1+z_2\simeq 1$ and   defined $\Qbar^2\equiv \zq\zaq Q^2$ together with
\begin{equation}
\label{eq: rel. soft: Omega}
\begin{aligned}
    \Omega^2&\equiv\frac{\zg}{\zq\zaq}\bigg(\zaq(\ktq^2+m^2)+\zq(\ktaq^2+m^2)+\Qbar^2\bigg)
    \simeq \frac{\zg}{\zq\zaq} \big(\Pt^2+\Qbar^2+m^2\big) \,.
\end{aligned}
\end{equation}
The structure of the last denominator in \eqn{eq: soft gluon: energy denominator qqg} shows that the scale $\Omega^2$ plays the role of an effective virtuality (``screening mass'') for the gluon emission. This scale is related to the hard scale in \eqn{Mh} via $\Omega^2=z_3 M_h^2$.

Using the Feynman rules of LCPT for evaluating the spinor matrix elements which appear in \eqs\eqref{eq: soft gluon: emission LCWF A} and \eqref{eq: soft gluon: emission LCWF B}, one finds the following result for the 
regular piece of the $q\bar q g$ LCWF~\cite{Lappi:2016oup, Beuf:2021qqa, Beuf:2022ndu}:%
\begin{multline}
\label{eq: rel. soft: regular LCWF}
    \Psi^{\gamma^* \rightarrow \qqg}_\text{reg}(\zq,\zaq,\zg,\kbq,\kbaq,\kbg)=\sum_{h'_1h'_2}2q^+ee_fgt^a\sqrt{\zq\zaq}\, \frac{2\polg^{*m}}{\kbg^2+\Omega^2}\delta_{\hq\hq'}\delta_{\haq\haq'} \\
    \cross \polp^i\Bigg\{ \frac{\Big(k_3+\frac{\zg}{\zq}k_2\Big)^m \Big( \varphi^{ij}(\zq,h'_1) (k_1+k_3)^j\delta_{h'_1-\haq} - \sqrt{2}m\varepsilon^{i*}_{h'_1} \delta_{h'_1\haq} \Big) }{ (\kbq+\kbg)^2 +\Qbar^2+m^2} \\
    - \frac{\Big(k_3+\frac{\zg}{\zaq} k_1\Big)^m\Big( \varphi^{ij}(\zq,\hq) k_1^j\delta_{\hq-h'_2} - \sqrt{2}m\varepsilon^{i*}_{h'_2} \delta_{\hq h'_2}\Big) }{\kbq^2+\Qbar^2+m^2} \Bigg\} \,.
\end{multline}
Although not explicitly shown by our above notations, this LCWF also depends upon the polarisation indices $\lambda$ (for the photon) and $\sigma$ (for the gluon) and upon the colour indices $a, \,\alpha,\,\beta$ for the gluon, quark, and antiquark, respectively. This wave function is multiplied by the product $\delta(1-z_1-z_2)\delta^{(2)}(\kbq+\kbaq+\kbg)$ of $\delta$--functions for longitudinal and transverse   momentum conservation, which in our conventions is included in the phase space measure, as discussed in more detail in Appendix~\ref{sec: LCPT}.

The first term in the parenthesis describes gluon emission from the quark, with $\kbq+\kbg = -\kbaq$. The second term, with a relative minus sign, corresponds to emission from the antiquark. The helicity structure of the massless part of the photon vertex is encoded in the function
\begin{equation}
    \label{eq: soft gluon: phi function helicity decomposition}
    \varphi^{ij}(z,h)=(2z-1)\delta^{ij} + ih\varepsilon^{ij} .
\end{equation} The mass terms involve the quark polarization $\varepsilon^*_h$, with helicity $h=\pm$. 

Let us now open a parenthesis on the polarization structure of the LCWF in terms of the light front helicities for gluons and massive quarks in the present framework. In LCPT, we work in terms of physical polarization states, where the most natural basis for the spin degrees of freedom is given by the light front helicity~\cite{Soper:1972xc}. The light front helicities of both gauge particles and fermions can be represented in terms of two-dimensional polarization vectors
\begin{equation}
\label{eq: intro: 2D circular polarization}
\begin{aligned}
    \boldsymbol{\varepsilon}_{\pm } = \frac{1}{\sqrt{2}}  \begin{pmatrix} \mp1\\-i\end{pmatrix}\,,
\end{aligned}
\end{equation}
where $\alpha=\pm$ are the two light front helicities. Note that especially for massive quarks it is more transparent to use circular rather than linear polarization states, since they have an easier interpretation in terms of Clebsch-Gordan coefficients and the usual algebra of angular momentum addition in quantum mechanics.
 The two dimensional transverse polarizations $\boldsymbol{\varepsilon}_{\alpha=\pm}$ satisfy:%
\begin{subequations}
\label{eq: intro: polarization identities}
\begin{align}
    \boldsymbol{\varepsilon}_\alpha \cdot \boldsymbol{\varepsilon}_\beta^* &= \delta_{\alpha\beta} \,, \\
    \sum_\alpha \boldsymbol{\varepsilon}_\alpha^i \boldsymbol{\varepsilon}_\alpha^{j*} &= \delta^{ij} \, , \\
    \sum_\alpha \alpha\boldsymbol{\varepsilon}_\alpha^i \boldsymbol{\varepsilon}_\alpha^{j*} &= -i\varepsilon^{ij}\,,
\end{align}
\end{subequations}
where $\varepsilon^{ij}$ is the Levi-Civita symbol $\varepsilon^{12}=-\varepsilon^{21}=1$.
For gauge particles we work in the light cone gauge $\varepsilon^+=0$, where the full 4-dimensional polarization vector of the transversely polarized virtual photons and gluons are expressed in terms of the 2-dimensional polarization vectors as
\begin{equation}
\label{eq: intro: circular polarization}
\begin{aligned}
    \varepsilon_{\alpha}^\mu(p)&=\bigg(0, \frac{\boldsymbol{\varepsilon}_\alpha^i p^i}{p^+}, \boldsymbol{\varepsilon}_\alpha \bigg).
\end{aligned}
\end{equation}

Returning to \eqn{eq: rel. soft: regular LCWF}, we note the different helicity structures in the massless and massive parts of the photon splitting vertex. The massless part $\delta_{\hq'-\haq} $ or $\delta_{\hq-\haq'}$ conserves the helicity on the quark line\footnote{Note that fermion helicity conservation in the massless part of a photon-to-fermion-antifermion splitting vertex corresponds to the fermion and antifermion having opposite helicity. In terms of angular momentum, the spin 1 of the transverse photon goes into the orbital angular momentum of the quark-antiquark pair, while the spins of the quark and antiquark are antiparallel. In the part of the vertex proportional to the mass, on the other hand, the vertex is not proportional to the transverse momentum, i.e. there is no orbital angular momentum, and consequently the spin 1 of the photon has to be made up of the parallel spins $1/2$ of the quark and antiquark.}, and the polarization vector of the photon is contracted with the relative transverse momentum in the splitting via the structure $\varphi^{ij}$. The massive part of the vertex, on the other hand, has a helicity flip structure $\delta_{\hq' \haq}$ or $\delta_{\hq \haq'}$. In the massive part of the vertex the polarization of the photon $\polp^i$ is contracted with the quark or antiquark polarization vector $\polp^i \varepsilon_{\hq'}^{i*} = \delta_{\lambda \hq'}$ or 
$\polp^i \varepsilon_{\haq'}^{i*} = \delta_{\lambda \haq'}$. This means that for a fixed photon helicity, only one of the two quark helicity states can contribute, although we keep the quark polarization vector explicit here because it makes it easier to track the algebra in the later stages of the calculation. As a consequence of the helicity structure we also already know at this level that there cannot be any interference between the massless and $\sim m$ parts of the wavefunction.  Consequently, since the interaction with the target is eikonal, there cannot be any terms linear in $m$ in the cross section. In the exact kinematics the corresponding expression is more complicated, since also the gluon emission vertex has a helicity flip part proportional to the mass~\cite{Beuf:2022ndu}. Here, this term does not contribute because it is suppressed at $z_3 \ll 1$ without a compensating large ratio of hard to semihard transverse momenta, but it will in the soft quark limit in Sec.~\ref{Sec: soft quark}. In any case, because states with different helicity quarks in the final state cannot interfere, the cross  section will only involve even powers of $m$.

\subsubsection{The correlation limit}

At this point, it is convenient to replace the transverse momenta $\kbq$ and $\kbaq$ of the hard
partons with their linear combinations $\Pb$ (the dijet relative momentum) and $\Kb$ (the dijet imbalance) introduced in Sect.~\ref{Sec:kinem}, that is,
\begin{align}
    \label{eq: definition relative momentum and imbalance}
    \Pb\equiv \frac{\zaq\kbq-\zq\kbaq}{\zq+\zaq}\simeq \zaq\kbq-\zq\kbaq \, , \qquad
    %\label{eq: soft gluon: imbalance}
    \Kb \equiv \kbq+\kbaq \,,
\end{align}
which implies $ \kbq\simeq\Pb+ \zaq\Kb=\Pb- \zaq\kbg$ and $\kbaq\simeq -\Pb+ \zq\Kb
=-\Pb- \zq\kbg$, where we also used the fact that $\Kb \simeq -\kbg$ since the total momentum transfer is small.
This change of variables is useful as it allows us to implement the simplifications associated with the our general assumptions $\Pt\simeq \ktq\simeq\ktaq \gg \ktg$ and $z_3\ll 1$. To deduce the relevant simplifications, consider first the terms independent of the quark mass in the accolade in \eqn{eq: rel. soft: regular LCWF}, that is (we omit irrelevant factors and denote $\Qtil^2\equiv\Qbar^2+m^2$):
\begin{align}\label{expansion z3}
\frac{\Big(k_3^m+\frac{\zg}{\zq}k_2^m\Big) k_2^j}{ \ktaq^2 +\Qtil^2}    + \frac{\Big(k_3^m +\frac{\zg}{\zaq} k_1^m \Big)k_1^j}{k_{1\perp}^2+\Qtil^2} \,.
\end{align}
We need to isolate the leading order contributions to \eqn{expansion z3} in the simultaneous expansion in powers of $z_3$ and of $\ktg/\Pt$. In general, and despite the fact that $z_3\ll 1$, one cannot ignore the pieces proportional to $z_3$ within the emission vertices  (e.g. one cannot approximate $k_3^m+ ({\zg}/{\zq})k_2^m\simeq k_3^m$) because we also have $ \ktg\ll \ktq, \ktaq$. The analysis is furthermore complicated by the fact that the dominant contributions to each of the two terms in \eqn{expansion z3} cancel in their sum, as we shall see.

For the contributions that are explicitly linear  in $z_3$, one can already replace $ \kbq\simeq\Pb
\simeq -\kbaq$, to deduce
\begin{align}\label{linz3}
\frac{\zg}{\zq}\frac{k_2^m k_2^j}{ \ktaq^2 +\Qtil^2}    + 
\frac{\zg}{\zaq} \frac{k_1^m k_1^j}{k_{1\perp}^2+\Qtil^2} \,\simeq\,
\frac{z_3}{z_1z_2}\frac{P^mP^j}{\Pt^2+\Qtil^2}\,=\,\Omega^2\frac{P^mP^j}{(\Pt^2+\Qtil^2)^2}\,,
\end{align}
where we also used \eqn{eq: rel. soft: Omega} to replace $z_3$ in the r.h.s. For the remaining terms in \eqn{expansion z3}, the would-be dominant terms cancel in the difference, as anticipated, so we must extract the sub-dominant contributions, with the following result:
\begin{align}\label{expden}
k_3^m\left[\frac{-P^j -\zq k_3^j}{(\Pb+ \zq\kbg)^2+\Qtil^2}+
\frac{P^j -\zaq k_3^j}{(\Pb- \zaq\kbg)^2+\Qtil^2}\right]
 \,\simeq\,- \frac{k_3^m k_3^{n}}{P_{\perp}^2 +\Qtil^2}
	\left( 
	\delta^{jn} - \frac{2 P^j P^n}{P_{\perp}^2 +\Qtil^2}
	\right)\,.
\end{align}
The two contributions in \eqs\eqref{linz3}-\eqref{expden} are parametrically of the same order when $\Omega^2\sim \ktg^2$, or $z_3\sim \ktg^2/M_h^2$, which is the most interesting regime for the problem at hand. If, however, $z_3$ is {\it much} smaller, i.e. $z_3\ll \ktg^2/M_h^2$, then the non-eikonal contribution
in \eqn{linz3} becomes negligible and we are left with the piece in \eqn{expden}~\cite{Iancu:2021rup}.

The corresponding treatment of the terms involving the quark mass  in \eqn{eq: rel. soft: regular LCWF} is analogous and results in
\begin{align}
    \frac{\Big(k_3^m+\frac{\zg}{\zq}k_2^m\Big)}{ \ktaq^2 +\Qtil^2}    - \frac{\Big(k_3^m +\frac{\zg}{\zaq} k_1^m \Big)}{k_{1\perp}^2+\Qtil^2} \simeq -\left[ \frac{2 k_3^m k_3^{n}+\delta^{mn} \Omega^2}{(P_{\perp}^2 +\Qtil^2)^2} \right] P^n \;.
\end{align}

Putting together the previous results, one finds the following approximation for the regular piece of the LCWF:
\begin{multline}
\label{eq: rel. soft: final regular contribution}
    \Psi^{\gamma^* \rightarrow \qqg}_{\text{reg}}(\zq,\zaq,\zg,\Pb,\kbg)=2q^+ee_fgt^a\sqrt{\zq\zaq}
    \varepsilon^i_\lambda\varepsilon_\sigma^{*m} \\
    \cross \Bigg\{\frac{2k_3^m k_3^n}{\ktg^2+\Omega^2} 
     \Bigg[  \varphi^{ij}(\zq,\hq)\frac{\delta_{\hq-\haq} }{\Pt^2 + \Qtil^2} \left( \delta^{jn} - \frac{2 P^j P^n}{\Pt^2 + \Qtil^2}\right) +  2\sqrt{2}m\varepsilon_{\hq}^{i*}  \frac{P^n \delta_{\hq\haq}}{\big(\Pt^2 + \Qtil^2\big)^2} \Bigg]\\
    -\frac{2\Omega^2 \delta^{mn}}{\ktg^2+\Omega^2}\frac{ P^jP^n \varphi^{ij}(\zq,\hq)  \delta_{\hq-\haq} - \sqrt{2}m\varepsilon^{i*}_{\hq}P^n \delta_{\hq\haq} }{( \Pt^2+\Qtil^2 )^2}\Bigg\} \,.
\end{multline}

The instantaneous contribution is computed in Appendix \ref{sec: instanteous contribution} with the following result
\begin{equation}
\label{eq: rel. soft: instantaneous contribution}
    \Psi^{\gamma^* \rightarrow \qqg} _{\text{inst}}(\zq,\zaq,\zg,\Pb,\kbg)=2q^+\frac{ ee_fgt^a}{\Pt^2+\Qtil^2}\frac{\Omega^2}{\ktg^2+\Omega^2}\sqrt{ \zq \zaq}\polp^i \polg^{*m}\varphi^{ij}(\zq,\hq)\delta^{jm} \delta_{\hq-\haq} \,.
\end{equation}
This depends upon the quark mass  only through the light cone energy denominators which result in  the two scales $\Omega^2$ and $\Qtil^2$.

It is now convenient to combine the last term within the parenthesis in the regular contribution, \eq\eqref{eq: rel. soft: final regular contribution}, with the instantaneous contribution, \eq\eqref{eq: rel. soft: instantaneous contribution}, to reconstruct the same tensorial structure as for the first term within the square brackets in 
\eq\eqref{eq: rel. soft: final regular contribution}:
\begin{align}
    -\frac{P^jP^m }{(\Pt^2+\Qtil^2)^2}
     + \frac{1}{2}\frac{\delta^{jm}}{\Pt^2+\Qtil^2}
    =\frac{1}{2}\frac{1}{\Pt^2+\Qtil^2}\Bigg(\delta^{jm}- \frac{2P^jP^m}{\Pt^2 + \Qtil^2}\Bigg) \,.
\end{align}

The total photon wavefunction in the approximations of interest is thus obtained as
\begin{align}
    \Psi^{\gamma^*\rightarrow \qqg}(\zq,\zaq,\zg,\Pb,\kbg)=2q^+ee_fgt^a2\sqrt{\zq\zaq}\polp^i \polg^{*m}\,H^{in}(\zq,\zaq,\Pb,Q,m)\frac{k_3^nk_3^m+\delta^{nm}\Omega^2/2}{\ktg^2+\Omega^2} \,,\label{PsiFull}
\end{align}
with the ``hard'' tensor $H^{in}$ defined as 
\begin{align}
    H^{in}(\zq,\zaq,\Pb,Q,m)
  \equiv   \varphi^{ij}(\zq,\hq)\frac{\delta_{\hq-\haq}  }{\Pt^2 + \Qtil^2} \left( \delta^{jn} - \frac{2 P^j P^n}{\Pt^2 + \Qtil^2}\right)+  2\sqrt{2}m\varepsilon_{\hq}^{i*}  \frac{P^n \delta_{\hq\haq} }{\big(\Pt^2 + \Qtil^2\big)^2}\,. 
    \label{eq: soft gluon: hard tensor}
\end{align}

It is finally convenient to separate the  traceless part of the
the second rank tensor encoding the dependence upon $\kbg$ in \eqn{PsiFull}
from its diagonal part. This  leads to the following, final, expression for the $q\bar q g$ LCWF:
\begin{multline}
    \Psi^{\gamma^*\rightarrow \qqg}(\zq,\zaq,\zg,\Pb,\kbg)   =
    2q^+ee_fgt^a2\sqrt{\zq\zaq} \polp^i \polg^{*m}\bigg[ \Psi^{im}_\po (\zq,\zaq,\zg,\Pb,\kbg) \\+ 
    \frac{H^{im}(\zq,\zaq,\Pb,Q,m)}{2} \bigg] \,,
    \label{eq: rel. soft: Pomeron and diagonal LCWF}
\end{multline}
with the following definition for the ``Pomeron'' wavefunction:
\begin{align}
    &\Psi^{im}_\po (\zq,\zaq,\zg,\Pb,\kbg)\equiv 
    H^{in}(\zq,\zaq,\Pb, Q, m)\frac{k_3^nk_3^m-\delta^{nm}\ktg^2/2}{\ktg^2+\Omega^2}
     \,.\label{eq:Psipo}
\end{align}

Importantly, equations like \eqn{eq: soft gluon: hard tensor} show that the quantity $\Qtil^2=z_1z_2Q^2+m^2$ acts as a ``screening mass'' for the distribution in $\Pt$, which ultimately implies that the typical values of $\Pt$ are of order $\Qtil$, as anticipated in the introduction to this section. So, the hardness condition $M_h^2\gg \Qs^2$ also implies that $P_\perp$ is hard ($P_\perp^2\gg \Qs^2$) except for (less interesting) rare configurations.

\subsubsection{Interaction with the target: the gluon dipole}
\label{subsubsection: interaction with the nucleus very soft gluon}
 
As discussed in Appendix~\ref{sec: LCPT}, the effects of the multiple scattering between the $q\bar q g$ system and the target can be resummed to all orders in the eikonal approximation, which is most conveniently formulated in the transverse coordinate representation. To that aim, we need the Fourier transform of the LCWF \eqref{eq: rel. soft: Pomeron and diagonal LCWF} to the transverse coordinate space, defined as
\begin{align}\label{FT}
    \Psi_{\alpha\beta}^a(\zq,\zaq,\zg,\xb,\yb,\zb)&=\int \frac{\der^2\kbq\der^2\kbaq\der^2\kbg}{(2\pi)^6}e^{-i\kbq\cdot\xb-i\kbaq\cdot\yb-i\kbg\cdot\zb}\Psi_{\alpha\beta}^a(\zq,\zaq,\zg,\kbq,\kbaq,\kbg),
    \end{align}
where we have explicity shown also the colour indices $a$, $\alpha$ and $\beta$ (for the gluon, quark and antiquark, respectively), as they will be important for what follows. In this representation, the effect of the collision with the nuclear shockwave is very simple: the colour matrix $t^a_{\alpha\beta}$ which encodes the colour structure of the amplitude in the absence of the interaction (see e.g. \eqn{eq: rel. soft: Pomeron and diagonal LCWF}) gets replaced by the more complicated operator 
\begin{align}\label{Tqqg}
t^a_{\alpha\beta}\,\rightarrow\,\mathcal{T}^a_{\alpha\beta}(\xb,\yb,\zb)\equiv 
\Big(U^{ab}(\zb)V(\xb) t^b V^\dagger(\yb)-t^a\Big)_{\alpha\beta}\,,
\end{align}
where $V(\xb)$, $V^\dagger(\yb)$ and $U(\zb)$ are Wilson lines describing the colour rotations of the quark, the antiquark, and the gluon, respectively (see \eqs\eqref{eq: LCPT: Definition Fund. WL}-\eqref{eq: LCPT: Definition Adj. WL} for explicit expressions). Notice that the $\mathcal{T}^a_{\alpha\beta}$ vanishes in the absence of the scattering, as required by unitarity.

More precisely, the scattering amplitude $\mathcal{T}^a_{\alpha\beta}(\xb,\yb,\zb)$ refers to the case where there is no constraint on the final state of the $q\bar q g$ system --- that is, to the calculation of the {\it total} cross section, which includes both elastic and inelastic interactions. 
Here, however, we are interested in the {\it diffractive} process, where the scattering is elastic --- meaning that the final $q\bar q g$ system is in a colour singlet state. The corresponding scattering amplitude is obtained from  \eqn{Tqqg} by projecting the (generally coloured) final state $\mathcal{T}^a_{\alpha\beta }\big |q_{\alpha}\bar{q}_{\beta} g_a\big\rangle$ onto the colour singlet state $t^a_{\alpha\beta }\big |q_{\alpha}\bar{q}_{\beta} g_a\big\rangle$ (see Appendix~\ref{sec: LCPT} for details, notably \eqs\eqref{eq: LCPT: projection}-\eqref{eq: LCPT: Fierz identity}). In turn, this projection amounts to replacing the colour operator in the r.h.s. of \eqn{Tqqg} by 
\begin{align}     \label{eq:qqg interaction}
\mathcal{T}^a_{\alpha\beta}(\xb,\yb,\zb)&\,\rightarrow\,t^a_{\alpha\beta}\,\frac{1}{\sqrt{\Cf\Nc}}\,
 \Tr\big[t^c \mathcal{T}^c(\xb,\yb,\zb)\big]\\*[0.2cm] \nonumber
 &\,=\,t^a_{\alpha\beta}\,\sqrt{\Cf\Nc}\big(S_{\qqg}(\xb,\yb,\zb) -1\big) \, ,
\end{align}
where we introduced  the $S$-matrix for the elastic scattering of the $q\bar q g$ system, defined as
\begin{equation}
\label{eq: soft gluon: s-matrix qqg}
    S_{\qqg}(\xb,\yb,\zb)\equiv \frac{1}{\Cf\Nc}\, U^{ba}(\zb)\Tr\big[t^bV(\xb)t^aV^\dagger(\yb)\big] \,.
\end{equation}
For {\it coherent} diffraction, the target must scatter elastically as well (it should not dissociate in the final state). This means that the CGC average over the target must be computed already at the level of the {\it amplitude}. This amounts to replacing the quantity \eqref{eq: soft gluon: s-matrix qqg}  by
\begin{align}
    \label{eq: soft gluon: averaged scattering qqg}
    \Sm_{\qqg}(\xb,\yb,\zb)&\equiv\frac{1}{\Cf\Nc}\big\langle U^{ba}(\zb)\Tr\big(t^bV(\xb)t^aV^\dagger(\yb)\big) \big\rangle \, ,
\end{align}
where the brackets denote the average over the  ensemble of target colour fields evolved up to the rapidity scale $Y_\po=\ln(1/x_\po)$ set by the rapidity gap.

So far, our treatement of the interaction with the target was exact (within the limits of the eikonal approximation, of course). In what follows, we describe the additional simplifications which occur for hard $q\bar q$ dijets, i.e. in the regime where $M_h^2\gg \Qs^2(\xpo, A)$, with $M_h$ defined in \eqn{Mh}. In this case, the LCWF takes the simplified form shown in \eqs\eqref{eq: rel. soft: Pomeron and diagonal LCWF}-\eqref{eq:Psipo} and the scattering operator simplifies as well, as we now explain.

The crucial observation is that, in the hard regime, the size $r\equiv |\xb-\yb|$ is very small, $r\sim 1/M_h$, so the $q\bar q g$ system effectively behaves like a {\it gluon-gluon dipole}~\cite{Wusthoff:1997fz,GolecBiernat:1999qd,Hebecker:1997gp,Buchmuller:1998jv,Hautmann:1998xn,Hautmann:1999ui,Hautmann:2000pw,Iancu:2021rup,Hatta:2022lzj,Iancu:2022lcw}: one gluon leg is the physical gluon, and the other leg is the $q\bar q$ pair, which after emitting the gluon remains in an adjoint colour state. The simplest way to see this is to notice that the phase of the exponentials in the Fourier transform \eqref{FT} can be rewritten as
\begin{equation}
\label{eq: soft gluon: exponent FT}
    \kbq\cdot\xb +\kbaq\cdot\yb +\kbg\cdot\zb = \Pb\cdot\rb + (\Kb + \kbg)\cdot\bb + \kbg\cdot\Rb \,.
\end{equation}
where $\rb$ represents the transverse size of the $q\bar q$ pair, $\bb$ is the center-of-energy of that pair, and  $\Rb$ denotes the transverse separation between the pair and the gluon:
\begin{align}
\rb&\equiv \xb-\yb \,, &
         \bb&\equiv\frac{\zaq\xb+\zq\yb}{\zq+\zaq} \,, &
         \Rb&\equiv \zb -\bb \,.
\end{align}
\eqn{eq: soft gluon: exponent FT} shows that the variables $\Pb$ and $\rb$ are conjugated with each other, and similarly  $\kbg$ and $\Rb$. Since $P_\perp$ is hard for the typical configurations, $P_\perp^2\sim M_h^2\gg \Qs^2$, the $q\bar q$ separation is correspondingly small: $r\sim 1/P_\perp\ll 1/Q_s$. Similarly, the fact that elastic scattering favors semi-hard gluon emissions with $\ktg\lesssim Q_s$ implies that the transverse size $R$ of the effective $gg$ dipole is relatively large, $R\sim 1/\ktg\gtrsim 1/Q_s$.

Based on the above, one can simplify the elastic $S$-matrix for the $q\bar q g$ system by approximating  $\xb\simeq\yb\simeq \bb$ in \eqn{eq: soft gluon: averaged scattering qqg}, to deduce
\begin{align}\label{eq: soft gluon: gluon-gluon dipole}
    \Sm_{\qqg}(\xb,\yb,\zb)\simeq \Sm_{\qqg}(\bb,\bb,\Rb+\bb)&=\frac{1}{\Cf\Nc}\big\langle  U^{ba}(\Rb+\bb)\Tr\big[t^bU^{\dagger ac}(\bb)t^c\big] \big\rangle \nonumber \\
    &=\frac{1}{\Nc^2-1}\big\langle  \Tr\big[U(\Rb+\bb)U^{\dagger }(\bb)\big] \big\rangle\equiv \Sm_g(\Rb) 
   .
\end{align}
(As before, we have neglected the inhomogeneity of the target in the transverse plane.)
The quantity $\Sm_g(\Rb)$ is recognised as the $S$-matrix for the elastic scattering of a gluon-gluon dipole of transverse separation $\Rb$. The corresponding approximation
\beq
\Sm_{\qqg}(\xb,\yb,\zb)-1 \simeq \Sm_g(\Rb)- 1\equiv -\mathcal{N}_g(\Rb)
\eeq
on the colour operator which enters the amplitude, cf. \eqn{eq:qqg interaction}, features the elastic scattering amplitude $\mathcal{N}_g(\Rb)$ of that dipole.

We are now prepared to combine the approximation \eqref{eq: rel. soft: Pomeron and diagonal LCWF} for the LCWF in the absence of the collision with the corresponding approximation \eqref{eq: soft gluon: gluon-gluon dipole} for the colour operator, to obtain our final result for the interacting amplitude in the correlation limit. 

First, we need to compute the Fourier transform of the LCWF in \eqref{eq: rel. soft: Pomeron and diagonal LCWF} to transverse coordinate space and then multiply the result with the  $gg$ dipole amplitude $\mathcal{N}_g(\Rb)$. Since the latter is independent of the size $\rb$ of the $q\bar q$ pair, there is actually no need to perform the Fourier transform from $\Pb$ to $\rb$. (The final result for the cross section is anyway needed in transverse momentum space.) Since moreover $\Kb + \kbg=0$ in the absence of the collision, it is clear that the only non-trivial Fourier transform is that from $\kbg$ to $\Rb$, which yields 
\begin{align}
    \Psi_\po^{im}(\zq,\zaq,\zg,\Pb,\Rb)&=\int \frac{\der^2\kbg}{(2\pi)^2}\,
    e^{i\kbg\cdot\Rb}\, \Psi_\po^{im}(\zq,\zaq,\zg,\Pb,\kbg) \nonumber\\*[0.2cm]
    &=H^{in}(\zq,\zaq,\Pb,Q,m)\int \frac{\der^2\kbg}{(2\pi)^2}\, e^{i\kbg\cdot\Rb}\, \frac{k_3^nk_3^m-\delta^{nm}\frac{\ktg^2}{2}}{\ktg^2+\Omega^2} \nonumber \\*[0.2cm]
    &=H^{in}(\zq,\zaq,\Pb,Q,m)\,\frac{1}{2\pi}\bigg(\frac{\delta^{nm}}{2}-\frac{R^nR^m}{\Rt^2} \bigg)\Omega^2K_2(\Omega \Rt) \,,
    \label{eq: rel. soft: Pomeron part}
\end{align}
with $K_2(\Omega \Rt)$ the modified Bessel function of second kind, order 2.
We have only shown the Pomeron piece of the LCWF in \eqn{eq: rel. soft: Pomeron and diagonal LCWF}, since the diagonal piece gives a vanishing contribution after adding the interaction with the target. Indeed,
its Fourier transform results in a $\delta$-function $\delta^{(2)}(\Rb)$, which vanishes after  multiplication with the amplitude $\mathcal{N}_g(\Rb)$, due to the colour transparency of a zero size dipole. 

Now it suffices to multiply the above result for the mixed-space wave function with the $gg$ dipole amplitude $\mathcal{N}_g(\Rb)$ and then perform the inverse Fourier transform from $\Rb$ to $\kbg$ in order fix the \emph{final} gluon momentum. The latter is still equal to minus the imbalance of the $q\bar{q}$ (recall that we consider elastic scattering off a homogeneous nucleus), for which we shall use the notation $-\Kb$, as before. 

To summarise, the amplitude for the diffractive production of 2+1 jets with a hard $q\bar q$ pair and a semi-hard gluon and in the leading twist approximation takes the following, {\it factorised}, form
\begin{align}\label{PsiPom}
    &\mathcal{A}^{im}_\po (\zq,\zaq,\zg,\Pb,\Kb)\,=\,
    H^{in}(\zq,\zaq,\Pb, Q, m) \, G^{nm}(\Kb,\Omega, Y_\po)\,, \end{align}
with the hard tensor $H^{in}$ defined in \eq\eqref{eq: soft gluon: hard tensor} and the semi-hard tensorial distribution
\begin{align}
  G^{nm}(\Kb,\Omega, Y_\po)&\equiv \frac{1}{2\pi}\int\der^2\Rb \, e^{i\Kb\Rb} \bigg(\delta^{nm}-\frac{2R^n R^m}{\Rt^2} \bigg) \Omega^2K_2(\Omega \Rt)\mathcal{N}_{g}(\Rt, Y_\po) \nonumber \\
  &=\bigg( \frac{K^nK^m}{\Kt^2}-\frac{\delta^{nm}}{2}\bigg) G(\Kt,\Omega, Y_\po) \,,
  \label{eq: rel. soft: semi-hard tensor}
\end{align}
which encodes the spatial distribution of the gluon emission by the small $q\bar{q}$ dipole as well as the scattering of the effective gluon-gluon dipole with the target. In writing \eqn{eq: rel. soft: semi-hard tensor}, we have also indicated the dependence upon the rapidity gap $Y_\po$ (as introduced by the high-energy evolution of the $gg$ dipole amplitude) and we assumed target isotropy in the transverse plane: $\mathcal{N}_{g}(\Rb) =\mathcal{N}_{g}(\Rt) $.
By contracting \eqn{eq: rel. soft: semi-hard tensor}
 with $K^nK^m/\Kt^2$, one obtains an integral representation for the scalar quantity $G(\Kt,\Omega, Y_\po)$:
\begin{equation}
\label{eq: rel. soft: scalar semihard tensor}
    G(\Kt,\Omega, Y_\po)=\Omega^2\int_0^\infty \der \Rt \Rt J_2(\Kt \Rt)K_2(\Omega \Rt)\mathcal{N}_g(\Rt,Y_\po) \,. 
\end{equation}
\subsection{TMD factorization of the cross section}

The structure of the Pomeron wavefunction in \eqn{PsiPom} is strongly suggestive of {\it TMD factorization}: the ``hard tensor'' $H^{in}(\zq,\zaq,\Pb, Q^2, m)$ encodes the kinematical variables associated with the ``hard projectile'' (the incoming photon and its $q\bar q$ fluctuation), while the ``semi-hard tensor'' $G^{nm}(\Kb,\Omega, Y_\po)$ describes the distribution of the momentum imbalance $\Kb$ acquired by the projectile via its scattering off the nuclear target.
Yet, this factorization is not yet complete: the tensor $G^{nm}(\Kb,\Omega, Y_\po)$ still depends upon $\Pt$ and the other ``hard'' variables via the quantity $\Omega$, cf. \eqn{eq: rel. soft: Omega}. In what follows, we explain how a complete factorization can be achieved and provide a physical interpretation for it, in terms of the Pomeron structure of the target.

Namely, we will argue that \eqn{PsiPom} is consistent with the following physical picture: the Pomeron --- a colourless fluctuation of the nuclear target with longitudinal fraction $x_\po$ (represented in \eqn{PsiPom} by the gluon-gluon dipole amplitude $\mathcal{N}_{g}(\Rt,Y_\po) $) --- splits into a gluon-gluon pair in a colour singlet state and with zero total transverse momentum. One of these gluons, with transverse momentum $\Kb$ and splitting fraction $x$ w.r.t. the Pomeron, propagates in the $t$-channel and gets absorbed by the $q\bar q$ pair from the projectile. The other gluon,  with transverse momentum $-\Kb$ and splitting fraction $1-x$, propagates in the $s$-channel and gives rise to the semi-hard jet in the final state.

The splitting fraction $x={\xqq}/{\xpo}$ has been introduced in \eqn{eq: intro: variable x}, which in the present notations (and to the accuracy of interest) can be rewritten as 
\begin{equation}
\label{xOmega}
    x\simeq\frac{\bar Q^2+\Pt^2+m^2}{\bar Q^2+\Pt^2+m^2+\frac{z_1z_2}{z_3}\Kt^2}=\frac{\Omega^2}{\Omega^2+\Kt^2}\quad\Rightarrow \quad   \Omega^2=\frac{x}{1-x}\Kt^2 \,.
\end{equation}
The last equality above allows one to express the  quantity $\Omega$  in terms of the {\it target} variables $x$ and $\Kt$, thus achieving a complete factorization in  \eqn{PsiPom}. This discussion sheds more  light on the physical interpretation of the semi-hard tensor in \eqn{eq: rel. soft: semi-hard tensor}: with $\mathcal{N}_{g}\to 1$, this tensor describes the wavefunction of the colourless gluon-gluon pair (the traceless tensor in the r.h.s. of \eqn{eq:Psipo}). After including $\mathcal{N}_{g}$, the tensor $G^{nm}(\Kb,\Omega, Y_\po)$ describes the formation of the $gg$ pair with time-like virtuality\footnote{If $k_s^\mu$ and $k_t^\mu$ denote  the 4-momenta of the gluons propagating in the $s$-channel and in the $t$-channel, respectively, then  $k_s^\mu=(z_3q^+, (1-x)x_\po P_N^-, -\Kb)$ and $k_t^\mu=(-z_3q^+, x x_\po P_N^-, \Kb)$ and one can easily check that $(k_s+k_t)^2=\Omega^2$.} $\Omega^2$ via the decay of the Pomeron.

Clearly, the factorization observed above at the level of the amplitude will also transmit to the elastic cross section, which is obtained by combining the prefactor 
from \eqn{eq: rel. soft: Pomeron and diagonal LCWF} with the tensor amplitude from \eqn{PsiPom}
\begin{equation}
    \mathcal{A}(\zq,\zaq,x,\Pb,\Kb) =\int \der^2\bb e^{-i(\Kb+\kbg)\bb}     ee_fg2\sqrt{\zq\zaq}\sqrt{\Cf\Nc} \polp^i \polg^{*m} \mathcal{A}^{im}_\po.
\end{equation}
We obtain the cross section by taking the modulus squared of this amplitude, including the phase space integration \eqref{eq: LCPT: PS} and performing the average/sum over polarisations:
\begin{multline}
\label{eq: soft quark: TMD factorized cross section}
    \frac{\der\sigma^{\gamma_T^*A\rightarrow \qqg A }_\mathrm{D}}{\der\zq\der\zaq\der\zg\der^2 \Kb\der^2 \Pb\der^2 \kbg}=\delta(1-\zq-\zaq)\delta^{(2)}(\Kb+\kbg)
   \frac{S_\perp\aem \as e_f^2 \Cf\Nc}{4\pi^4\zg}\\*[0.2cm]
   \cross\Bigg((\zq^2+\zaq^2) \frac{\Pt^4 + \Qtil^4}{\big(\Pt^2 + \Qtil^2\big)^4}+ 2m^2  \frac{\Pt^2}{ \big(  \Pt^2+\Qtil^2\big)^4}\Bigg) [G(\Kt,\Omega, Y_\po)]^2 \,,
\end{multline}
where $S_\perp$ is the transverse area of the nuclear target, assumed to be homogeneous.

This is not yet our final result: the r.h.s. of \eqn{eq: soft quark: TMD factorized cross section} still depends upon $z_3$ (the longitudinal fraction of the gluon w.r.t. the photon), but in order for the TMD factorization to be manifest, one should eliminate $z_3$ in terms of the target variables, $x$ or $x_\po$. This can be easily done by using \eqn{xOmega} together with the expression \eqref{eq: rel. soft: Omega} to deduce
\begin{align}
\zg=\frac{x}{1-x}\,\frac{K_{\perp}^2}{Q^2+\frac{P_{\perp}^2+m^2}{z_1 z_2}}\quad\Longrightarrow\quad
  \frac{\der \zg}{ \zg}=\frac{1}{1-x}
  \frac{\der x}{x}\,.
\end{align}
After this change of variable and performing the trivial integration over $\kbg$,
we finally obtain the following, TMD-factorised, expression for the
diffractive cross section,
\begin{equation}
\label{eq: rel. soft: final result compacted}
    \frac{\der\sigma^{\gamma_T^*A\rightarrow \qqg A }_\mathrm{D}}{\der \zq\der \zaq\der^2 \Kb\der^2 \Pb\der \ln(1/x)}=\mathcal{H}(\zq, \zaq,\Pt,Q, m)\,\frac{\der x\mathcal{G}_\po (x,\xpo,\Kt)}{\der^2 \Kb} \,.
\end{equation}
The hard factor:
\begin{align}
    \mathcal{H}(\zq, \zaq,\Pt,Q, m)\equiv\delta(1- \zq- \zaq)
   \aem \as e_f^2\Bigg[\big( \zq^2+ \zaq^2  \big) \frac{\Pt^4 + \Qtil^4}{\big(\Pt^2 + \Qtil^2\big)^4}+ 2m^2  \frac{\Pt^2}{ \big( \Pt^2+\Qtil^2\big)^4}\Bigg],
\end{align}
 describes the formation of the $\qbarq$ pair from the virtual photon fusing with gluon in the Pomeron. 

The semi-hard factor,
\begin{equation}\label{GDTMD}
    \frac{\der x\mathcal{G}_\po (x,\xpo,\Kt)}{\der^2 \Kb}\equiv \frac{S_\perp(\Nc^2-1)}{4\pi^3} \frac{[G(x,\Kt,Y_\po)]^2}{2\pi(1-x)}  \,,
\end{equation}
is the unintegrated gluon distribution of the Pomeron (a.k.a. the gluon diffractive TMD). Physically, it represents the number density of gluons in the Pomeron wavefunction with transverse momentum $\Kt$ and longitudinal momentum fraction $x$ with respect to the Pomeron. It is here understood that $G(x,\Kt,Y_\po)
\equiv  G(\Kt,\Omega, Y_\po)$ with $G(\Kt,\Omega, Y_\po)$ defined in \eqn{eq: rel. soft: scalar semihard tensor} and $\Omega$ related to $x$ and $\Kt$ as shown in \eqn{xOmega}.
\eqn{GDTMD} depends upon $Y_\po$ only via the high-energy evolution of the gluon dipole amplitude $\mathcal{N}_g(\Rt,Y_\po)$. Notice that $x$ and $x_\po$ are not independent variables, since
 $x={\xqq}/{\xpo}$, with  $\xqq$ fully determined by the hard kinematics and the total energy (cf. \eqn{eq: intro: variable xqq}).
 
 The gluon DTMD in \eqn{GDTMD} is independent of the quark mass $m$ and hence it is the same as originally obtained in Refs.~\cite{Iancu:2022lcw,Hatta:2022lzj}. In Appendix~\ref{sec:gDTMD ms} we write the corresponding expression in momentum space, which was first presented in Ref.~\cite{Hatta:2022lzj}. For completeness, in these momentum space expressions, we also show the dependence on the total momentum transferred from the target $\Deltab$, which was assumed to vanish in the present analysis.

\section{Quark diffractive TMD}
\label{Sec: soft quark}
In this section we calculate the diffractive parton-level cross section in the correlation limit where the quark is the soft parton while the antiquark and gluon are hard, with transverse momentum $\ktaq^2 \simeq\ktg^2,\, Q^2\sim M_h^2 \gg \ktq^2,\,m^2 $ and $\ktq^2 \sim \Qs^2$. In contrast with the previous diffractive scattering with a soft gluon emission, the quark mass cannot longer play the role of a hard scale. Instead, its dependence will be encoded in the quark DTMD distribution which we will factorize from the hard factor in the differential cross section. 
\subsection{The photon light cone wavefunction and target interaction}
\subsubsection{Gluon emission from antiquark}
\begin{figure*}[tbp!]
\centering{
\includegraphics[width=0.4\textwidth]{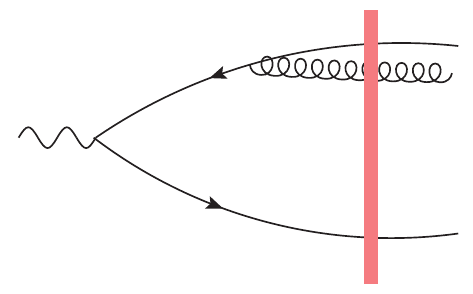}
\begin{tikzpicture}[overlay]
\node[anchor=south west] at (-6cm,3.7cm) {\hypertarget{diag:c}{(c)}};
    \draw[dash pattern=on 4pt off 6pt,line width=0.7pt, -](-4.1cm,3.8cm) -- (-4.1cm,-0.2cm);
    \draw[dash pattern=on 4pt off 6pt,line width=0.7pt, -](-2cm,3.8cm) -- (-2cm,-0.2cm);
    \draw[dotted, line width=0.9pt, -to](-5.7cm,0.3cm)-- (-4.1cm,0.3cm);
    \draw[dotted, line width=0.9pt, -to](-5.7cm,0cm) -- (-2cm,0cm);
    \node[anchor=south east] at (-3.3cm,2.8cm) {$2'$};
    \node[anchor=south east] at (-0.3cm,3.2cm) {$2$};
    \node[anchor=south east] at (-0.3cm,2.15cm) {$3$};
    \node[anchor=south east] at (-0.3cm,0.8cm) {$1$};
    \node[anchor=south east] at (-4.1cm,0.3cm) 
         {ED$_{\gamma \rightarrow q\bar{q}'}$};
    \node[anchor=south east] at (-2.1cm,0cm) {ED$_{\gamma \rightarrow\qqg}$};
\end{tikzpicture}
\rule{2.5em}{0pt}
\includegraphics[width=0.4\textwidth]{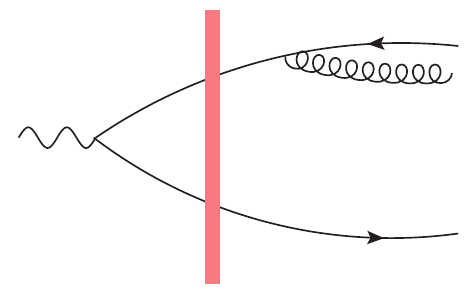}
\begin{tikzpicture}[overlay]
\node[anchor=south west] at (-6cm,3.7cm) {\hypertarget{diag:d}{(d)}};
    \draw[dash pattern=on 4pt off 6pt,line width=0.7pt, -](-4.1cm,3.8cm) -- (-4.1cm,-0.2cm);
    \draw[dash pattern=on 4pt off 6pt,line width=0.7pt, -](-2.8cm,3.8cm) -- (-2.8cm,-0.2cm);
    \draw[dotted, line width=0.9pt, -to](-5.7cm,0.3cm)-- (-4.1cm,0.3cm);
    \draw[dotted, line width=0.9pt, -to](0cm,0.3cm) -- (-2.8cm,0.3cm);
    \node[anchor=south east] at (-2.8cm,2.9cm) {$2'$};
    \node[anchor=south east] at (-0.3cm,3.2cm) {$2$};
    \node[anchor=south east] at (-0.3cm,2.15cm) {$3$};
    \node[anchor=south east] at (-0.3cm,0.8cm) {$1$};
    \node[anchor=south east] at (-4.1cm,0.3cm) 
         {ED$_{\gamma \rightarrow q\bar{q}'}$};
    \node[anchor=south east] at (-0.2cm,-0.4cm) {$-$ED$_{\bar{q}'\rightarrow\bar{q}g}$};
\end{tikzpicture}
}
\caption{Gluon emission from the antiquark, (c) before and (d) after the interaction with the target represented as an orange band. Particle momenta are labeled using numbers as a shorthand notation. The quark is soft $\zq \ll 1$ and semihard $\ktq \sim \Qs$. The dotted arrow indicates the time evolution from the asymptotic state to the intermediate state  where the energy denominator is evaluated. 
}
\label{fig: quark DTMD: soft gluon antiquark}
\end{figure*}
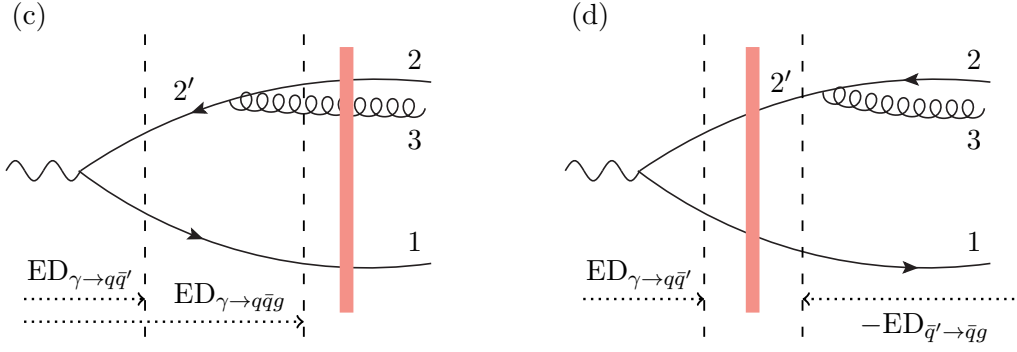
First we calculate the scattering amplitude where the gluon (now hard $\ktg^2 \sim Q^2)$ is emitted from the antiquark. \fig\ref{fig: quark DTMD: soft gluon antiquark} shows the amplitude for the two possible time orderings: gluon emission before \hyperlink{diag:c}{(c)} and after \hyperlink{diag:d}{(d)} the interaction with the shockwave\footnote{The LCWF in diagram \hyperlink{diag:c}{(c)} is the same as in \hyperlink{diag:b}{(b)}, but we will here evaluate it in a different kinematical regime.}. The corresponding light cone wavefunctions are
\begin{align}
\label{eq: soft quark: LCWF diagram (c)}
    \Psi^{\hyperlink{diag:c}{(c)}}&=\sum_{\bar{q}'}ee_f\frac{\bar{u}(\vec{k}_1) \slashed{\varepsilon}_\lambda(q)v(\vec{k}'_2)}{\EDqqp}(-gt^a)\frac{\bar{v}(\vec{k}_2')\slashed{\varepsilon}^*_\sigma(\vec{k}_3)v(\vec{k}_2)}{\EDqqg} \,, \\
\label{eq: soft quark: LCWF diagram (d)}
   \Psi^{\hyperlink{diag:f}{(d)}}&=\sum_{\bar{q}'}ee_f\frac{\bar{u}(\vec{k}_1) \slashed{\varepsilon}_\lambda(q)v(\vec{k}'_2)}{\EDqqp}(-gt^a)\frac{\bar{v}(\vec{k}_2')\slashed{\varepsilon}^*_\sigma(\vec{k}_3)v(\vec{k}_2)}{-(\EDaqg)} \,, 
\end{align}
where the first vertex corresponds to the splitting of the transverse photon and the second vertex represents the gluon emission by the antiquark. 
Strictly speaking $\Psi^{\hyperlink{diag:c}{(c)}}$ is the $\gamma^* \to \qqg$ LCWF, and $\Psi^{\hyperlink{diag:d}{(d)}}$ the product of a $ \gamma^* \to \qbarq$ LCWF before the shockwave and a conjugate $\qqg \to \qbarq$ LCWF after it, following the terminology used in Appendix~\ref{sec: LCPT}. 
The energy denominators are always computed with respect to the asymptotic state. 
In the diagram this is reflected by the direction of the dotted arrows.
The gluon emission after the shockwave actually corresponds to a (complex conjugate) gluon absorption LCWF, where the energy decreases.
For convenience, we write the associated merging energy denominator as minus the corresponding splitting one $-(\EDaqg)$.

The phase-space integral for the intermediate antiquark propagator is
\begin{equation}
    \sum_{\bar{q}'}=\int\frac{\der^3 \Vec{k'_2}}{2k'^+_2(2\pi)^3}(2\pi)^3\delta(\Vec{k'_2}-\Vec{k_2}-\Vec{k_3})=\frac{1}{2(\zaq+\zg)q^+}=\frac{1}{2q^+} \,.
\end{equation}

We are now interested in the correlation limit where the $\Bar{q}g$ system is hard, with relative momentum $\Pt^2 \gg \Qs^2$, whereas the quark is soft $(\zq\ll1)$ and  semi-hard $(\ktq^2\sim \Qs^2)$. In this configuration, the transverse separation between the hard $\bar{q}g$ pair and the soft quark $\Rt\sim 1/\Qs$ is much larger than the transverse separation $\rt\sim 1/\Pt$ between the antiquark and the gluon. The distance $\rt$ is so small, that the gluon and quark pair scatters in the same way as its parent antiquark $\Bar{q}'$. In other words, the transverse size $\rt$ is much smaller than the correlation length $1/\Qs$ of the colour fields in the target. Hence, there is no difference between the two time orderings of the gluon emission shown in \fig\ref{fig: quark DTMD: soft gluon antiquark}, i.e,  the partonic $\qqg$ system scatters as a $q\Bar{q}$ dipole of size $\Rt$. Therefore, we can add the amplitudes in \eqs\eqref{eq: soft quark: LCWF diagram (c)} and \eqref{eq: soft quark: LCWF diagram (d)} which involve the following linear combination of energy denominators, which in fact can be seen as a consequence of the orthogonality relation \eq\eqref{eq: LCPT: relation wavefunctions}, 
\begin{align}
    &\frac{1}{\EDqqp}\bigg(\frac{1}{\EDqqg}-\frac{1}{\EDaqg}\bigg) \nonumber\\
    &=\frac{1}{q^- - k_1^{-}  - k'^{-}_2}\bigg(\frac{1}{q^- - k_1^{-}  -  k^{-}_2 - k_3^{-} }-\frac{1}{ k'^{-}_2 -  k^{-}_2 - k_3^{-} } \bigg) \nonumber\\
    &=-\frac{1}{ k'^{-}_2 -  k^{-}_2 - k_3^{-} }\frac{1}{q^- - k_1^{-}  -  k^{-}_2 - k_3^{-} } \,. \label{eq: soft quark: sum ED antiquark emission}
\end{align}
In this case, unlike in the soft gluon limit earlier, the hard  gluon is not resolved by the target. Thus both the $\qbarq$ and $\qqg$ states scatter with the same S-matrix, and the S-matrix can be factorized from the wavefunctions. After this factorization, we are in fact  now using the orthogonality relation \eq\eqref{eq: LCPT: relation wavefunctions} to combine the LCWFs of diagrams~\hyperlink{diag:c}{(c)} and \hyperlink{diag:d}{(d)} into the LCWF $(\Psi^{\qqg\rightarrow\gamma^*})^\dagger$. 

By using the hard relative momentum of the antiquark-gluon system
\begin{equation}
\label{eq: relative momentum}
\begin{aligned}
    \Pb=\frac{\zaq\kbg-\zg\kbaq}{\zaq+\zg} \,,
\end{aligned}
\end{equation}
together with the diffractive condition $\kto=\kbq+\kbaq+\kbg=0$, we can make the following change of variables to conveniently use $\Pb$ and $\kbq$ as independent momenta: 
\begin{align}
\label{eq: soft quark: change to P momentum}
    \kbaq&=-\Pb -\frac{\zaq}{\zaq+\zg}\kbq \,, & 
    \kbg&=\Pb -\frac{\zg}{\zaq+\zg}\kbq \,.
\end{align}
With the above change of variables, the energy denominators in the second line of \eq\eqref{eq: soft quark: sum ED antiquark emission} can be written as 
\begin{align}
     k'^{-}_2 -  k^{-}_2 - k_3^{-}  &= \frac{\ktaq^{'2}+m^2}{2k_2^{'+}}-\frac{\ktaq^2+m^2}{2k_2^+}-\frac{\ktg^2}{2k_3^+} =\frac{-1}{2q^+\zaq\zg}(\Pt^2+\zg^2m^2) \,, \\
    q^- - k_1^{-}  -  k^{-}_2 - k_3^{-}  &=-\frac{Q^2}{2q^+}-\frac{\ktq^{2}+m^2}{2k_1^{+}}-\frac{\ktaq^2+m^2}{2k_2^+}-\frac{\ktg^2}{2k_3^+} =\frac{-1}{2q^+}\Bigg(\frac{\ktq^2+\omega^2}{\zq}\Bigg) \,,
\end{align}
where we defined
\begin{align}
    &\omega^2\equiv \zq\Bigg(\frac{\Pt^2}{\zaq\zg}+Q^2\Bigg)+m^2 \,. 
\label{eq: soft quark: ED omega}
\end{align}
The LCWF in the absence of the scattering is obtained as (in momentum space)
\begin{multline}
\label{eq: soft quark: LCWF E&F momentum space}
    \Psi^{\hyperlink{diag:c}{(c)}+\hyperlink{diag:d}{(d)}}(\zq,\zaq,\zg,\kbq,\Pb)=\sum_{h'_2\alpha'_2}(-2q^+)ee_fgt\sqrt{\zq\zaq}\polp^i\polg^{m*}\\
    \cross \frac{\phi^{ij}(0,\hq) k_1^j\delta_{\hq-h'_2} - \sqrt{2}m\varepsilon^{*i}_{\hq}\delta_{\hq h'_2}}{\ktq^2+\omega^2} \frac{P^n\tau^{mn}(\zaq,\haq)\delta_{\haq\haq'} - \zg^2 \sqrt{2}m \varepsilon^{*n}_{\haq}\delta^{mn}\delta_{\haq-\haq'} }{\Pt^2+\zg^2m^2} \,,
\end{multline}
where the functions $\phi^{ij}(z,h)$ and $\tau^{mn}(z,h)$ are
\begin{align}
    \label{eq: splitting helicity structure aq}
    \phi^{ij}(\zq,\hq)&=(2\zq-1)\delta^{ij} + i\hq\varepsilon^{ij} \,, \\
    \tau^{mn}(\zaq,\haq)&=\big(1+\zaq\big)\delta^{mn} - i\haq(1-\zaq)\varepsilon^{mn} \,.
    \label{eq: emission helicity structure q}
\end{align}
To include the effect of the scattering, we need to Fourier transform the light cone wavefunction in \eq\eqref{eq: soft quark: LCWF E&F momentum space} to mixed space. With $\xb$, $\yb$ and $\zb$ the transverse positions of respectively the final quark, antiquark and gluon, we define: $\rb$ as the transverse separation of the $\barqg$ pair, $\Rb$ the overall size of the $q\bar{q}g$ system and $\bb$ the transverse position of the intermediate antiquark, so that  
\begin{align}
 \rb &\equiv \zb - \yb \,, & & \Rb\equiv \xb - \bb \,, & & \bb\equiv \frac{\zaq\yb+\zg\zb}{\zaq+\zg} \,.
\end{align}
By inverting the above relations, we get 
\begin{align}
\label{eq: soft quark: change of coordinate variables}
    \xb&=\Rb + \bb \,,& & \yb = \bb -\frac{\zg}{\zaq + \zg}\rb \,, & & \zb = \bb +\frac{\zaq}{\zaq + \zg}\rb \,.
\end{align}
The recoil of the antiquark when emitting the gluon is small  $|\yb-\bb|\sim \rt$.
Imposing $\kto \simeq 0$ and performing the change of variables in \eq\eqref{eq: soft quark: change of coordinate variables}, the exponent of the Fourier transform to mixed space takes the form
\begin{equation}
\label{eq: change of variables exponential}
\begin{aligned}
    \kbq\cdot\xb + \kbaq\cdot\yb + \kbg\cdot\zb &=\kbq\cdot\Rb + \Pb\cdot\rb \,.
\end{aligned}
\end{equation}

The quark at large distance $\Rb$ sees the gluon and antiquark at the same transverse position:
\begin{equation}
    \yb \simeq \zb \simeq \bb = \xb -\Rb \,.
\end{equation}
Then, the colour algebra describing the interaction between the incoming photon state and the target is given by
\begin{align}
    &\sum_{\substack{\alpha_1\alpha_2,\beta_1\beta_2,a,b}}\frac{(t^b_{\beta_1\beta_2})^*}{\sqrt{\Cf\Nc}}t^a_{\alpha_1\alpha_2}\Big(U^{ba}(\zb)V(\xb)_{\beta_1\alpha_1} V^\dagger(\yb)_{\alpha_2\beta_2} -\delta_{\beta_1\alpha_1}\delta_{\beta_2\alpha_2}\delta_{ab}  \Big) \nonumber\\
    &=\sqrt{\Cf\Nc}\bigg(\frac{1}{\Nc}\Tr\big(V(\xb)V^\dagger(\xb-\Rb)\big)  - 1\bigg) \nonumber \\
    &=\sqrt{\Cf\Nc}\big(S(\Rb)  - 1\big)=-\sqrt{\Cf\Nc}N(\Rb) \,.
    \label{eq: soft quark: scattering barq}
\end{align}
Here, $S(\Rb)\equiv (1/\Nc)\Tr \big(V(\xb)V^\dagger(\yb)\big)$ (with $\Rb=\xb-\yb$) is the $S$-matrix for the elastic scattering of a quark-antiquark dipole, $N(\Rb)=1-S(\Rb)$ is the corresponding scattering amplitude --- both evaluated for a given configuration of the colour fields in the target. In coherent diffraction, the average over the target colour fields must be performed at the amplitude level:
\begin{equation}
\label{eq: soft quark: scattering N}
    \langle N(\Rb) \rangle =\mathcal{N}(\Rb) \,.
\end{equation}
The interaction described with \eq\eqref{eq: soft quark: scattering barq} only depends on the size $\Rt$ of the $q\Bar{q}$ dipole. This implies that the scattering cannot change the hard momentum $\Pb$ (momentum variable dual to $\rb$) which is the relative momentum of the $\Bar{q}g$ pair. The scattering only modifies the imbalance $\Kb=\kbaq+\kbg$, or equivalently, the momentum $-\kbq$ of the semihard quark. We do not need to Fourier transform from $\Pb$ to $\rb$ because the LCWF dependence in $\Pb$ remains unchanged after adding the collision. It suffices to perform the Fourier transform from $\kbq$ to $\Rb$, multiply by the scattering amplitude, and then transform back from $\Rb$ to $-\Kb$. Therefore, we first write the Fourier transform to mixed space of the light cone wavefunction \eqref{eq: soft quark: LCWF E&F momentum space}:
\begin{equation}
\begin{aligned}
\label{eq: soft quark: LCWF E&F mixed space}
    \Psi^{\hyperlink{diag:c}{(c)}+\hyperlink{diag:d}{(d)}}(\zq,\zaq,\zg,\Rb,\Pb)&=2q^+ee_fgt^a\sqrt{\zq\zaq}\frac{\varepsilon^i_\lambda}{2\pi}\\
    &\cross\frac{\varepsilon^{*m}(P^n\tau^{mn}(\zaq,\haq)\delta_{\haq\haq'} - \zg^2 \sqrt{2}m \varepsilon^{*n}_{\haq}\delta^{mn}\delta_{\haq-\haq'} )}{\Pt^2+\zg^2m^2}\\
    &\cross\Big( \phi^{ij}(0,\hq) \frac{iR^j}{\Rt}\omega K_1(\omega \Rt)\delta_{\hq-\haq'} 
    - \sqrt{2}m\varepsilon_{\hq}^{i*}K_0(\omega  \Rt)\delta_{\hq\haq'}\Big) \,.
\end{aligned}
\end{equation}
We multiply the above expression by the scattering amplitude $-\mathcal{N}(\Rb)$ in \eq\eqref{eq: soft quark: scattering N} and then we Fourier transform the result back to momentum space. The Fourier transform of the massless term in \eq\eqref{eq: soft quark: LCWF E&F mixed space} including the effect of the interaction gives
\begin{equation}
    -\frac{i}{2\pi}\int \der^2\Rb e^{i\Kb\cdot\Rb}\frac{R^j}{\Rt}\omega K_1(\omega \Rt)\mathcal{N}(\Rb) =\frac{K^j}{\Kt}\frac{\mathcal{Q}_1(\omega,\Kt)}{\omega} \,,
\end{equation}
where
\begin{equation}
    \label{eq:Q1-def}
    \mathcal{Q}_1(\omega,\Kt)\equiv \omega^2\int_0^\infty \der \Rt\Rt J_1(\Kt\Rt)K_1(\omega \Rt)\mathcal{N}(\Rt) \;.
\end{equation}
The Fourier transform of the massive term in \eq\eqref{eq: soft quark: LCWF E&F mixed space} with the addition of the scattering effect reads
\begin{equation}
\begin{aligned}
    -\frac{1}{2\pi}\int \der^2\Rb e^{i\Kb\cdot\Rb}K_0(\omega \Rt)\mathcal{N}(\Rb) 
    &=-\frac{\mathcal{Q}_0(\omega,\Kt)}{\omega^2} \,,
\end{aligned}
\end{equation}
thus
\begin{equation}
\label{eq:Q0-def}
\begin{aligned}
    \mathcal{Q}_0(\omega,\Kt) =  \omega^2 \int_0^{\infty} \der \Rt \Rt J_0(\Kt \Rt) K_0(\omega \Rt)\mathcal{N}(\Rt)  \,.
\end{aligned}
\end{equation}
Therefore, we can write the amplitude represented in \fig\ref{fig: quark DTMD: soft gluon antiquark} as
\begin{multline}
\label{eq: soft quark: amplitude qbar}
    \mathcal{A}_{\bar{q}}(\zq,\zaq,\zg,\Pb,\Kb)=2q^+ee_fg\sqrt{\zq\zaq}\varepsilon^i_\lambda \varepsilon^{m*}_{\sigma}\frac{P^n\tau^{mn}(\zaq,\haq)\delta_{\haq\haq'}-\zg^2\sqrt{2}m\varepsilon^{n*}_{\haq}\delta^{mn}\delta_{\haq-\haq'}}{\Pt^2+\zg^2m^2} \\
    \cross\Bigg[ \phi^{ij}(0,\hq) \Bigg(\frac{K^j}{K}\frac{\mathcal{Q}_1(\omega,\Kt)}{\omega}\Bigg)\delta_{\hq-\haq'} 
    + \sqrt{2}m\varepsilon_{\hq}^{i*}\frac{\mathcal{Q}_0(\omega,\Kt)}{\omega^2}\delta_{\hq\haq'}\Bigg] \,.
\end{multline}
Here the functions $\phi^{ij}(z,h)$ and $\tau^{mn}(z,h)$ are defined in \eqs\eqref{eq: splitting helicity structure aq} and \eqref{eq: emission helicity structure q}, respectively, and the variable $\omega$ in \eq\eqref{eq: soft quark: ED omega}.
\subsubsection{Gluon emission from quark}
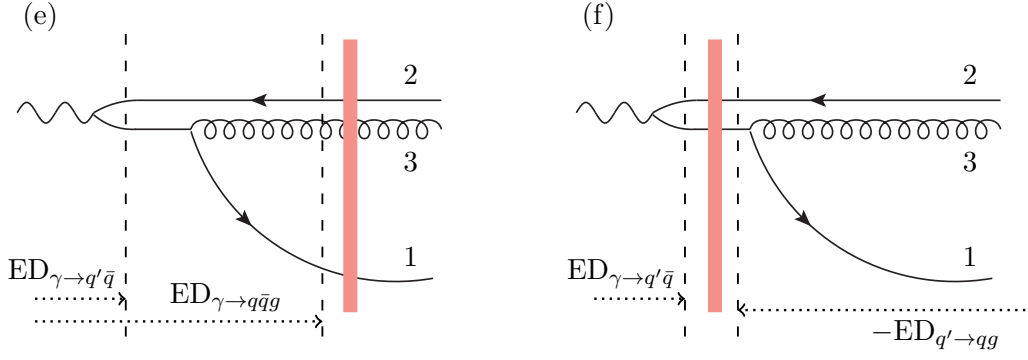
\begin{figure*}[tbp!]
\centering{
\includegraphics[width=0.4\textwidth]{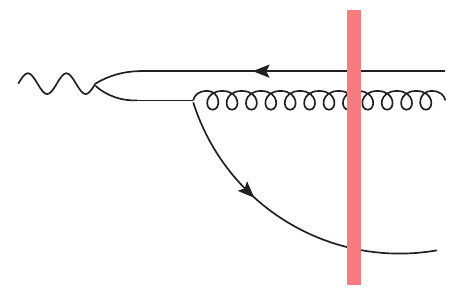}
\begin{tikzpicture}[overlay]
\node[anchor=south west] at (-6cm,3.7cm) {\hypertarget{diag:e}{(e)}};
    \draw[dash pattern=on 4pt off 6pt,line width=0.7pt, -](-4.5cm,3.8cm) -- (-4.5cm,-0.2cm);
    \draw[dash pattern=on 4pt off 6pt,line width=0.7pt, -](-1.9cm,3.8cm) -- (-1.9cm,-0.2cm);
    \draw[dotted, line width=0.9pt, -to](-5.7cm,0.3cm)-- (-4.5cm,0.3cm);
    \draw[dotted, line width=0.9pt, -to](-5.7cm,0.0cm) -- (-1.9cm,0.0cm);
        \node[anchor=south east] at (-4.5cm,0.3cm) 
         {ED$_{\gamma \rightarrow q'\bar{q}}$};
    \node[anchor=south east] at (-2.3cm,0cm) {ED$_{\gamma \rightarrow \qqg}$};
         \node[anchor=south east] at (-0.5cm,3cm) {$2$};
         \node[anchor=south east] at (-0.5cm,1.85cm) {$3$};
         \node[anchor=south east] at (-0.5cm,0.6cm) {$1$};
\end{tikzpicture}
\rule{2.5em}{0pt}
\includegraphics[width=0.4\textwidth]{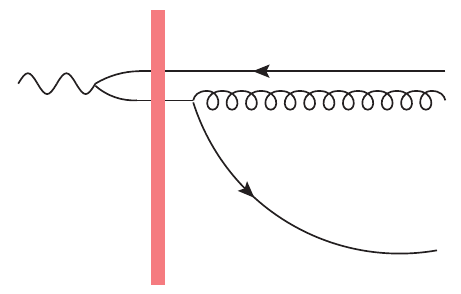}
\begin{tikzpicture}[overlay]
\node[anchor=south west] at (-6cm,3.7cm) {\hypertarget{diag:f}{(f)}};
    \draw[dash pattern=on 4pt off 6pt,line width=0.7pt, -](-4.5cm,3.8cm) -- (-4.5cm,-0.2cm);
    \draw[dash pattern=on 4pt off 6pt,line width=0.7pt, -](-3.8cm,3.8cm) -- (-3.8cm,-0.2cm);
    \draw[dotted, line width=0.9pt, -to](-5.7cm,0.3cm)-- (-4.5cm,0.3cm);
    \draw[dotted, line width=0.9pt, -to](0cm,0.15cm) -- (-3.8cm,0.15cm);
    \node[anchor=south east] at (-4.5cm,0.3cm) 
         {ED$_{\gamma \rightarrow q'\bar{q}}$};
    \node[anchor=south east] at (-0.2cm,-0.5cm) {$-\EDqg$};     
         \node[anchor=south east] at (-0.5cm,3cm) {$2$};
         \node[anchor=south east] at (-0.5cm,1.85cm) {$3$};
         \node[anchor=south east] at (-0.5cm,0.6cm) {$1$};
\end{tikzpicture}
}
\caption{Gluon emission from the quark before (e) and after (f) the interaction with the target (orange band). 
}
\label{fig: quark DTMD: soft gluon quark}
\end{figure*}
In \fig\ref{fig: quark DTMD: soft gluon quark} we represent the amplitudes corresponding to the two time orderings where the gluon emission from the quark can occur either before \hyperlink{diag:e}{(e)} or after \hyperlink{diag:f}{(f)} the interaction\footnote{The LCWF in diagram \hyperlink{diag:e}{(e)} is again the same as in \hyperlink{diag:a}{(a)}, but redrawn here since we will here evaluate it in a different kinematical regime.}. The corresponding wavefunctions are
\begin{align}
\label{eq: soft quark: LCWF diagram (e)}
    \Psi^{\hyperlink{diag:e}{(e)} }&=\sum_{q'}ee_f\frac{\bar{u}(\vec{k}'_1) \slashed{\varepsilon}_\lambda(q)v(\vec{k}_2)}{\EDqpq}gt^a\frac{\bar{u}(\vec{k}_1)\slashed{\varepsilon}^*_\sigma(\vec{k}_3)u(\vec{k}'_1)}{\EDqqg} \,, \\
\label{eq: soft quark: LCWF diagram (f)}
   \Psi^{\hyperlink{diag:f}{(f)} }&=\sum_{q'}ee_f\frac{\bar{u}(\vec{k}'_1) \slashed{\varepsilon}_\lambda(q)v(\vec{k}_2)}{\EDqpq}gt^a\frac{\bar{u}(\vec{k}_1)\slashed{\varepsilon}^*_\sigma(\vec{k}_3)u(\vec{k}'_1)}{-(\EDqg )} \,.
\end{align}

Note that the gluon emission by a quark is defined with an overall positive sign with respect to the gluon emission from the antiquark, described with \eqs\eqref{eq: soft quark: LCWF diagram (c)} and \eqref{eq: soft quark: LCWF diagram (d)}. Additionally, the gluon emission before \eqref{eq: soft quark: LCWF diagram (e)} and after \eqref{eq: soft quark: LCWF diagram (f)} the scattering has a relative minus sign,  since the diagram with gluon emission after the shockwave actually corresponds to gluon absorption in the language of LCWFs, and in an absorption the LC energy decreases rather than increasing.  We can anticipate that the amplitude~\hyperlink{diag:f}{(f)} will not contribute to the cross section, since the scattering of a small $\qbarq$ dipole is suppressed.

The phase-space integral for the intermediate quark in \eqs\eqref{eq: soft quark: LCWF diagram (e)} and \eqref{eq: soft quark: LCWF diagram (f)} is
\begin{equation}
    \sum_{q'}=\int\frac{\der^3 \Vec{k'_1}}{2k'^+_1(2\pi)^3}(2\pi)^3\delta(\Vec{k'_1}-\Vec{k_1}-\Vec{k_3})=\frac{1}{2(\zq+\zg)q^+}=\frac{1}{2q^+\zg} \,.
\end{equation}

We are interested in the special configuration where the antiquark and gluon are hard and back-to-back, $\ktaq \simeq \ktg \sim\Pt \gg \Qs$,  whereas the final quark is semi-hard $\ktq \sim \Qs$ and soft $\zq \ll1$. Since the hard gluon is emitted by the quark, this means that the intermediate quark $q'$ must be hard as well, with transverse momentum $\kbq'=\kbq+\kbg \simeq\kbg$. Thus, the hard quark-antiquark dipole with transverse size $\rt\sim 1/\Pt$ scatters weakly due to colour transparency, and diagram\,\hyperlink{diag:f}{(f)} is power suppressed. In contrast, the emission of a soft quark at some large distance $\Rt$ makes the $\qqg$ system to behave effectively as a large quark-antiquark dipole of size $\Rt$. Therefore, the only contribution to the scattering amplitude comes from diagram\,\hyperlink{diag:e}{(e)}. 

Using the notation $\omega$ defined in \eq\eqref{eq: soft quark: ED omega}, the energy denominators in \eq\eqref{eq: soft quark: LCWF diagram (e)} can be written as
\begin{align}
    \mathrm{ED}_{\gamma \rightarrow\qqg}&=q^- - k_1^{-}  - k_2^- - k_3^{-}  =-\frac{1}{2q^+}\bigg(\frac{\ktq^2+\omega^2}{\zq}\bigg) \,,\\
    \mathrm{ED}_{\gamma \rightarrow q'\bar{q}}&=q^--  {k'_1}^{-}  - k_2^-=-\frac{1}{2q^+\zaq\zg}\Big(\Qhat^2+P_{\perp}^2 + m^2\Big) \,.
\end{align}
Here we define $\Qhat^2=\zaq\zg Q^2$. Then, the light cone wavefunction for diagram\,\hyperlink{diag:e}{(e)} is 
\begin{multline}
\label{eq: soft quark: LCWF diagram E}
    \Psi^{\hyperlink{diag:e}{(e)} }(\zq,\zaq,\zg,\Pb,\kbq)=\frac{1}{\zg}2q^+ee_fgt\sqrt{\zq\zaq}\polp^i \polg^{m*} \frac{\Big(  \tilde{\phi}^{ij}(\zaq,h'_1) P^j\delta_{h'_1-\haq} - \sqrt{2}m\varepsilon_{h'_1}^{i*}\delta_{h'_1\haq}\Big)}{P_{\perp}^2+\Qhat^2+m^2} \\
    \cross \frac{ \Big(-k^n_1\tilde{\tau}^{mn}(\zaq,\hq)\delta_{\hq\hq'}-\zg\sqrt{2}m \varepsilon^{n*}_{\hq}\delta^{mn}\delta_{\hq-\hq'}\Big)}{\ktq^2+\omega^2} \,,
\end{multline}
where the functions $\tilde{\varphi}^{ij}(z,h)$ and $\tilde{\tau}^{mn}(z,h)$ are
\begin{align}
\label{eq: quark phi}
    \tilde{\phi}^{ij}(\zaq,h'_1)& =(1-2\zaq)\delta^{ij} + ih'_1\varepsilon^{ij} \,, \\
    \tilde{\tau}^{mn}(\zaq,\hq)&=(1-\zaq)\delta^{mn} -i\hq(1-\zaq) \varepsilon^{mn} \,.
    \label{eq: quark tau}
\end{align}

In the light cone wavefunction \eqref{eq: soft quark: LCWF diagram E}, we have substituted $\kbaq \simeq \Pb$ and $\bKb \simeq-\kbq$, where $\bKb$ is the relative momentum of the quark and gluon:
\begin{equation}
\label{eq: soft quark: imbalance barK}
\begin{aligned}
    \bKb&=\frac{\zq\kbg-\zg\kbq}{\zq+\zg} \,.
\end{aligned}
\end{equation}
Using that the total transverse momentum transferred from the target $\kto=0$, we can write the change of variables
\begin{align}
\label{eq: change of variables P}
    \kbq&=-\bKb -\frac{\zq}{\zq+\zg}\kbaq \,, \\
    \kbg&=\bKb-\frac{\zg}{\zq+\zg}\kbaq \,.
\end{align}

Similarly to the diffractive process with a soft gluon emission in Section~\ref{Sec: soft gluon}, we define in coordinate space the separation $\rb$ between the antiquark and the intermediate quark, the dipole size $\Rb$ of the final quark and gluon, and $\bb$ as the position of the intermediate quark: 
\begin{align}
    \rb&\equiv\yb - \bb \,, & & \Rb \equiv \xb - \zb \,, & & \bb \equiv \frac{ \zq\xb+  \zg\zb}{ \zq +  \zg} \,.
\end{align}
The inverse transformation is given by
\begin{align}
\label{eq: change variables 2}
    \xb&=\bb + \frac{ \zg}{ \zq +  \zg}\Rb \,,
    & & \yb = \bb +\rb \,,
    & & \zb = \bb -\frac{ \zq}{ \zq +  \zg}\rb \,.
\end{align}
With the above change of variables, the phase space involved in the Fourier transform from momentum to coordinate space of the light cone wavefunction \eq\eqref{eq: soft quark: LCWF diagram E} becomes 
\begin{equation}
\label{eq: change exponent 2}
    \kbq\cdot\xb + \kbaq\cdot\yb + \kbg\cdot\zb = \kbaq\cdot\rb + \kbq\cdot\Rb \simeq \Pb\cdot\rb + \kbq\cdot\Rb \,.
\end{equation}

The next step is to include the effect of the scattering, given by \eq\eqref{eq: soft quark: scattering barq}, since the quark eventually sees the two other final partons at $\zb\simeq\yb\simeq\bb$. Analogously to the previous calculation, where the gluon is emitted by the antiquark, we Fourier transform only with respect to the transverse momentum $\kbq$.

Finally, the amplitude corresponding to diagram\,\hyperlink{diag:e}{(e)} reads
\begin{multline}
\label{eq: soft quark: amplitude q}
    \mathcal{A}_{q} ( \zq, \zaq, \zg,\Pb,\Kb)=\frac{1}{ \zg}2q^+ee_fg\sqrt{ \zq \zaq}\frac{\varepsilon^i_\lambda\Big(  \tilde{\phi}^{ij}( \zaq,h'_1) P^j\delta_{h'_1-\haq} - \sqrt{2}m\varepsilon_{h'_1}^{i*}\delta_{h'_1\haq}\Big)}{{\Pt^2+\Qhat^2+m^2}}\\
    \cross \varepsilon^{m*}_{\sigma}\Bigg[\Bigg(\frac{K^n}{\Kt}\frac{\mathcal{Q}_1(\omega,\Kt)}{\omega}\Bigg)(-\tilde{\tau}^{mn}( \zaq,\hq))\delta_{\hq\hq'}+\frac{\mathcal{Q}_0(\omega,\Kt)}{\omega^2} \zg\sqrt{2}m\varepsilon^{n*}_{\hq}\delta^{mn}\delta_{\hq-\hq'} \Bigg].
\end{multline}
Here the functions $\tilde{\phi}^{ij}(z,h)$ and $\tilde{\tau}^{mn}(z,h)$ are defined in \eqs \eqref{eq: quark phi} and \eqref{eq: quark tau}, respectively, and the functions $\mathcal{Q}_1(\omega,\Kt)$ and $\mathcal{Q}_0(\omega,\Kt)$ in respectively \eqs\eqref{eq:Q1-def} and \eqref{eq:Q0-def}.
\subsection{TMD factorization of the cross section}
The differential cross section 
\begin{equation}
\label{eq: soft quark: prel. diff cross section}
\begin{aligned}
   \frac{\der\sigma^{\gamma_T^*A\rightarrow \qqg A }_\mathrm{D}}{\der \zq\der \zaq\der \zg\der^2\Pb\der^2 \Kb}&=\delta(1-\zaq-\zg)
   \frac{S_\perp\aem\as e_f^2\Cf\Nc }{2(2\pi)^4\zg}\sum_{\sigma,\lambda,h}\big|\mathcal{A}(\zq,\zaq,\zg,\Pb,\Kb)\big|^2 \,,
\end{aligned}
\end{equation}
is obtained from the squared amplitude, whose contributions are given in \eqs\eqref{eq: soft quark: amplitude qbar} and \eqref{eq: soft quark: amplitude q}:
\begin{equation}
\label{eq: soft quark: amplitude square}
    |\mathcal{A}(\zq,\zaq,\zg,\Pb,\Kb)|^2 = |\mathcal{A}_{\bar{q}}|^2 + |\mathcal{A}_{q}|^2 +2\Re\big(\mathcal{A}_{q}\mathcal{A}_{\bar{q}}^*\big) \,.  
\end{equation}
We then separately calculate each contribution in \eq\eqref{eq: soft quark: amplitude square} to the differential cross section \eqref{eq: soft quark: prel. diff cross section}. Starting from the direct term from the gluon emission by the antiquark, we get 
\begin{align}
\label{eq: soft quark: direct antiquark}
   \frac{\der\sigma_{\bar{q}}}{\der \zq\der \zaq\der \zg\der^2 \Kb\der^2 \Pb}&=
   \delta_z\frac{S_\perp\aem\as e_f^2\Cf\Nc }{4\pi^4}\frac{(1+\zaq^2)}{\zg}\frac{1}{\Pt^2}\bigg(\frac{\mathcal{Q}^2_1(\omega,\Kt)}{\omega^2}+\frac{m^2\mathcal{Q}^2_0(\omega,\Kt) }{\omega^4} \bigg) \;.
\end{align}
Here we have used the shorthand notation for the longitudinal momentum conserving delta function $\delta_z\equiv(1-\zaq-\zg)$, the functions $\mathcal{Q}_1(\omega,\Kt)$ and $\mathcal{Q}_0(\omega,\Kt)$ respectively defined in \eqs\eqref{eq:Q1-def} and \eqref{eq:Q0-def} and the variable $\omega$ in \eq\eqref{eq: soft quark: ED omega}. We recognize in \eqref{eq: soft quark: direct antiquark} the DGLAP splitting function 
\begin{equation}
    P_{gq}(\zg)=\Cf \frac{1+(1-\zg)^2}{\zg} \,,
\end{equation}
and the characteristic transverse momentum spectrum $1/\Pt^2$ for a hard splitting $\bar{q}'\rightarrow\bar{q}g$ that occurs in the final state.

The second term in \eq\eqref{eq: soft quark: amplitude square} represents the contribution from gluon emission by the quark:
\begin{align}
\label{eq: soft quark: direct quark}
   \frac{\der\sigma_{q}}{\der \zq\der \zaq\der \zg\der^2 \Kb\der^2 \Pb}&=
   \delta_z\frac{S_\perp\aem\as e_f^2\Cf\Nc }{4\pi^4}\frac{(\zg^2+\zaq^2)}{\zg} \nonumber \\
   &\hspace{3.4cm} \cross\frac{\Pt^2}{(\Pt^2+\Qhat^2)^2}\bigg(\frac{\mathcal{Q}^2_1(\omega,\Kt)}{\omega^2}+\frac{m^2\mathcal{Q}^2_0(\omega,\Kt)}{\omega^4} \bigg) \;,
\end{align}
where $\Qhat^2=\zaq\zg Q^2$. 

Finally, the last term in \eq\eqref{eq: soft quark: amplitude square}, corresponding to the interference, contributes with a negative sign as 
\begin{align}
\label{eq: soft quark: interference}
   \frac{\der\sigma_{q\bar{q}}}{\der \zq\der \zaq\der \zg\der^2 \Kb\der^2 \Pb}&=-
   \delta_z\frac{2S_\perp\aem\as e_f^2\Cf\Nc }{4\pi^4}\frac{\zaq^2}{\zg}\frac{1}{\Pt^2+\Qhat^2}\bigg(\frac{\mathcal{Q}^2_1(\omega,\Kt)}{\omega^2}+\frac{m^2\mathcal{Q}^2_0(\omega,\Kt) }{\omega^4}\bigg) .
\end{align}

In order to obtain the TMD factorization of the differential cross section, we need to transfer the semi-hard quark from the LCWF of the projectile to that of the target. In this target picture, the Pomeron splits into a colourless $q\bar q$ pair: a $t$-channel antiquark with splitting fraction $x$, which is absorbed by the produced quark-gluon pair, and a $s$-channel quark with splitting fraction $1-x$ which emerges in the final state as the semi-hard jet. For this picture and the associated TMD factorization to become manifest, one must replace the longitudinal variable $z_1$ in the previous results by $x$. This is easily done by using the conservation of the  minus light cone energy: the minus component of the $t$-channel antiquark, equal to $x \xpo \PN$,  must compensate the minus light cone energy of the virtual photon, $q^-=-Q^2/2q^+$, and put the $\barqg$ pair on shell. This implies $x \xpo = \xbarqg$ where
\begin{align}
    x_{\bar{q}g}&=\frac{1}{2q^+\PN} \Big(\frac{\Pt^2}{\zaq\zg}+\frac{m^2}{\zaq}+Q^2 \Big) \,,\\
    x_{\po}&=\frac{1}{2q^+\PN} \Big(\frac{\Pt^2}{\zaq\zg}+\frac{\Kt^2+m^2}{\zq}+Q^2 \Big) \,.
\end{align}
Then,
\begin{equation}
\label{eq: soft quark: variable x}
    x=\frac{x_{\bar{q}g}}{\xpo}=\frac{ \frac{\Pt^2}{\zaq\zg}+\frac{m^2}{\zaq}+Q^2 }{\frac{\Pt^2}{\zaq\zg}+\frac{\Kt^2+m^2}{\zq}+Q^2 } \,.
\end{equation}
By solving the above expression for $\zq$, we find 
\begin{equation}
\label{eq: soft quark: z1 -> x}
    \zq=\frac{x (\Kt^2+m^2)}{(1-x)\left(\frac{\Pt^2}{\zaq\zg} +Q^2\right)+\frac{m^2}{\zaq}} \approx \frac{x}{(1-x)}\frac{ (\Kt^2+m^2)}{\left(\frac{\Pt^2}{\zaq\zg} +Q^2\right)} \,. 
\end{equation}

Here, the soft quark condition $\zq\ll 1$ for generic values of $x$, requires $m^2+K^2_\perp\ll (\Pt^2/\zaq\zg) +m^2/\zaq+Q^2$, therefore both $m^2$ and $\Kt^2$ should be taken to be semi-hard. This shows that, in this process, the quark mass acts as a semi-hard scale, and not as a part of the hard scale as was the case for the gluon diffractive TMD (recall \eqn{Mh}).

\eqs\eqref{eq: soft quark: variable x} and \eqref{eq: soft quark: ED omega} can be combined to yield
\begin{equation}
   x=\frac{\omega^2 -m^2 + \frac{\zq}{\zaq}m^2}{ \omega^2  + \Kt^2}\simeq\frac{\omega^2-m^2}{\omega^2+\Kt^2} \quad\Rightarrow\quad
    \omega^2=\frac{m^2+x\Kt^2}{1-x} \,.
    \label{eq: soft quark: omega}
\end{equation}
Using \eq\eqref{eq: soft quark: z1 -> x} one can also change the measure
for the cross section, from $z_1$ to $x$:
\begin{equation}
    \frac{\der\zq}{\zq}=\frac{1}{1-x}\frac{\der x}{x} \quad\Rightarrow\quad 
    \frac{\der\sigma}{\der \ln(1/x)}=\frac{\der \sigma}{\der \zq} \frac{x}{(1-x)^2}\frac{\Kt^2+m^2}{\frac{\Pt^2}{\zaq\zg}+Q^2} \,.
\end{equation}
By applying the change to target variables to the individual contributions in \eqs\eqref{eq: soft quark: direct antiquark}, \eqref{eq: soft quark: direct quark} and \eqref{eq: soft quark: interference} and subsequently summing them, we obtain the differential cross section for diffractive $q\bar{q}g$ production with a soft quark. This can be factorized as 
\begin{equation}
\label{eq: soft quark: factorized cross section}
\begin{aligned}
   \frac{\der\sigma^{\gamma_T^*A\rightarrow \qqg A }_\mathrm{D}}{\der\zaq\der\zg\der^2 \Kb\der^2 \Pb\der\ln(1/x)}&=\frac{4\pi^2\aem}{Q^2}e_f^2 \mathcal{H}(\zaq,\zg,\Pt,Q)\frac{\der x q_\po (x,\xpo,\Kt,m)}{\der^2\Kb} \,.
\end{aligned}
\end{equation}
In the above result, the hard factor $\mathcal{H}$ is mass-independent:

\begin{equation}
\begin{aligned}
    \mathcal{H}(\zaq,\zg,\Pt,Q)&=\delta_z \frac{\as\Cf}{2\pi^2}\frac{1}{\zg}\frac{\Qhat^2\big[(\Pt^2+\Qhat^2)^2+\Pt^4\zg^2 +\Qhat^4\zaq^2\big]}{\Pt^2(\Pt^2+\Qhat^2)^3} \,,
\end{aligned}
\end{equation}
and it agrees with the expression obtained in Ref.\,\cite{Hauksson:2024bvv} for the analogous diffractive $\qqg$ production with massless quarks. 

The mass-dependent semi-hard factor in \eq\eqref{eq: soft quark: factorized cross section} is the quark diffractive TMD:
\begin{multline}
\label{eq:softquark:qDTMD}
    \frac{\der x q_\po (x,\xpo,\Kt,m)}{\der^2\Kb}= \frac{S_\perp \Nc}{8\pi^4}\frac{x}{1-x}\frac{\Kt^2+m^2}{(x\Kt^2+m^2)} \\
    \cross \bigg( \mathcal{Q}_1^2(x,\xpo,\Kt)+\frac{(1-x)m^2}{x\Kt^2+m^2}\mathcal{Q}^2_0(x,\xpo,\Kt)\bigg) \,,
\end{multline}
where the functions $\mathcal{Q}_1(x,\xpo,\Kt)$ and $\mathcal{Q}_0(x,\xpo,\Kt)$ are defined in \eqs\eqref{eq:Q1-def} and \eqref{eq:Q0-def}, respectively, with the change of variables in \eq\eqref{eq: soft quark: omega}. As follows from \eq\eqref{eq: soft quark: z1 -> x}, the quark mass $m$ is now treated as a semi-hard scale, in contrast with the diffractive scattering with soft gluon emission studied in Section \ref{Sec: soft gluon}. Setting the quark mass to zero, we recover the quark DTMD in Ref.\,\cite{Hauksson:2024bvv}. The quark DTMD can also be written in momentum space where Eq.\,\eqref{eq:softquark:qDTMD} can be generalized to include the $\Deltab$ dependence, as discussed in appendix~\ref{sec:DTMD-ms}. 

For the physical discussion to follow, it is useful to observe that the quark distribution per unit transverse area represents the quark (or antiquark) occupation number inside the Pomeron summed over colour and helicity states~\cite{Hauksson:2024bvv} (below, $\bb$ is the transverse coordinate of the measure quark, or impact parameter):
\begin{align}\label{occup}
\frac{1}{S_\perp} \frac{\der x q_\po (x,\xpo,\Kt)}{\der^2\Kb}\,=\,x\frac{\der N_q}{ \der x \der^2\bb \der^2\Kb}\,\equiv \,n_q(x,\xpo,\Kt)\,.
\end{align}
By integrating this quantity over $\ln(1/x)$ (at fixed $\xbj \equiv x\xpo $) one obtains the diffractive occupation number of the quark in the nucleus \cite{Mueller:2024frs}. Notice that the dependence upon the quark mass $m$ is generally kept implicit in our notations. When we will need to emphasise this dependence, we shall denote the quark occupation number as $n_q(x,\xpo,\Kt; m)$.
\subsection{Heavy-quark DTMD at tree-level}
\label{sec:Phys}

In this section, we shall explore the effects of including the quark mass $m$ on the quark diffractive TMD, as derived in the previous section. It is intuitively clear that $m$ will act as an infrared regulator in the quark distribution at low transverse momenta $\Kt$, thus competing with gluon saturation. It is therefore interesting to understand the interplay between these two mechanisms. In turn, this interplay must depend upon the ratio between $m$ and the saturation momentum $\Qs$, in a way that we would like to clarify  in what follows.

To that aim, we shall rely on the tree-level result in \eqn{eq:softquark:qDTMD} together with the semi-classical expression for dipole scattering amplitude $\mathcal{N}(R_\perp)$ provided by  the  McLerran-Venugopalan model~\cite{McLerran:1993ni,McLerran:1993ka}, that is
\begin{align}\label{MV}
    \mathcal{N}(R_\perp)&=
    1-\exp\left(-\frac{1}{4}Q_A^2R_\perp^2\ln\left(\frac{1}{R_\perp\Lambda}+e\right)\right)\,.  
\end{align}
Here, $Q_A^2$ denotes the colour charge density squared of the valence quarks per unit transverse area and $\Lambda$ is the QCD confinement scale. The expression in the exponent represents the scattering amplitude in the single scattering (2-gluon exchange) approximation. The saturation momentum $\Qs$ is conveniently defined as the value of $2/R_\perp$ at which the exponent is equal to one (meaning that multiple scattering starts to be important). This condition yields
\begin{align}\label{Qsat}
\Qs^2 = Q_A^2 \,\ln\left(\frac{\Qs}{2\Lambda}+e\right)\,.
\end{align}
Notice that, due to the logarithm in the r.h.s., $\Qs^2$ can be significantly larger than the original scale $Q_A^2$. The following, piece-wise, approximation to \eqref{MV} will also be useful:
 \begin{align}\label{MVapprox}
    \mathcal{N}(R_\perp)&\simeq\left\{
    \begin{array}{ll}
    \frac{1}{4}Q_A^2R_\perp^2\ln \frac{1}{R_\perp\Lambda} & \quad \textrm{when $\Qs R_\perp\ll 1$,}\\*[0.2cm]
    1-\exp\left(-\frac{1}{4}\Qs^2R_\perp^2\right) & \quad \textrm{when $\Qs R_\perp\gtrsim 1$,}
    \end{array}\right.
\end{align}
with the expression in the second line known as the GBW model~\cite{GolecBiernat:1999qd}. The strong scattering regime at $\Qs R_\perp\gtrsim 1$, where $\mathcal{N}(R_\perp)\sim \order{1}$ and multiple scattering becomes important, corresponds to strong colour fields in the structure of the Pomeron, hence to {\it gluon saturation}. Via \eqn{eq:softquark:qDTMD}, this implies a similar property for the  {\it diffractive quark distribution}  at sufficiently low transverse momenta and for not too large quark mass.

Before we discuss the novel aspects introduced by the quark mass, let us briefly summarise the corresponding results in the massless case ($m=0$)\,\cite{Hauksson:2024bvv}:

\texttt{(i)} For sufficiently large transverse momenta $\Kt^2\gg \tilde{Q}_\mathrm{s}^2(x)\equiv (1-x)\Qs^2$, the quark DTMD abruptly  falls, according to a power law $1/\Kt^4$. This reflects the weakness of the elastic scattering for a small dipole, cf. the first line in Eq.~\eqref{MVapprox}.

\texttt{(ii)} At lower momenta $\Kt^2\lesssim \tilde{Q}_\mathrm{s}^2(x)$, the distribution flattens out and eventually saturates, due to the importance of multiple scattering for a relatively large dipole. In particular, when $\Kt\to 0$, the quark occupation number \eqref{occup} reaches a value that is independent of $\Qs$ and equal to $n_0\equiv \Nc x/(8\pi^4)$. This is {\it quark saturation} inside the Pomeron.

\texttt{(iii)} The fact that the transition  between the dilute and the dense regimes occurs at the $x$-dependent, {\it effective}, saturation momentum $ \tilde \Qs(x)$ is because of the interplay between the two scales which control the integral over $R_\perp$ in \eqn{eq:Q1-def} (for a given value of $\Kt$): the saturation scale $\Qs$ (as implicit in the dipole amplitude) and the virtuality  $\omega$ (which in the massless case reads $\omega^2=x\Kt^2/(1-x)$).

As we will now argue, this scale  $ \tilde Q_s(x)$ is also the relevant  scale to be compared with the quark mass $m$ when the quark is heavy. For more clarity, let us first consider the $\Kt\to 0$ limit of \eqn{eq:softquark:qDTMD}. In this limit, and for a non-zero quark mass, the only non-vanishing contribution\footnote{When $\Kt\to 0$ but $m\ne 0$, the contribution involving $\mathcal{Q}_1$ vanishes since $J_1(KR)\to 0$ in that limit. Note that this would not be true when $m=0$; in that case, the whole contribution to the quark diffractive TMD comes from the $\mathcal{Q}_1$-piece and remains non-zero when  $\Kt\to 0$\,\cite{Hauksson:2024bvv}.}
is that proportional to  $\mathcal{Q}_0^2$, which reads
\begin{align}
\lim\limits_{\Kt\to 0} n_q(x,\xpo,\Kt)=\frac{\Nc x}{8\pi^4}\left[\int_0^\infty\der u u K_0(u)\,\mathcal{N}\left(\frac{u\sqrt{1-x}}{m}\right)\right]^2\,.\label{eq:qDTMD-small-Kt}
\end{align}
The above integral is dominated by $u\sim\order{1}$, due to the modified Bessel function which vanishes exponentially when $u\gg 1$. Hence the argument $R_\perp= u\sqrt{1-x}/m$ of the dipole amplitude obeys  $\Qs R_\perp \sim \Qs\sqrt{1-x}/m=\tilde \Qs(x)/m$. So, the integral can explore both the weak-scattering ($\mathcal{N}\ll 1$) and the strong-scattering ($\mathcal{N}\sim 1$) regimes, depending upon the value of the ratio $m/\tilde \Qs(x)$. 

When $m\lesssim \tilde \Qs(x)$, one has $\Qs R_\perp \gtrsim 1$ and the integral is sensitive to multiple scattering (cf. the second line in \eqn{MVapprox}). In particular, for a very light quark with $m\ll \tilde \Qs(x)$, the integral  can be evaluated using the  ``black disk'' limit $\mathcal{N}= 1$ and then the final result is then the same as in the massless limit (since 
$\int_0^\infty\der u u K_0(u)=1$). This shows that the two limits $\Kt\to 0$ and $m\to 0$ commute with each other, which was not {\it a priori} obvious --- indeed, by reversing the order of limits one selects either the first term (which involves $\mathcal{Q}_1^2$), or the second term (proportional to $\mathcal{Q}_0^2$), in the second line of \eqn{eq:softquark:qDTMD}.

In the opposite situation of a very heavy quark, $m\gg \tilde \Qs(x)$, the integral is controlled by a single hard scattering ($\Qs R_\perp\ll 1$), where the first line in \eqn{MVapprox} applies. Then a simple scaling argument demonstrates that the $\Kt\to 0$ limit of the quark diffractive TMD is strongly suppressed  w.r.t. the massless case, by a factor $\tilde{Q}_\mathrm{s}^4/m^4$. In that case, the quarks with $\Kt=0$ are still in a dilute regime.

These qualitative considerations can be made more precise by using the GBW version of the MV model (cf. the second line of \eqn{MVapprox}), for which the integral in \eqn{eq:qDTMD-small-Kt}
 can be performed analytically, to yield
 \begin{align}
\lim\limits_{\Kt\to 0}n_q(x,\xpo,\Kt)=n_0\left(1-te^{t}\textrm{E}_1(t)\right)^2\,,\quad t\equiv \frac{m^2}{(1-x)\Qs^2},
\end{align}
with $\textrm{E}_1(t)\equiv \int_t^\infty\der z \,e^{-z}/z$ the exponential integral function. Using the known approximations for this function at either low, or large, values of its argument, one can confirm our previous qualitative findings. In particular, for a light quark ($t\ll 1$), one finds
 \begin{align}
\lim\limits_{\Kt\to 0}n_q(x,\xpo,\Kt)\simeq n_0\left(1-t
\ln\frac{e^{-\gamma_E}}{t}\right)^2\quad \mbox{for}\quad t\ll 1,
\end{align}
where $\gamma_E\approx 0.577216 $ is the Euler-Mascheroni constant. Hence,  when taking the limit $m\to 0$, the massless occupation number at $\Kt\to 0$, that is, $n_q(x,\xpo,\Kt\to 0; m=0)=N_c x/(8\pi^4)$,
is approached from below.

It is straightforward to generalise the above conclusions to generic values of $\Kt$. Consider first the tail at large transverse momenta $\Kt\gg m,\,\tilde \Qs(x)$. In that regime, the quark mass should play no role and indeed the dominant behaviour is the same as in the massless case. The dominant contribution comes from the term proportional to $\mathcal{Q}_1^2$ in \eqn{eq:softquark:qDTMD} and can be evaluated by using the two-gluon exchange approximation for the dipole scattering amplitude\footnote{Indeed, the condition  $\Kt\gg \tilde \Qs(x)$ restricts the integral over $R$ to relatively small values $R\ll 1/\Qs$. When $x\ll 1$, this constraint is introduced by the oscillatory Bessel function $J_1(\Kt R)$ in \eqn{eq:Q1-def}, while for $1-x\ll 1$, it rather comes via the modified Bessel function $K_1(\omega R)$.}, cf.  the first line of  \eqn{MVapprox}. A straightforward calculation yields (see  Appendix~\ref{sec:DTMD-ms} for details and Ref.\,\cite{Hauksson:2024bvv} for a similar result):
\begin{align} 
   n_q(x,\xpo,\Kt)\underset{\Kt\gg\, m,\,\tilde \Qs}\simeq 
   \frac{\Nc x(1-x)^2}{8\pi^4}\frac{Q_A^4}{\Kt^4}\left[\frac{1}{2}+ {x}\ln\left(\frac{K_\perp^2}{c_1^2x(1-x)\Lambda^2}\right)\right]^2,
   \label{eq:DTMD-high-kt}
\end{align}
with $c_1=2e^{3/2-\gamma_E}$.  The expression within the square brackets is the sum of two contributions: a constant piece and a piece which is enhanced by the transverse logarithm $\ln(\Kt^2/\Lambda^2)$, but also multiplied by a factor of $x$. Since $\Kt^2\gg \Lambda^2$, the second piece dominates for {\it generic} values of $x$ (like $x\sim 1/2$), but not also for very small values $x\ll 1$, where the constant piece gives the leading contribution. For what follows, it is useful to understand the physical origin of these two contributions. This emerges from the explicit calculations in Appendix~\ref{sec:DTMD-ms}, with conclusions that we summarise here, for convenience.

As previously mentioned,  \eqn{eq:DTMD-high-kt} has been obtained by evaluating the dipole elastic amplitude
$ \mathcal{N}(R_\perp)$ in the single scattering approximation. This quantity is built with the propagator $\mathcal{D}(\qb)$ of the pair of gluons exchanged in the $t$-channel (one gluon with transverse momentum $\qb$ and the other one with $-\qb$):
\begin{align}\label{ND}
    \mathcal{N}(R_\perp)\simeq   =
     \int\der^2\qb \,\mathcal{D}(\qb) \left[1-e^{i\qb\cdot\Rb}\right]\quad\mbox{with}\quad 
     \mathcal{D}(\qb)=\frac{1}{2\pi}\,\frac{Q_A^2}{q_\perp^4}\,.
\end{align}
(It is easy to check that the above integral over $\qb$ yields the expression in the first line of  \eqn{MVapprox} in the leading logarithmic approximation.) When this expression is used within the integral \eqref{eq:Q1-def} for
 $\mathcal{Q}_1(\omega,\Kt)$ (with $\Kt\gg m,\,\tilde \Qs(x)$), one can identify two physical mechanisms for the production of a quark-antiquark pair by the Pomeron, corresponding to different values for $q_\perp$:
 
 \texttt{(i)} {\it Hard scattering: $q_\perp\gtrsim \Kt$.} This is the situation where the transverse momenta, $\qb$ and $-\qb$, carried by the two $t$-channel gluons are directly transmitted to the produced quark and antiquark. The respective integral over $\qb$ reads
 \begin{align}\label{HScatt}
     \int\der^2\qb \,\mathcal{D}(\qb)\,\Theta(q_\perp-\Kt)\simeq\,\frac{Q_A^2}{2\Kt^2}\,,
\end{align}
and is dominated by its lower limit $\Kt$. This contribution yields the constant term within the square brackets in \eqn{eq:DTMD-high-kt}.
 
 \texttt{(ii)} {\it Hard splitting: $q_\perp\ll \Kt$.} In this case, the gluons accumulate at relatively low transverse momenta $q_\perp\ll \Kt$ to create the (inclusive) gluon PDF $x_\po G(x_\po,\Kt^2)$ on the resolution scale $\Kt^2$. Then a (colour singlet) pair of gluons from this distribution creates a quark-antiquark pair via a hard splitting. In this case the integral over $\qb$ becomes
 \begin{align}\label{HSplitt}
     \int\der^2\qb \,\qb^2\,\mathcal{D}(\qb)\,\Theta(\Kt-q_\perp)\simeq\,\frac{Q_A^2}{2}\,\ln\frac{\Kt^2}{\Lambda^2}\,.
\end{align}
The quantity $\qb^2\,\mathcal{D}(\qb)\sim {Q_A^2}/{q_\perp^2}$ within the integrand is recognised as the bremsstrahlung spectrum of the valence quarks, while the result of the integration is (up to a factor) the gluon PDF  $x_\po G(x_\po,\Kt^2)$ (see \eqn{eq:MVPDF} in Appendix~\ref{sec:DTMD-ms}). This is the origin of the second term within the square brackets in \eqn{eq:DTMD-high-kt}. In this case, the factor $1/\Kt^2$ in the amplitude comes from the hard splitting.

When decreasing $\Kt$, the transition from the tail  in $1/\Kt^4$ to a relatively flat distribution occurs at the largest among the intrinsic scales $\tilde \Qs(x)$ and $m$.  When the saturation scale is much larger than the quark mass, $\tilde \Qs(x)\gg m$, the mechanism responsible for the flattening of the spectrum is gluon saturation/strong dipole scattering. We then have a genuine saturation plateau at $\Kt\lesssim \tilde \Qs(x)$, where the quark occupation number inside the Pomeron is of order $n_0$, which is as large as possible.

On the other hand, for a very heavy quark with $ m\gg \tilde \Qs(x)$, the integrals over the dipole size $R$ in \eqs\eqref{eq:Q1-def} and \eqref{eq:Q0-def} are cut off by the modified Bessel functions already at very small values $R_\perp^2 \sim (1-x)/m^2\ll 1/\Qs^2$, for which the scattering is weak and can be evaluated in the single-scattering approximation (the first line in Eq.~\eqref{MVapprox}).  There is still a tail $1/\Kt^4$ at asymptotically large momenta $\Kt\gg m$, but this flattens out at lower momenta $\Kt\lesssim m$, where the quark occupation numbers are roughly independent of $\Kt$ and very small --- of relative order
 $\tilde Q_s^4/m^4\ll 1$ w.r.t. to the massless case. 
 
 For this case too we have been able to obtain an analytic result, valid when $ m,\,\Kt \gg \tilde Q_s(x)$ and for generic values of $x$, but without assuming any hierarchy  between $m$ and $K_\perp$. When  specialised to the MV model \eqref{MV}, this reads (see Appendix~\ref{sec:DTMD-ms} for its derivation):
\begin{align}
   n_q(x,\xpo,\Kt)\underset{\tilde Q_s\ll \, m,\,\Kt}{\sim }\frac{N_cx(1-x)^2Q_A^4}{32\pi^4}\frac{[(m^2+2x\Kt^2)^2+m^2\Kt^2]}{(\Kt^2+m^2)^4}\,\ln^2\frac{\Kt^2+m^2}{\Lambda^2}
    \,.\label{eq:qDTMD-small-Qs}
\end{align}
When $\Kt\ll m$, one finds an occupation number of order $\tilde Q_s^4/m^4$, in agreement with the previous discussion. When $\Kt\gg m$,  one recovers the second term within the square brackets in  \eq\eqref{eq:DTMD-high-kt} --- the one that is enhanced by the transverse logarithm and describes a hard splitting. (The other term, which describes a hard scattering and would dominate at small $x$, is not properly accounted by the approximations leading to \eqn{eq:qDTMD-small-Qs}.) The above result confirms the fact that
the transition between the $1/\Kt^4$ tail at  large $\Kt$ and the ``saturation plateau'' at low $\Kt$ occurs around $\Kt\sim m$. 

\begin{figure}[t]
\centering
\includegraphics[width=0.49\textwidth,page=1]{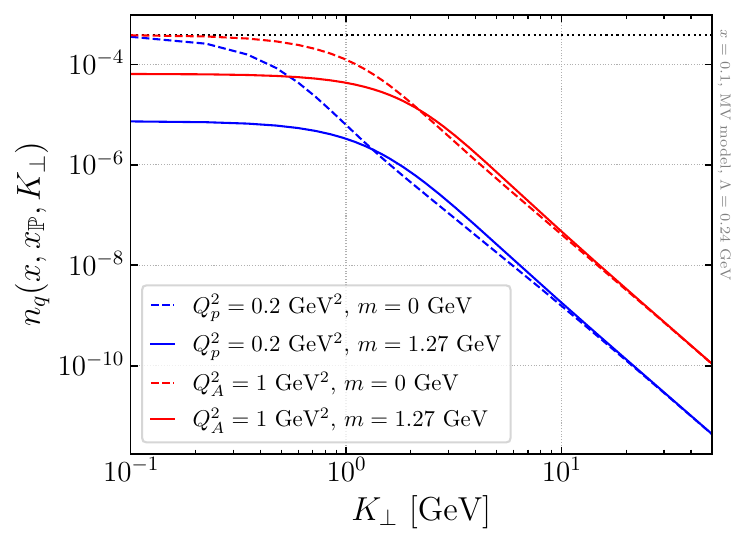}\hfill
\includegraphics[width=0.49\textwidth,page=2]{diags/plot-qTMD-mass.pdf}
  \caption{(Left) Quark occupation number inside the Pomeron in the MV model for massless (dashed curves) and massive quarks (plain curves) for two different values of the saturation scale $\Qs$ corresponding to a proton (blue curves) and a nucleus (red curves) target. (Right) Ratio of the quark occupation number inside the Pomeron between nucleus and proton, normalized by $A^{2/3}$. Note that the quark occupation number $n_q$ is defined as the quark DTMD per unit of transverse area of the target, so at the level of the quark DTMD the effective normalization is actually $A^{4/3}$.}  
  \label{fig:plot-qTMD-MV}
\end{figure}

To render these considerations fully quantitative for generic values of $\Kt$ and for the interesting case of charm-quark production, where the quark mass $m=1.27$~GeV is comparable with $\Qs$, we have numerically computed the tree-level quark DTMD in \eqn{eq:softquark:qDTMD}  within the MV model, with the results shown in the left plot in Fig.\,\ref{fig:plot-qTMD-MV}. To illustrate the effects of the quark mass, we show results for both $m=0$ and $m=1.27$~GeV. Also, to better distinguish the effects of gluon saturation, we have shown results obtained with two values for the parameter $Q_A^2$ of the MV model, cf. \eqn{MV} --- a larger value $Q_A^2=1$~GeV$^2$ corresponding to a heavy nucleus ($A$) and a smaller one $Q_p^2=0.2$~GeV$^2$  for the case of a ``proton'' ($p$). (Notice that $Q_A^2=A^{1/3}Q_p^2$ with $A^{1/3}=5$.) The respective saturation momenta in the MV model (cf. \eqn{Qsat}) will be denoted as $Q_{sA}$ and $Q_{sp}$.
The value of $x$ is fixed to 0.1. 

As anticipated, a main effect of the non-zero quark mass is a substantial reduction in the height of the ``saturation plateau'' at relatively low transverse momenta. This reduction is considerably stronger for the case of the proton target, since the ratio $\tilde Q_{sp}/m$ is considerably smaller than for the nucleus. Also the fact that the quark mass is comparable to the nuclear saturation momentum explains why the transition between the ``saturation plateau''  and the tail $1/\Kt^4$ occurs at similar scales in the massive case, while in the massless case the respective scales are strongly separated (roughly by a factor $A^{1/3}=5$).

By plotting the quark diffractive TMD divided by the target area in Fig.\,\ref{fig:plot-qTMD-MV} (left), we have removed the trivial $A$-dependence associated with $S_\perp(A)\propto A^{2/3}$. But even with this normalisation, the results for the nucleus and for the proton differ by a large factor, which at sufficiently large $\Kt$ appears to be roughly the same for $m=0$ and for the massive case. The large-$\Kt$  behaviour can 
be easily understood with reference to \eqn{eq:DTMD-high-kt}, whose r.h.s. is independent of the quark mass and proportional to $Q_A^4$. Hence, when $\Kt\gg m,\,\tilde Q_s(x)$, we expect the nucleus-to-proton ratio for the results in the left plot of Fig.\,\ref{fig:plot-qTMD-MV} to be close to $Q_A^4/Q_p^4=A^{2/3}=25$. In order to verify this expectation and also to better visualise the $A$-dependence at lower values for $\Kt$, we have shown in the right plot in Fig.\,\ref{fig:plot-qTMD-MV} the following ratio (the variables $x$ and $x_\po$ are left implicit in what follows):
\begin{align}\label{RpA}
    R_{pA}(\Kt) \equiv \frac{1}{A^{2/3}}\,\frac{n_q^A(x,\xpo,\Kt)}{n_q^p(x,\xpo,\Kt)}\,,
\end{align}
which at large $\Kt$ should approach to one according to our previous arguments. Besides the two curves for the quark DTMD, one for $m=0$ (dashed blue curve) and another one for $m=1.27$ GeV (red curve), we also show the similar ratio for the gluon DTMD in grey. Remarkably, the quark DTMD  features a {\it Cronin-like peak}, which is more pronounced when the mass of the quark decreases. Such a peak has been originally observed (and explained)  in relation with the {\it inclusive gluon TMDs} (the WW TMD and the dipole one), as computed in the MV model~\cite{Kharzeev:2003wz,Iancu:2004bx}. Its emergence is by no mean generic or automatic, and as a matter of fact there is no peak for the gluon {\it diffractive} TMD, as shown by the grey curve in Fig.\,\ref{fig:plot-qTMD-MV} (right).
So, it would be interesting to understand both the origin of this peak in the quark diffractive TMD at small $x$ and the difference w.r.t. the gluon DTMD.

Since the Cronin peak in the quark case is more pronounced when $m=0$, let us first consider the case of a massless quark. In that case, the transition between the saturation plateau at relatively low transverse momenta and the $1/\Kt^4$ tail at high momenta occurs at $\Kt\sim \tilde Q_s(x)$ for both the nucleus and the proton, but the respective saturation scales are widely separated, of course. For $\Kt\lesssim \tilde Q_{sp}(x)$, the quark occupation numbers inside the Pomeron are of order one for both the nucleus and the proton; hence, the ratio Eq.\,\eqref{RpA} reduces to the overall normalisation factor $A^{-2/3}$. For intermediate momenta $\tilde Q_{sp}(x)\ll \Kt\lesssim \tilde Q_{sA}(x)$, the quark distribution is still in a dense regime for the nucleus, but it already exhibits the dilute tail $\propto 1/\Kt^4$ in the case of the proton. Finally, in the high-$\Kt$ regime at $\Kt\gg \tilde Q_{sA}(x)$, the tail in \eqn{eq:DTMD-high-kt} applies to both $A$ and $p$ and the ratio $R_{pA}$ is expected to approach to one, as previously mentioned. To summarise, 
\begin{align}
	\label{Rpw}
	R_{pA}(K_{\perp}) \simeq
	\begin{cases}    
    \displaystyle{\ \frac{1}{A^{2/3}}}
        &\quad \text{for} \quad
    K_\perp^2\!\lesssim \tilde Q_{sp}^2,
    \\*[0.3cm]
    \displaystyle{\ \frac{1}{A^{2/3}}\,\frac{4K_{\perp}^4}{Q_{p}^4}}
    &\quad \text{for} \quad
    \tilde{Q}_{sp}^2 \ll K_\perp^2\!\lesssim \tilde Q_{sA}^2,
    \\*[0.3cm]
    \displaystyle{\ 1} 
        &\quad \text{for} \quad
    K_\perp^2\!\gg \tilde Q_{sA}^2.
    \end{cases}
   \end{align}
  In obtaining the intermediate line, we have used approximations valid at $x\ll 1$ for both the nucleus and the proton, in line with the fact that the plots in Fig.\,\ref{fig:plot-qTMD-MV} refer to $x=0.1$. Specifically, for the proton we kept only the first term within the square brackets in \eqn{eq:DTMD-high-kt}, that is\footnote{In the subsequent discussion of the case $x\ll 1$, we shall not distinguish between $\Qs$ and $\tilde Q_s$.},
   \begin{align} 
  n_q^p(x,\xpo,\Kt; m=0)\simeq 
   \frac{n_0}{4}\frac{Q_p^4}{\Kt^4}  \quad\mbox{when}\quad 
\Kt\gg\, Q_{sp} \quad \mbox{and} \quad
   x\ll1\,. \label{eq:high-kt-low-x}
\end{align}
For the nucleus, which is at saturation in the relevant range in $\Kt$, we have used the following result
  \begin{align} 
  n_q^A(x,\xpo,\Kt, m=0)\simeq 
  n_0\,e^{-2\Kt^2/Q_{sA}^2} \quad\mbox{for}\quad 
\Kt\lesssim Q_{sA} \quad \mbox{and} \quad
   x\ll1\,, \label{eq:sat-low-x}
\end{align}
which has been obtained in Ref.\,\cite{Hauksson:2024bvv} by using the GBW model (a good approximation to the MV model in the saturation region, cf.~\eqn{MVapprox}). The Gaussian in the r.h.s. expresses the momentum broadening of the $\Kt$-distribution due to multiple soft scattering off the gluons at saturation. We have not explicitly indicated this Gaussian in \eqn{Rpw}, to simplify writing, but it will be taken into account in our subsequent arguments.

The piece-wise approximation in \eqn{Rpw}  allow us to explain the gross features visible in the right plot in Fig.\,\ref{fig:plot-qTMD-MV}, and notably the Cronin peak. From the figure we see that, in the massless case, the peak is located at a value $\Kt\sim 1~\mbox{GeV}$ which is close to the nuclear saturation momenta. To study that region, we  use the intermediate line in  \eqn{Rpw}:
   \begin{align} \label{Cronin}
   R_{pA}(K_{\perp}\sim Q_{sA}) \sim \,
   \frac{4e^{-2}}{A^{2/3}}\,\frac{Q_{sA}^4}{Q_{p}^4}\,= \,
   4e^{-2}\ln^2\left(\frac{Q_{sA}}{2\Lambda}+e\right)\,,
   \end{align}
 where we also used \eqn{Qsat} for the saturation momentum in the MV model.  Since $Q_{sA}^2\gg \Lambda^2$, the final result is larger than one, thus explaining the Cronin peak. 
 
 One may think that the emergence of the logarithm $\ln(Q_{sA}/\Lambda)$ in the above estimate for $R_{pA}$ is merely an ``accident'' of the peculiar definition of the saturation momentum in the MV model, \eqn{Qsat}, but in fact this has a profound physical origin, that we briefly explain now. To that aim, we adapt the respective discussion in  Ref.\,\cite{Iancu:2004bx}, that was originally formulated for the gluon WW TMD. Within the MV model, the effect of multiple scattering off uncorrelated colour sources is to redistribute the quarks in $\Kt$-space: quarks that would lie in the bins at $\Kt\to 0$ in the power-law spectrum $1/\Kt^4$ in the absence of rescattering are pushed towards bins at higher $\Kt$~\cite{Kharzeev:2003wz}. Yet, due to the Gaussian nature of the emerging spectrum, as illustrated in \eqn{eq:sat-low-x}, most of these displaced quarks remain within the saturation domain at $\Kt\lesssim Q_s$, where they create a dense system with occupation numbers of order $n_0$. In the higher bins at $\Kt\gg Q_s$, the quark occupancy is essentially unchanged and can be read off from \eqn{eq:DTMD-high-kt}. For $x\ll 1$, this gives a quark occupation number $\sim n_0(Q_A^4/\Kt^4)$. If extrapolated to lower momenta $\Kt\sim Q_{sA}$, this becomes $n_0(Q_A^4/Q_{sA}^4)= n_0/[\ln(Q_{sA}/\Lambda)]^2 \ll 1$, which is much smaller than the corresponding value at saturation $n_0$. We thus find that the gluon occupation numbers predicted by \eqs\eqref{eq:sat-low-x} and respectively Eq.\,\eqref{eq:DTMD-high-kt} cannot be continuously 
 connected with each other. This shows that there should be a rather abrupt decrease in the quark occupancy  around $\Qs$, which is not correctly captured by our analytic approximations\footnote{For the case of the gluon WW TMD, this property has been quantitatively demonstrated in~\cite{Iancu:2004bx}.} but is included in our numerical estimates for $R_{pA}$. This mismatch does not affect the analytic argument in  \eqn{Cronin} since in that case one compares the gluon occupancy in the nucleus near the edge  of the saturation domain (where \eqn{eq:sat-low-x} applies)  to that in the proton in bins well above the respective saturation scale $Q_{sp}$ (where \eqn{eq:DTMD-high-kt} is indeed justified).
 
So far, we have concentrated on the case $x\ll 1$, as this matches our numerical results. But on the basis of our analytic results, one can also predict the trend with increasing $x$. For generic values like $x\sim 1/2$, the second term in \eqn{eq:DTMD-high-kt} will dominate over the first one since it is enhanced by the transverse logarithm. But this also means that, when computing  $R_{pA}$ for $\Kt\sim \tilde Q_{sA}(x)$, the r.h.s. of  \eqn{eq:high-kt-low-x} for the proton will acquire an additional factor $\ln^2(\Kt^2/\Lambda^2) \sim \ln^2(\tilde Q_{sA}/\Lambda^2)$. After including this factor in the estimate \eqref{Cronin}, it is clear that the logarithmic enhancement in $R_{pA}$ for $\Kt\sim \tilde Q_{sA}(x)$ will roughly cancel between the additional logarithm squared from the proton distribution and the one produced by the relation \eqref{Qsat} between $Q_{sA}$ and $Q_A$. This discussion predicts the disappearance of the Cronin peak with increasing $x$. We have verified this prediction via numerical calculations and found that the peak indeed disappears when $x\ge 0.3$.

This last argument also explains why there is no Cronin peak for the gluon DTMD. In that case too, the $1/\Kt^4$ tail at large $\Kt$ involves two contributions~\cite{Iancu:2022lcw, Hauksson:2024bvv}, as  in \eqn{eq:DTMD-high-kt}: one that corresponds to a hard splitting and is enhanced by the transverse logarithm $\ln(\Kt^2/\Lambda^2)$ (from the gluon PDF), and another one which is generated by a gluon exchange with momentum $\sim \Kt$ and shows no logarithmic enhancement. However, in the gluon case, none of these terms is suppressed at small $x$, so the terms involving the transverse logarithm dominates at all values of $x$, thus preventing the emergence of a Cronin peak.

Returning to the case of the quark DTMD, it is quite easy to understand the main effects of the quark mass on the $R_{pA}$ ratio, as visible too in Fig.\,\ref{fig:plot-qTMD-MV}. As already mentioned, a non-zero mass  suppresses the height of the ``saturation plateau'' and this suppression (which is proportional to $(Q_s/m)^4$) is considerably stronger for a proton than for a nucleus. Hence, after including the quark mass, the ratio  $R_{pA}$ gets considerably enhanced (as compared to the massless case) at very low momenta $\Kt\lesssim Q_{sp}$. Another effect of the mass is to remove the suppression of the ``hard splitting contribution'' to the quark DTMD (the contribution proportional to the transverse logarithm squared, cf. \eqn{eq:qDTMD-small-Qs}) at small $x$. Accordingly, this contribution starts to dominate over the ``hard scattering'' piece, with the effect that the Cronin peak gets strongly suppressed --- in agreement with the numerical results in Fig.\,\ref{fig:plot-qTMD-MV}. We expect the peak to be totally washed out when further increasing the mass.

\section{Summary}
\label{Sec: summary}

In this paper, we have studied TMD factorization within the Colour Glass Condensate effective theory for diffractive dijet production in photon-nucleus scattering at leading order, including the effects of non-zero quark masses. Our main results are as follows: (i) for dijet processes sensitive to the gluon diffractive TMD, the gluon DTMD is identical to that obtained in the massless case; the mass dependence enters only through the hard factor and can therefore be regarded as a hard scale from the perspective of TMD factorization; (ii) for dijet processes sensitive to the quark diffractive TMD, the distribution acquires a mass-dependent correction, which we compute here for the first time, while the hard factor can be evaluated in standard perturbative QCD with massless quarks. In this latter case, the quark mass should be treated as a semi-hard scale, analogous to the saturation scale, making it particularly interesting to investigate the interplay between saturation and mass effects in the quark diffractive TMD distribution.

More precisely, after introducing the kinematics for diffractive dijet production in photon–nucleus interactions in Section~\ref{Sec:kinem}, we have first investigated in Section~\ref{Sec: soft gluon}, within the CGC framework, the diffractive production of two massive quarks accompanied by an unmeasured soft gluon --- whose presence in the final state is necessary in order to have a leading-twist contribution to the cross section, as first noted in Ref.\,\cite{Iancu:2021rup}. Analogous to the massless case studied in Refs.\,\cite{Iancu:2022lcw,Hauksson:2024bvv}, the CGC cross section in the limit $M_h \gg K_\perp, Q_s$ where $M_h^2= Q^2+M_{q\bar q}^2$ factorizes in terms of the gluon DTMD at the scale $K_\perp$, which encodes the transverse momentum dependence of the gluon distribution inside the Pomeron. In this case, mass effects only modify the hard factor multiplying the gluon DTMD that enters the cross section. This hard factor is identical to the one computed in standard collinear factorization for the $\gamma^*g \to q\bar{q}$ process at leading order; in particular, it coincides with that for inclusive back-to-back massive $q\bar{q}$ production. Although this result is expected, the LCPT calculation within the CGC framework clarifies the regime of masses where TMD factorization is valid: in the absence of any other hard scale like $P_\perp$ or $Q$, the mass of the heavy quarks should be (much) larger than the total transverse momentum of the pair and the saturation scale. The TMD-factorized expression for diffractive massive $q\bar{q}$ pair production obtained in this work can be used for phenomenological studies of diffractive double open-heavy-flavor production in UPCs or at the future EIC. As mentioned in the introduction, another interesting process is forward diffractive $J/\Psi$ or $\Upsilon$ production at small transverse momenta in UPCs or at the EIC, where the mass of the charm or bottom quarks becomes the hard scale of the process, since the two heavy quarks are typically produced with a small relative transverse momentum. Consequently, the applicability of TMD factorization is expected to be more favorable for bottomonia than for charmonia. The phenomenology of this process, that we leave for future work, can be achieved in a joint CGC + non-relativistic QCD framework as investigated in Refs.\,\cite{Kang:2013hta,Cheung:2024qvw}.

At the same $\as$ order in pQCD, there is also the possibility of diffractively producing a massive quark (respectively antiquark)-gluon pair, accompanied by an unmeasured, soft, massive antiquark (respectively quark) that is integrated out, thereby rendering the corresponding diffractive dijet cross section leading twist. This final state has been considered in Section~\ref{Sec: soft quark}. In this case, the CGC expression for the diffractive dijet cross section factorizes in terms of the quark DTMD, but only if the mass is treated as a semi-hard scale, with $m^2 \ll  Q^2 + M_{qg}^2$, and without any hierarchy among $m$, $K_\perp$, and $\Qs$. Since $m$ is semi-hard, it does not appear in the expression of the hard factor, which is therefore identical to that obtained in collinear factorization at moderate $x$, computed at leading order for the partonic process $\gamma^* q \to qg$ with massless quarks. Instead, the quark mass modifies the quark DTMD itself, which acquires an explicit mass dependence. The expressions obtained within the CGC framework for the quark DTMD including mass corrections are novel. In particular, the formula in momentum space including the dependence on the total momentum transfer from the target can be found in Appendix~\ref{sec:DTMD-ms}.

In order to unravel the interplay between gluon saturation dynamics and mass effects in the quark DTMD, we have performed a tree-level analysis of the massive quark DTMD in the MV model. We show that at low transverse momenta, the quark mass plays a role analogous to the saturation scale in taming the growth of the quark occupation number in the Pomeron. This opens the possibility of probing the saturation regime by measuring the differential suppression of the quark DTMD for charm versus bottom quarks. 
Another interesting phenomenon that arises as the quark mass increases is the progressive suppression of the Cronin peak in the (properly normalized) ratio of the diffractive quark occupation number in a large nucleus to that in a proton. We emphasize, however, that these results are expected to be significantly modified by quantum evolution, in particular by CSS evolution.\footnote{The effects of quantum evolution on the gluon diffractive TMD have been considered in Ref.\,\cite{Iancu:2025jsu}.} In particular, it will be important to investigate the role of the Sudakov factor, including quark-mass effects, in the further suppression of the Cronin peak, along the lines of recent studies of inclusive heavy-meson pair correlation including soft gluon resummation~\cite{Marquet:2025jdr,Gao:2026azd}.

On the phenomenological side, the heavy-quark DTMD can be accessed by measuring forward dijet correlations with one $D$- or $B$-tagged jet and a light jet in diffractive events in UPCs or at the future EIC, or more simply through diffractive SIDIS with a $D$ or $B$ meson in the final state. Overall, our study lays the groundwork for a systematic description of processes that factorize in terms of diffractive gluon and quark DTMDs at small $x$, and that involve quarkonia or open heavy flavor in the final state, both at the LHC and at the future EIC.

\acknowledgments

We thank Nestor Armesto, Thierry Gousset, Cyrille Marquet and Dionysios Triantafyllopoulos for useful discussions. 

P.~C. is funded by the Agence Nationale de la Recherche under
grant ANR-25-CE31-5230 (TMD-SAT).
P.G.E. and T.L. have been supported by the Research Council of Finland, the Centre of Excellence in Quark Matter (projects 346324 and 364191)
and by the European Research Council (ERC, grant agreements No. ERC-2023-101123801 GlueSatLight and No. ERC-2018-ADG-835105 YoctoLHC). P.G.E. acknowledges  support from the Finnish Cultural Foundation. F.S. is supported by the Laboratory Directed Research and Development of Brookhaven National Laboratory and RIKEN-BNL Research Center, as well as the National Science Foundation (NSF)
within the framework of the JETSCAPE collaboration,
under grant number OAC-2514008 (CSSI:C-SCAPE). P.C., E.I. and F.S. acknowledge the Saturated Glue (SURGE) Topical Theory Collaboration, funded by the U.S. Department of Energy, Office of Science, Office of Nuclear Physics.
P.G.E gratefully acknowledges the Theory Group at Subatech for funding a research visit during the final stages of this work. F.S. also acknowledges the U.S. Department of Energy, Office of Science, Office of Nuclear Physics under the umbrella of the Quark-Gluon Tomography (QGT) Topical Collaboration with Award DE-SC0023646.
The content of this article does not reflect the official opinion of the European Union and responsibility for the information and views expressed therein lies entirely with the authors.

\appendix

\section{Scattering off an external field in Light Cone Perturbation Theory}
\label{sec: LCPT}

In this Appendix we first introduce the general notation of Light Cone Perturbation theory for scattering off a classical field \cite{Bjorken:1970ah}. Based on the general formulae, we will then focus on diffractive $\qqg$ production at leading-order in $\as$. 

The incoming and outgoing states called ``dressed states'' cross the shockwave (i.e. interact with the target field) at light cone time $ x^+=0$. These dressed states are obtained from the free particle states at $x^+\rightarrow  \mp \infty$ with the time-evolution operator $U_\mathrm{I}(x^+,x_0^+)$ in the interaction picture: 
\begin{subequations}
\label{eq: LCPT: dressed states}
\begin{align}
\label{eq: LCPT: in}
    |\phi\rangle_\mathrm{D}^{\mathrm{in}}&=U_\mathrm{I}(0,-\infty)|\phi\rangle\,, \\
\label{eq: LCPT: out}
    |\phi\rangle_\mathrm{D}^{\mathrm{out}}&=U_\mathrm{I}(0,\infty)|\phi\rangle = U_\mathrm{I}^\dagger(\infty,0)|\phi\rangle\,.
\end{align}
\end{subequations}
In the free theory, the interaction picture is equivalent to the Heisenberg picture where the free-theory eigenstates $|\phi\rangle$ do not depend on time. The interaction-picture dressed states are time dependent, but they are needed at the time $x^+=0$ where the target shockwave is placed. 

Both the free states and dressed states form an orthonormal basis in the space of Fock states since the time-evolution operator is unitary. Fock states are specified by the number of particles, their quantum numbers and their momenta, which we write here with a general notation $(a,b,n, \dots)$. For example, for single particle states, the orthonormalization convention is chosen as \cite{Bjorken:1970ah},   
\begin{equation}
    \big\langle a(\vec{p},\lambda_p,i_p) \big| b(\vec{q},\lambda_q,i_q ) \big\rangle = 2q^+(2\pi)^3 \delta^3(\vec{p}-\vec{q}\,)\delta_{i_p,i_q} \delta_{\lambda_p,\lambda_q} \,,
\end{equation}
where $\lambda_{p,q}$ is the light cone helicity and $i_{p,q}$ the colour index of the particles. The three dimensional delta function is defined as $\delta^{(3)}(\vec{p}-\vec{q}\,)=\delta({p}^+-{q}^+)\delta^{(2)}(\boldsymbol{p}-\boldsymbol{q})$.

For a theory with a light cone Hamiltonian $\hat{H}=\hat{H}_0+\hat{V}$, the multiparticle dressed states in \eq\eqref{eq: LCPT: dressed states} can be written as a Dyson perturbative series of the free particle states \cite{Brodsky:1997de, Li:2022tmi}:
\begin{multline}
    \label{eq: LCPT: Definition Fock state}
    |a\rangle_\mathrm{D}^{\mathrm{in,out}} =\Bigg[\prod_{i=1}^{N_a}\sqrt{Z_{i}}\Bigg]\Bigg[  |a\rangle + {\sum_{n}}'|n\rangle \frac{\langle n | \hat{V}| a\rangle}{E_a-E_n \pm i\delta} \\ 
    + {\sum_{n,m}}'|n\rangle\frac{\langle n | \hat{V}| m\rangle}{E_a-E_n \pm i\delta}\frac{\langle m | \hat{V}| a\rangle}{E_a-E_m \pm i\delta} + \dots\Bigg] \;.
\end{multline}
Here the energies $E$ are the eigenvalues of the free-part of the Hamiltonian, e.g. $\hat{H}_0|a\rangle=E_a|a\rangle$. The different sign of $i\delta$ between the in- and out- states is determined by the condition that these states must reduce to the free states in the past/future. This means that the evolution from the asymptotic state to the interaction at $x^+=0$ is forward/backward in time. That is, the time evolution operators are complex conjugates of each other \cite{Bjorken:1970ah}.  

In order to prevent the energy denominators to become zero in the perturbative series \eqref{eq: LCPT: Definition Fock state}, the intermediate states $|n\rangle$ where each individual particle has the same momentum as the asymptotic state $|a\rangle$  
are excluded of the sum denoted with a prime. These contributions are then included in single-particle renormalization factors $\sqrt{Z}$ for each of the $N_a$ particles in the state $|a\rangle$. These factors are determined from the normalization of the single-particle states and they can be chosen to be real. Although it would be possible to let this divergence to be regulated by $\pm i\delta$ \cite{Chen:1995pa}, it is more convenient to define the renormalization coefficients for the states, so that for this case the $i\delta$ term is irrelevant.
In some cases it is possible for the energy denominator to become zero for a state that is not identically the same as the incoming state, corresponding to contributions that are still there in the sum with a prime. These can correspond to e.g. collinear singularities associated with the emission of a massless particle  or to cuts that can give an imaginary part to a scattering amplitudes. The former must be regularized in some conventional way, while the latter are regularized by the $i\delta$ term, whose sign determines the sign of the imaginary part of the amplitude originating from this pole.
The leading order calculation in $\as$ presented in this manuscript is, however, safe from these singularities. We refer the reader to Ref.\,\cite{Beuf:2024msh} for a detailed discussion of this divergence in diffractive DIS at next-to-leading order in $\as$.  

In the Fock state expansion \eqref{eq: LCPT: Definition Fock state}, we use the following shorthand notation for the momentum space integration of the intermediate fluctuations together with overall momentum-conserving delta functions: 
\begin{multline}
    \label{eq: LCPT: Definition phase space}{\sum_{n}}'\coloneqq\sum_{\substack{\mathrm{helicity}\\\mathrm{colour}}}\prod_{i=1}^{N_n} \int \frac{\der k_i^+}{2\pi}\frac{\theta(k_i^+)}{2k_i^+}(2\pi)\delta\bigg(\textstyle\sum\limits_{j=1}^{N_a} k_j^+-\textstyle{\sum\limits_{k=1}^{N_n}k_k^+\bigg)}\\
   \cross\displaystyle\prod_{l=1}^{N_n}\displaystyle\int \frac{\der^2\boldsymbol{\mathrm{k}}_l}{(2\pi)^2}(2\pi)^2\delta^2\bigg(\textstyle\sum\limits_{m=1}^{N_a}\boldsymbol{\mathrm{k}}_m-\textstyle\sum\limits_{r=1}^{N_n}\boldsymbol{\mathrm{k}}_r\bigg) \,.
\end{multline}

The LCWFs are defined as the coefficients in the Fock state expansion \eqref{eq: LCPT: Definition Fock state}. They can be computed in a perturbative expansion
\begin{equation}
    \label{eq: LCPT: Definition LCWF}
    \Psi^{a\rightarrow n}=\frac{\langle n | \hat{V}| a\rangle}{E_a-E_n \pm i\delta} 
    + {\sum_{m}}'\frac{\langle n | \hat{V}| m\rangle}{E_a-E_n \pm i\delta}\frac{\langle m | \hat{V}| a\rangle}{E_a-E_m \pm i\delta} + \dots \;.
\end{equation}
Because the evolution  in light cone time $t \equiv x^+$ is generated by the light cone energy operator $E\equiv k^-$,  the energy denominator $\mathrm{ED}_{a\rightarrow n}=E_a - E_n $ has  a physical interpretation  as the inverse lifetime of the fluctuation from $a$ to $n$. If this fluctuation is a splitting, the energy increases and the energy denominator is negative. The matrix elements $\langle n | \hat{V}| a\rangle$ are vertices in Feynman rules, with $\hat{V}$ the interaction part of the Hamiltonian. For example, the matrix element for the leading order process where a photon fluctuates into a quark pair is written as
\begin{equation}
    \big\langle q(\vec{p}_1)\,\bar{q}(\vec{p_2})\, \big| \hat{V}\big|\,\gamma(\vec{q}\,)\big\rangle = (2\pi)^3 \delta^{3}(\vec{q}-\vec{p_1}-\vec{p_2})ee_f\bar{u}(\vec{p_1})\slashed{\varepsilon}(\vec{q}\,)v(\vec{p_2}) .
\end{equation}
Here $e$ is the electromagnetic coupling constant, $e_f$ the fractional charge of the quarks, and $\varepsilon$ the polarization vector of the photon.
Spinor matrix elements like $\bar{u}(\vec{p_1})\slashed{\varepsilon}(\vec{q}\,)v(\vec{p_2})$ for light front helicity polarization states in light cone gauge have been evaluated explicitly for the light front helicity states in many references, for example, in Ref.\,\cite{Lappi:2016oup}. A general definition also applicable in $2-2\varepsilon$ transverse dimensions can be found in Appendix C in Ref.\,\cite{Beuf:2022ndu}. 

The cross section for a general scattering process is given by
\begin{equation}
    \frac{\der \sigma^{\mathrm{in}\rightarrow \mathrm{out}}}{\der (\textrm{P.S.})_\mathrm{out} } =  \big|\mathcal{A}^{\mathrm{in}\rightarrow \mathrm{out}}\big|^2  \,,
\end{equation}
where the invariant phase space measure is an integral over the single-particle phase space elements,
\begin{equation}
\label{eq: LCPT: PS}
    \der (\textrm{P.S.})_\mathrm{out}= 2\sum_{i=1}^{N_\mathrm{in}}k_i^+ (2\pi)\delta\bigg(\textstyle\sum\limits_{j=1}^{N_{\mathrm{in}}} k_j^+-\textstyle{\sum\limits_{k=1}^{N_{\mathrm{out}}}}k_k^+\bigg)\displaystyle\prod_{l=1}^{N_\mathrm{out}} \int \frac{\der k_l^+ \der^2\boldsymbol{\mathrm{k}}_l}{2k_l^+(2\pi)^3} \,.
\end{equation}

The scattering amplitude $\mathcal{A}^{\mathrm{in}\rightarrow \mathrm{out}}$ together with a longitudinal momentum-conserving delta function is related to the matrix element of the asymptotic states with the target field scattering operator $\hat{S}$ as
\begin{equation}
    \label{eq: LCPT: Definition scattering amplitude}
    \prescript{\mathrm{out}}{\mathrm{D}}{\big\langle} b \,\big| \hat S - 1 \big| \, a \big\rangle_\mathrm{D}^{\mathrm{in}} =2\sum_{i=1}^{N_\mathrm{in}}k_i^+  (2\pi)\delta\bigg(\textstyle\sum\limits_{j=1}^{N_{\mathrm{in}}} k_j^+-\textstyle{\sum\limits_{k=1}^{N_{\mathrm{out}}}}k_k^+\bigg)i \mathcal{A}^{\mathrm{in}\rightarrow \mathrm{out}} .
\end{equation}

Let us now focus on the diffractive massive quark production at leading order in $\as$ considered in this work. For this process, the asymptotic states of the incoming virtual photon and the outgoing quark-antiquark-gluon are written as   
\begin{subequations}
\begin{align}
\label{eq: LCPT: Fock state photon}
    \big|\gamma^*\big\rangle^{\mathrm{in}}_\mathrm{D} &=  \big|\gamma^*\big\rangle + {\sum_{q\bar{q}}}'\Psi ^{\gamma^* \rightarrow q\bar{q}}\big|q_1\bar{q}_2\big\rangle + {\sum_{\qqg}}'\Psi^{\gamma^* \rightarrow q\bar{q}g}\big|q_1\bar{q}_2g_3\big\rangle + \dots  , 
    \\    \label{eq: LCPT: Fock state 3-parton}
     \big|q_1\bar{q}_2g_3\big\rangle^{\mathrm{out}}_{\mathrm{D}} &= \big|q_1\bar{q}_2g_3\big\rangle+ {\sum_{q\bar{q}}}'\Psi^{\qqg \rightarrow q\bar{q}}\big|q_1\bar{q}_2\big\rangle + {\sum_{\gamma^*}}'\Psi^{\qqg\rightarrow \gamma^*}\big|\gamma^*\big\rangle +\cdots .
\end{align}
\end{subequations}
Here we have dropped the renormalization factors $Z_{\gamma^*}=1+\mathcal{O}(\aem^2)$ and $Z_gZ_{q^2}=1+\mathcal{O}(\as)$. The latter has a correction related to the self-energy loop corrections of the $\qbarq$ final state, which do not contribute to the tree level $\qqg$ wavefunction. The light cone wavefunctions in \eqs\eqref{eq: LCPT: Fock state photon} and \eqref{eq: LCPT: Fock state 3-parton} can be related through the  
orthogonality of the states in the  in- and out-bases: 
\begin{equation}
\prescript{\text{out}}{\mathrm{D}}{\big\langle}q_1\bar{q}_2g_3 \big|\gamma^*\big\rangle_{\mathrm{D}}^{\text{out}} 
=
\prescript{\text{in}}{\mathrm{D}}{\big\langle}q_1\bar{q}_2g_3 \big|\gamma^*\big\rangle_{\mathrm{D}}^{\text{in}} 
=0
\end{equation}
This allows us to obtain very useful constraints for the wavefunctions: 
\begin{align}
    0=&\Big[ \big\langle q_1\bar{q}_2g_3\big|+ \big(\Psi^{\qqg \rightarrow q\bar{q}})^\dagger \big\langle q_1\bar{q}_2\big| +\big(\Psi^{\qqg \rightarrow \gamma^*})^\dagger \big\langle \gamma^*\big| \Big]\\
    &\cross\Big[\big|\gamma^*\big\rangle + \Psi^{\gamma^* \rightarrow q\bar{q}}\big|q_1\bar{q}_2\big\rangle + \Psi^{\gamma^* \rightarrow \qqg}\big|q_1\bar{q}_2g_3\big\rangle \Big] \nonumber \\
    \Psi ^{\gamma^* \rightarrow \qqg} =& - \big(\Psi ^{\qqg \rightarrow q\bar{q}})^\dagger\Psi ^{\gamma^* \rightarrow q\bar{q}}-(\Psi^{\qqg \rightarrow \gamma^*})^\dagger \,.
    \label{eq: LCPT: relation wavefunctions}
\end{align}
In this paper we are working at tree level where, as discussed above, the $i\delta$ regulator in the light cone energy denominators does not matter. As a consequence the in- and out-wavefunctions are the same. Thus at the tree level, this relation also applies when mixing wavefunctions before and after the cut, i.e. in- and out-states.
In fact we have used this relation in both the soft gluon and soft quark contributions to the cross section. The $1\leftrightarrow 3$ wavefunctions $\Psi ^{\gamma^* \rightarrow \qqg}$ and 
$ \Psi^{\qqg\rightarrow \gamma^*}$ include both regular emission and instantaneous contributions
\begin{equation}
    \Psi ^{\gamma^* \rightarrow \qqg} = \Psi ^{\gamma^* \rightarrow \qqg}_{\text{reg}}+ \Psi ^{\gamma^* \rightarrow \qqg}_{\text{inst}} \,.
\end{equation}
At tree level it is easy to see directly from the expressions that the instantaneous contributions are related by 
\begin{equation}
    \Psi ^{\gamma^* \rightarrow \qqg}_{\text{inst}} = -(\Psi^{\qqg \rightarrow \gamma^*}_{\text{inst}})^\dagger \,,
\end{equation}
and cancel separately in \eq\eqref{eq: LCPT: relation wavefunctions}. Thus, at the order in which we are working, the orthogonality relation~\eqref{eq: LCPT: relation wavefunctions} also applies to the regular emission contributions by themselves.

For the two diffractive processes shown in \fig\ref{fig: general diffractive scattering}, we will calculate the scattering amplitude $\mathcal{A}^{\gamma^*\rightarrow\qqg}$ from the scattering matrix element in \eq\eqref{eq: LCPT: Definition scattering amplitude}, which is now written as
\begin{equation}
\label{eq: LCPT: NLO amplitude}
    \prescript{\mathrm{out}}{\mathrm{D},\mathrm{s}}{\big\langle q_1\Bar{q}_2g_3} \big | \hat{S}-1 \big| \gamma^*\big\rangle_\mathrm{D}^{\mathrm{in}} = 2 {q}^+(2\pi) \delta(k_1^+ + k_2^+ + k_3^+ - {q}^+)i \mathcal{A}^{\gamma^*\rightarrow \qqg}\,.
\end{equation}

The subscript ``$s$'' in the matrix element above denotes the colour-singlet normalization of the outgoing state in a diffractive process:
\begin{equation}
\label{eq: LCPT: projection}
     \big| \qqg\big\rangle^{\text{out}}_{\text{D,s}}=\frac{t^a_{\alpha_1\alpha_2}}{\sqrt{\Cf\Nc}}|q_{\alpha_1}\bar{q}_{\alpha_2} g_a\big\rangle_{\mathrm{D}}^{\text{out}} \,,
\end{equation}
where $\Nc$ is the number of colours and $\Cf$ the Casimir operator in the fundamental representation, which satisfy $2\Cf\Nc=\Nc^2-1$. The generators of the fundamental representation of 
SU(3) are hermitean, $(t^a)^*_{ij}=t^a_{ji}$, and normalized such that $\text{tr}(t^at^b)=\frac{1}{2}\delta^{ab}$. When summing over an adjoint colour index, we will also use the Fierz identity:  
\begin{equation}
\label{eq: LCPT: Fierz identity}
    t^a_{ij}t^a_{k\ell}=\frac{1}{2}\delta_{i\ell}\delta_{jk}-\frac{1}{2\Nc}\delta_{ij}\delta_{k\ell} \,.
\end{equation}

The scattering operator $\hat{S}$ acting on the incoming state in mixed space $(z,\boldsymbol{\mathrm{x}})$ results in a Wilson line for each parton in the state. The Wilson line is a path-ordered exponential which describes the colour rotation of the probe when scatters the small-$x$ gluons of the target nucleus, represented as a classical gluon field $A^-$. In this interaction, the transverse coordinate $\boldsymbol{\mathrm{x}}$ of the parton as well as its polarization/helicity are conserved, which is known as the eikonal approximation. For quark and gluon probes, the Wilson lines are given by 
\begin{subequations}
\begin{align}
    \label{eq: LCPT: Definition Fund. WL}    V(\boldsymbol{\mathrm{x}})&=\mathcal{P}\text{exp}\bigg(-ig\int\der x^+ t^a A^-_a (x^+,0,\boldsymbol{\mathrm{x}})  \bigg) , \\
    \label{eq: LCPT: Definition Adj. WL}    U(\boldsymbol{\mathrm{x}})&=\mathcal{P}\text{exp}\bigg(-ig\int\der x^+ T^a A^-_a (x^+,0,\boldsymbol{\mathrm{x}})  \bigg) ,
\end{align}
\end{subequations}
where $T^a$ are the generators of the adjoint representation of $\mathrm{SU}(3)$. The Wilson lines in \eqs\eqref{eq: LCPT: Definition Fund. WL} and \eqref{eq: LCPT: Definition Adj. WL} are related by $U^{\dagger ab }t^b=Vt^aV^\dagger$.

\section{Instantaneous contribution}
\label{sec: instanteous contribution}
The light cone wavefunction for the instantaneous vertex with a relatively soft gluon emission by the quark, represented in diagram \hyperlink{diag:g}{(g)} is \cite{Hanninen:2017ddy,Beuf:2017bpd,Beuf:2022ndu}
\begin{equation}
\label{eq: soft gluon: LCWF diagram i}
    \Psi^{\hyperlink{diag:g}{(g)}}_{\text{inst}}=\frac{ee_fgt^a_{\alpha_0\alpha_1}}{\mathrm{ED}_{\gamma \rightarrow \qqg}}\bar{u}(\vec{k}_1) \slashed{\varepsilon}^*_{\sigma}(\vec{q})\frac{\gamma^+}{2(k^+_1+k^+_3)}\slashed{\varepsilon}_\lambda v(\vec{k}_2) \,.
\end{equation}
Here the energy denominator is the same as \eqref{eq: soft gluon: energy denominator qqg}:
\begin{equation}
    \text{ED}_{\gamma \rightarrow \qqg}=\frac{-1}{2q^+ \zg}\big(\ktg^2 + \Omega^2 \big) \,.
\end{equation}

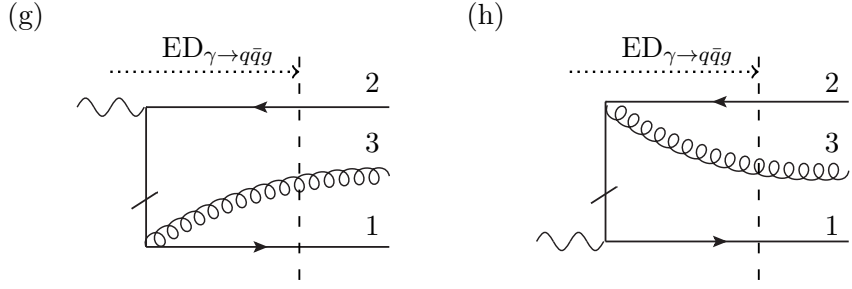
\begin{figure*}[tbp!]
\centering{
\includegraphics[width=0.3\textwidth]{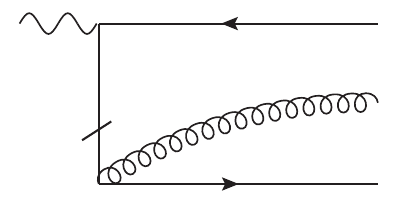}
\begin{tikzpicture}[overlay]
\node[anchor=south west] at (-5.5cm,2.9cm) {\hypertarget{diag:g}{(g)}};
    \draw[dash pattern=on 4pt off 6pt,line width=0.7pt, -](-1.5cm,2.7cm) -- (-1.5cm,-0.3cm);
    \draw[dotted, line width=0.9pt, -to](-4cm,2.5cm) -- (-1.5cm,2.5cm);
    \node[anchor=south east] at (-1.7cm,2.45cm) {ED$_{\gamma \rightarrow\qqg}$};
         \node[anchor=south east] at (-0.3cm,2.1cm) {$2$};
         \node[anchor=south east] at (-0.3cm,1.3cm) {$3$};
         \node[anchor=south east] at (-0.3cm,0.2cm) {$1$};
\end{tikzpicture}
\rule{3em}{0pt}
\includegraphics[width=0.3\textwidth]{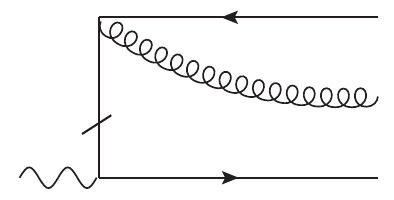}
\begin{tikzpicture}[overlay]
\node[anchor=south west] at (-5.5cm,2.9cm) {\hypertarget{diag:h}{(h)}};
    \draw[dash pattern=on 4pt off 6pt,line width=0.7pt, -](-1.5cm,2.7cm) -- (-1.5cm,-0.3cm);
    \draw[dotted, line width=0.9pt, -to](-4cm,2.5cm) -- (-1.5cm,2.5cm);
    \node[anchor=south east] at (-1.7cm,2.45cm) {ED$_{\gamma \rightarrow\qqg}$};
         \node[anchor=south east] at (-0.3cm,2.1cm) {$2$};
         \node[anchor=south east] at (-0.3cm,1.3cm) {$3$};
         \node[anchor=south east] at (-0.3cm,0.2cm) {$1$};
\end{tikzpicture}
}
\caption{Instantaneous diagrams contributing to the $\qqg$-component of the photon light cone wavefunction with soft gluon emission for generic $\beta$ values, with $\zg \sim \Qs^2 /h^2$.}
\label{fig: soft gluon: instantaneous}
\end{figure*}

By using the relation $ \slashed{\varepsilon}^*_{\sigma}(q)\gamma^+\slashed{\varepsilon}_\lambda=-\varepsilon_\lambda^i\varepsilon_\sigma^{*j}\gamma^+\gamma^j\gamma^i$ \cite{Beuf:2022ndu}, we can rewrite the light cone wavefunction \eqref{eq: soft gluon: LCWF diagram i} as
\begin{equation}
\begin{aligned}
    \Psi^{\hyperlink{diag:g}{(g)}}_{\text{inst}}(\zq,\zaq,\zg,\Pt,\ktg)&=
    -2q^+ \frac{ ee_fgt^a}{\ktg^2+\Omega^2}\frac{\zg\sqrt{ \zq \zaq}}{ \zq}\big(\delta^{ij} -i\hq\varepsilon^{ij} \big)\varepsilon^i_\lambda\varepsilon_\sigma^{j*}\delta_{\hq-\haq} \,.
\end{aligned}
\end{equation}
Similarly, the light cone wavefunction for the gluon emission by the antiquark is
\begin{equation}
\begin{aligned}
    \Psi^{\hyperlink{diag:h}{(h)}}_{\text{inst}}(\zq,\zaq,\zg,\Pt,\ktg)=&+2q^+ \frac{ ee_fgt^a}{\ktg^2 + \Omega^2 }\frac{\zg\sqrt{ \zq \zaq}}{ \zaq}\big(
    \delta^{ij} +i\hq\varepsilon^{ij} \big)\varepsilon^i_\lambda\varepsilon_\sigma^{j*}\delta_{\hq-\haq} \,.
\end{aligned}
\end{equation} 
Then, the total instantaneous contribution as a sum of the two diagrams in \fig\ref{fig: soft gluon: instantaneous} is 
\begin{equation}
\begin{aligned}
    \Psi_{\text{inst}}(\zq,\zaq,\zg,\Pt,\ktg)&= \Psi^{\hyperlink{diag:g}{(g)}}_{\text{inst}}+ \Psi^{\hyperlink{diag:h}{(h)}}_{\text{inst}}\\
    &=+2q^+\frac{ ee_fgt^a}{\ktg^2+\Omega^2}\frac{\Omega^2}{\Pt^2+\Qtil^2}\sqrt{ \zq \zaq}\varphi^{ij}(\zq,\hq)\varepsilon^i_\lambda\varepsilon_\sigma^{m*}\delta^{mj}\delta_{\hq-\haq}  \,,
\end{aligned}
\end{equation}
where we have used \eq\eqref{eq: rel. soft: Omega} to replace the factor $\zg/\zq\zaq$. 
\section{Heavy quark DTMD in momentum space}
\label{sec:DTMD-ms}

In this appendix, we present and discuss the momentum-space expression for the heavy-quark diffractive TMD distribution, and provide additional analytic results for the quark DTMD in the MV model based on this formulation.

\subsection{Diffractive SIDIS with a heavy quark}

Instead of taking the Fourier transform at the level of \eqs\eqref{eq:Q1-def}-\eqref{eq:Q0-def}, it is physically more enlightening to derive the momentum space version of the quark DTMD by considering the diffractive SIDIS process in the limit $Q^2\gg \Kt^2$, where $K_\perp$ refers to the transverse momentum of the measured quark. It also illustrates the universality of the CGC expression for the (massive) quark DTMD. We will also use this calculation to generalize the results of the main text to the case of non-zero total transverse momentum transfer, $\Deltab$, as in Ref.\,\cite{Rodriguez-Aguilar:2024efj}. In the colour dipole picture and CGC framework, the diffractive production of a massive quark with transverse momentum $\Kb$ at LO reads (see e.g. Refs.\,\cite{Iancu:2020jch,Mueller:2024frs})
\begin{align}
            &\frac{\der\sigma_{\rm D}^{\gamma^*_TA\to q(\bar q)A}}{\der^2\Kb\der z_1\der z_2}=\frac{\aem e_f^2\Nc}{(2\pi)^4}  \delta_z  
        \int\der^2\mathbf{b}\int\der^2\Rb\der^2\Rb' e^{-i \Kb\cdot(\Rb-\Rb') }  \nonumber\\
    &\times \left\{2 \left[z_1^2 + z_2^2 \right]  \frac{\Rb \cdot \Rb'}{R_\perp R'_\perp}  \bar{Q}^2 K_1(\bar{Q} R_\perp)K_1(\bar{Q} R'_\perp)+2m^2K_0(\bar QR_\perp)K_0(\bar QR_\perp')\right\}\nonumber\\
    &\times \left\langle S_{q\bar q}(\mathbf{b}+\Rb/2,\mathbf{b}-\Rb/2)-1\right\rangle\left\langle S_{q\bar q}^*(\mathbf{b}-\Rb/2+\Rb',\mathbf{b}-\Rb/2)-1\right\rangle \,,
    \label{eq:diffractive-SIDIS-ip}
\end{align}
with the effective virtuality of the process $\bar Q^2=z_1z_2Q^2+m^ 2$. The transverse coordinate variable $\mathbf{b}$ physically represents the impact parameter.
To get this expression, one starts from the amplitude for $q\bar q$ production, projects on the colour singlet $q\bar q$ state and integrate over the transverse momentum of the unmeasured antiquark. For simplicity, we focus here on the coherent diffraction case where the CGC average is performed at the level of the amplitude. The result we shall obtain below can easily be generalized to the incoherent case --- see also Ref.\,\cite{Rodriguez-Aguilar:2024efj} for a study of incoherent diffractive quark production using coordinate space expressions for the quark DTMD --- by appropriately taking the CGC average at the level of the cross section. 

We now introduce the double Fourier transform of the dipole $S$-matrix with respect to the dipole size and the dipole impact parameter:
\begin{align}
    \mathcal{D}(\mathbf{q},\xpo,\mathbf{\Delta})&\equiv \int\frac{\der^2\mathbf{r}}{(2\pi)^2}\int\frac{\der^2\mathbf{b}}{(2\pi)^2}e^{-i\mathbf{q}\cdot\mathbf{r}-i\mathbf{\Delta}\cdot\mathbf{b}}\frac{1}{N_c}\left\langle\textrm{Tr}\left(V(\mathbf{b}+\mathbf{r}/2)V^\dagger(\mathbf{b}-\mathbf{r}/2)\right)\right\rangle_{\xpo} \,.\label{eq:DF-ms}
\end{align}
Expressing the dipole $S$-matrices in terms of $\mathcal{D}$, 
\begin{align}
   \langle S_{q\bar q}(\mathbf{x}_1,\mathbf{x}_2)\rangle=\int\der^2\qb\int\der^2\mathbf{\Delta} \ e^{i\qb\cdot(\mathbf{x}_1-\mathbf{x}_2)}e^{i\mathbf{\Delta}\cdot(\mathbf{x}_1+\mathbf{x}_2)/2} \  \mathcal{D}(\mathbf{q},\xpo,\mathbf{\Delta}) \,,
\end{align}
the equation~\eqref{eq:diffractive-SIDIS-ip} is written in momentum space as 
\begin{align}
        &\frac{\der\sigma_{\rm D}^{\gamma^*_TA\to q(\bar q)A}}{\der^2\Kb\der z_1\der z_2}=\frac{\aem e_f^2\Nc}{(2\pi)^4}  \delta_z  
        \int\der^2\mathbf{b}\int\der^2\Rb\der^2\Rb' e^{-i \Kb\cdot(\Rb-\Rb') }  \nonumber\\
    &\times \left\{2 \left[z_1^2 + z_2^2 \right]  \frac{\Rb \cdot \Rb'}{R_\perp R'_\perp}  \bar{Q}^2 K_1(\bar{Q} R_\perp)K_1(\bar{Q} R'_\perp)+2m^2K_0(\bar QR_\perp)K_0(\bar QR_\perp')\right\}\nonumber\\
    &\times \int\der^2\qb\der^2\mathbf{\Delta}\int\der^2\qb'\der^2\mathbf{\Delta}'e^{i(\mathbf{\Delta}-\mathbf{\Delta}')\cdot\mathbf{b}}\Big[e^{i\qb\cdot \Rb} -e^{-i\mathbf{\Delta}\cdot\Rb/2}\Big]\Big[e^{-i\qb'\cdot \Rb'-i\mathbf{\Delta}'\cdot (\Rb'/2-\Rb/2)} -e^{i\mathbf{\Delta}'\cdot\Rb/2}\Big]\nonumber\\
    &\times\mathcal{D}(\mathbf{q},\xpo,\mathbf{\Delta})\mathcal{D}^*(\mathbf{q}',\xpo,\mathbf{\Delta}')\,.
    \label{eq:qDTMD-step2}
\end{align}
We can now perform the integral over the impact parameter $\mathbf{b}$ which identifies $\mathbf{\Delta}=\mathbf{\Delta}'$ in the amplitude and complex conjugate amplitude. Note that in order to obtain the above expression, we have expressed the $-1$ terms (no scattering contribution) in the colour structures of Eq.\,\eqref{eq:diffractive-SIDIS-ip} using the identity 
\begin{align}
    1&=\int\der^2\qb \int\der^2\mathbf{\Delta} \ f(\mathbf{\Delta})\mathcal{D}(\mathbf{q},\xpo,\mathbf{\Delta})\,,
\end{align}
for any function $f$ such that $f(\boldsymbol{0})=1$, owing to the unitarity of the Wilson lines. The choice of the functions $f(\mathbf{\Delta})=e^{i\mathbf{\Delta}\cdot(\mathbf{b}-\Rb/2)}$ for the amplitude and $f(\mathbf{\Delta'})=e^{-i\mathbf{\Delta'}\cdot(\mathbf{b}-\mathbf{R}/2)}$ for the complex conjugate amplitude colour structures enables one to factorize the expression given in Eq.\,\eqref{eq:qDTMD-step2} into
\begin{align}
        &\frac{\der\sigma_{\rm D}^{\gamma^*_TA\to q(\bar q)A}}{\der^2\Kb\der z_1\der z_2}=\frac{\aem e_f^2\Nc}{(2\pi)^2}  \delta_z  
        \int\der^2\Rb\der^2\Rb' e^{-i \Kb\cdot(\Rb-\Rb') }  \nonumber\\
    &\times \left\{2 \left[z_1^2 + z_2^2 \right]  \frac{\Rb \cdot \Rb'}{R_\perp R'_\perp}  \bar{Q}^2 K_1(\bar{Q} R_\perp)K_1(\bar{Q} R'_\perp)+2m^2K_0(\bar QR_\perp)K_0(\bar QR_\perp')\right\}\nonumber\\
    &\times \int\der^2\mathbf{\Delta}\int\der^2\qb\int\der^2\qb'\left[e^{i(\qb+\mathbf{\Delta}/2)\cdot \Rb} -1\right]\left[e^{-i(\qb'+\mathbf{\Delta}/2)\cdot \Rb'} -1\right]\mathcal{D}(\mathbf{q},\xpo,\mathbf{\Delta})\mathcal{D}^*(\mathbf{q}',\xpo,\mathbf{\Delta}') \,.
\end{align}
One can then recast the previous coordinate space expression in momentum space:
\begin{align}
    \frac{\der\sigma_{\rm D}^{\gamma^*_TA\to q(\bar q)A}}{\der^2\Kb\der z_1\der z_2}&=2\alpha_{em} e_f^2 N_c \delta_z \int\der^2\mathbf{\Delta } \left[(z_1^2+z_2^2)|\mathcal{A}(\Kb,\mathbf{\Delta},\bar Q,m)|^2+|\mathcal{B}(\Kb,\mathbf{\Delta},\bar Q,m)|^2\right]
\end{align}
with
\begin{align}
 \mathcal{A}(\Kb,\Deltab,\bar Q,m)&\equiv \int\der^2\qb \mathcal D(\qb,x_\po) \int\frac{\der^2\Rb}{2\pi} e^{-i\Kb\cdot\Rb}\left(e^{i(\qb+\Deltab/2)\cdot\Rb}-1\right) \frac{\Rb^i}{R_\perp}\bar QK_1(\bar QR_\perp)\\
 &=-i\int\der^2\qb \ \mathcal D(\qb,x_\po,\mathbf{\Delta})\left[\frac{(\Kb-\qb-\mathbf{\Delta}/2)^i}{(\Kb-\qb-\mathbf{\Delta}/2)^2+\bar Q^2}-\frac{\Kb^i}{K_\perp^2+\bar Q^2}\right]\label{eq:A_difSIDIS-ip}\\
  \mathcal{B}(\Kb,\Deltab,\bar Q,m)&\equiv \int\der^2\qb \mathcal D(\qb,x_\po) \int\frac{\der^2\Rb}{2\pi} e^{-i\Kb\cdot\Rb}\left(e^{i(\qb+\Deltab/2)\cdot\Rb}-1\right) mK_0(\bar QR_\perp)\\
  &=\int\der^2\qb \ \mathcal D(\qb,x_\po,\mathbf{\Delta})\left[\frac{m}{(\Kb-\qb-\mathbf{\Delta}/2)^2+\bar Q^2}-\frac{m}{K_\perp^2+\bar Q^2}\right]\label{eq:B_difSIDIS-ip}
\end{align}
At this stage, we have not made any approximation so our result still contains all $K_\perp/Q$ power corrections at small $x$. Before taking the TMD limit $Q^2\gg K_\perp^2$ in diffractive SIDIS, we have to express the diffractive variables $\beta$ and $x_\po$ in terms of the final state kinematics. To do so, one needs to know the transverse momentum of the antiquark which has already been integrated out. However, it is possible to effectively introduce back this transverse momentum using that the total transverse momentum of the $q\bar q$ pair is given by the transverse momentum transferred from the target, i.e.~$\mathbf{\Delta}=\mathbf{k}_{1}+\mathbf{k}_{2}=\Kb+\mathbf{k}_2$. Thus, the transverse momentum of the integrated antiquark is $\mathbf{k}_2=\mathbf{\Delta}-\Kb$. Then, using plus longitudinal momentum conservation $ x_\po P^-+q^-=k_1^- + k_2^-$, we get
\begin{align}
    x_\po =\frac{1}{2P\cdot q}\frac{z_2\Kb^2+z_1(\Kb-\mathbf{\Delta})^2+\bar Q^2}{z_1z_2}\,,
\end{align}
and since $\beta=x_{\rm Bj}/x_\po$,
\begin{align}
     \beta = \frac{\bar Q^2-m^2}{z_2\Kb^2+z_1(\Kb-\mathbf{\Delta})^2+\bar Q^2}\,.
\end{align}
The diffractive SIDIS cross section reads then
\begin{align}
     \frac{\der\sigma_{\rm D}^{\gamma^*_TA\to qA}}{\der^2\Kb\der \ln(1/\beta)}&=\int_0^1\der z_1\int_0^1\der z_2 \ \beta\delta\left(\beta-\frac{\bar Q^2-m^2}{\Kt^2+\bar Q^2}\right) \frac{\der\sigma_{\rm D}^{\gamma^*_TA\to q(\bar q)A}}{\der^2\Kb\der z_1\der z_2} \,.
\end{align}
In the limit $Q^2\gg \Kt^2,m$, the cross section is dominated by aligned jet configurations $z_1\ll 1$ or $1-z_1\ll 1$, and the delta function can be approximated by
\begin{align}
    \delta\left(\beta-\frac{\bar Q^2-m^2}{z_2\Kb^2+z_1(\Kb-\mathbf{\Delta})^2+\bar Q^2}\right)&\approx\frac{(\Kb-\Deltab)^2+m^2}{(1-\beta)^2Q^2}\delta\left(z_1-1+\frac{\beta}{1-\beta}\frac{(\Kb-\Deltab)^2+m^2}{Q^2}\right)\nonumber\\
    &+\frac{K_\perp^2+m^2}{(1-\beta)^2Q^2}\delta\left(z_1-\frac{\beta}{1-\beta}\frac{\Kt^2+m^2}{Q^2}\right)\,.
\end{align}
For a quark measured in the current fragmentation region, we only keep the first $\delta$ function with $z_1$ close to 1 (it would also be interesting to study the diffractive contribution in the target fragmentation region, following Ref.\,\cite{Caucal:2025qjg}).
It is then easy to integrate over $z_1$ and $z_2$: the two delta functions fix $z_2=1-z_1$ and $\bar Q=(\beta (\Kb-\Deltab)^2+m^2)/(1-\beta)$ in the \eqs\eqref{eq:A_difSIDIS-ip}-\eqref{eq:B_difSIDIS-ip} so that we get in the end a TMD factorized result
\begin{align}
     \frac{\der\sigma_{\rm D}^{\gamma^*_TA\to qA}}{\der^2\Kb\der \ln(1/\beta)}
     &=\frac{4\pi^2\aem e_f^2}{Q^2}\times \frac{\der x q_\po (x,\xpo,\Kb)}{\der^2\Kb} \,.
\end{align}
One recognizes in this expression the hard factor $\sim 1/Q^2$, which is the same as for SIDIS, while the quark DTMD in momentum space eventually reads
\begin{align}
     \frac{\der x q_\po (x,\xpo,\Kb)}{\der^2\Kb}=\frac{\Nc x}{2\pi^2}&\int\der^2\mathbf{\Delta}\int\der^2\qb\int\der^2\qb' \ \mathcal{D}(\qb,x_\po,\mathbf{\Delta})\mathcal{D}^*(\qb',x_\po,\mathbf{\Delta})\nonumber\\
     &\times\mathcal{T}_q(\Kb,\qb,\qb',\mathbf{\Delta};x,m)\label{eq:qDTMD-momentum-space-ip}
\end{align}
with, following the notations of Ref.\,\cite{Hatta:2022lzj},
\begin{align}
    \mathcal{T}_q(\Kb,\qb,\qb',\mathbf{\Delta};x,m)&=T_q(\Kb,\qb,\qb',\mathbf{\Delta};x,m)+T_q(\Kb,\boldsymbol{0}_\perp,\boldsymbol{0}_\perp,\mathbf{\Delta};x,m)\nonumber\\
    &-T_q(\Kb,\boldsymbol{0}_\perp,\qb',\mathbf{\Delta};x,m)-T_q(\Kb,\qb,\boldsymbol{0}_\perp,\mathbf{\Delta};x,m)\\
    T_q(\Kb,\qb,\qb',\mathbf{\Delta};\beta,m)&=\frac{((\Kb-\mathbf{\Delta})^2+m^2)}{\left[(1-\beta)(\Kb-\qb-\mathbf{\Delta}/2)^2+\beta(\Kb-\mathbf{\Delta})^2+m^2\right]}\nonumber\\
    &\times \frac{(m^2+(\Kb-\qb-\mathbf{\Delta}/2)\cdot(\Kb-\qb'-\mathbf{\Delta}/2))}{\left[(1-\beta)(\Kb-\qb'-\mathbf{\Delta}/2)^2+\beta(\Kb-\mathbf{\Delta})^2+m^2\right]}
\end{align}
This equation generalizes the formula from Ref.\,\cite{Hatta:2022lzj} to the massive quark case. It also extends the coordinate-space result for the incoherent quark DTMD~\cite{Rodriguez-Aguilar:2024efj} by accounting for quark mass corrections, provided one performs the CGC average outside the product between the two dipole operators and subtracts the coherent case.

Note however that our result differs from that in Ref.\,\cite{Hatta:2022lzj} even for $m=0$: the $\mathbf{\Delta}$ dependence not only appears inside the dipole scattering amplitude but also in the function $\mathcal{T}_q$. It will be interesting to investigate further the phenomenological consequences of this dependence, in particular for what concerns the azimuthal correlations between $\mathbf{\Delta}$ and $\Kb$~\cite{Hatta:2016dxp}.
For the particular case $m=0$ and $\Deltab=\mathbf{0}$, we recover the result obtained in Ref.\,\cite{Hatta:2022lzj}. 

If one assumes that the dipole $S$-matrix depends on the dipole size $\rb$ only, then $\mathcal{D}(\qb,x_\po,\Deltab)=\delta(\Deltab)\mathcal{D}(\qb,x_\po)$\footnote{Here, we slightly abuse the notation by using $\mathcal{D}(\qb,x_\po)$ to denote the Fourier transform of the dipole $S$-matrix with respect to the dipole size only.} and therefore, using
\begin{align}
\delta(\mathbf{0})=\int\frac{\der^2\mathbf{b}}{(2\pi)^2}e^{i\mathbf{0}_\perp\cdot\mathbf{b}_\perp}=\frac{S_\perp}{(2\pi)^2}\,,\label{eq:deltaofzero}
\end{align}
we find
\begin{align}
    \frac{\der x q_\po (x,\xpo,\Kb)}{\der^2\Kb}=\frac{S_\perp \Nc x}{8\pi^4}\int\der^2\qb\int\der^2\qb' \ \mathcal{D}(\qb,x_\po)\mathcal{D}^*(\qb',x_\po)\mathcal{T}_q(\Kb,\qb,\qb',\mathbf{0};x,m)\label{eq:qDTMD-momentum-space}\,.
\end{align}
This result is exactly equivalent to the coordinate space expression \eq\eqref{eq:softquark:qDTMD} obtained in the main text.

\subsection{Asymptotic behaviours}

We wish now to obtain the small and large $\Kt$ limits of the heavy-quark DTMD based on \eq\eqref{eq:qDTMD-momentum-space}. For simplicity and to connect the results with those discussed in the main text, we neglect $\Deltab$. For $\Kt\to 0$, one readily gets
\begin{align}
    \lim\limits_{\Kt\to 0}\frac{1}{S_\perp}\frac{\der x q_\po (x,\xpo,\Kb)}{\der^2\Kb}&=\frac{\Nc x}{8\pi^4}\left[\int\der^2\qb\frac{(1-x)\qb^2}{(1-x)\qb^2+m^2}\mathcal{D}(\qb,x_\po)\right]^2 \,,
\end{align}
which is equivalent to \eq\eqref{eq:qDTMD-small-Kt}. 

The dilute limit $Q_s\ll m,\Kt$ is more subtle. One particular contribution is obtained by expanding the kernel $\mathcal{T}_q(\Kb,\qb,\qb';x,m)$ for $||\qb||,||\qb'||\ll \Kb,m$ such that
\begin{align}
    \frac{1}{S_\perp}\frac{\der x q_\po (x,\xpo,\Kb)}{\der^2\Kb}\sim \frac{\Nc x(1-x)^2}{8\pi^4}\frac{[(m^2+2x\Kt^2)^2+m^2\Kt^2]}{(\Kt^2+m^2)^4}\left[\int\der^2\qb \ \qb^2\mathcal{D}(\qb,x_\po)\right]^2 .\label{eq:qDTMD-hard-contrib1}
\end{align}
In the MV model, the integral over $\qb$ is divergent because of the perturbative tail $1/\qb^4$ of the Fourier transform of the dipole $S$-matrix and the integral over $\qb$ must be cut-off in the UV by a scale of the order of $\Kt^2+ m^2$. Defining the (tree-level) gluon inclusive PDF as
\begin{align}
    x_\po G(x_\po,\mu)=\frac{\Nc S_\perp}{2\pi^2\as}\int\der^2\qb \ \qb^2\mathcal{D}(\qb,x_\po)\Theta(\mu^2-\qb^2)\,,
\end{align}
one can relate the dilute limit of the quark DTMD to the gluon PDF as
\begin{align}
    \frac{1}{S_\perp}\frac{\der x q_\po (x,\xpo,\Kb)}{\der^2\Kb}\sim \frac{\as^2x(1-x)^2}{2N_c}\frac{[(m^2+2x\Kt^2)^2+m^2\Kt^2]}{(\Kt^2+m^2)^4}\frac{\left[x_\po G(x_\po,\mu)\right]^2}{S_\perp^2} \,,\label{eq:diffTMD-dilute}
\end{align}
with $\mu\sim \Kt^2+m^2$. In particular, in the MV model one finds
\begin{align}
\frac{x_\po G(x_\po,\mu)}{S_\perp}\,=\,\frac{\Nc Q_A^2}{4\pi^2\as}\,\ln\frac{\mu^2}{\Lambda^2}\,,
\label{eq:MVPDF}
\end{align}
which together with \eqn{eq:diffTMD-dilute} leads to \eqn{eq:qDTMD-small-Qs} in the main text.

As discussed in Section~\ref{sec:Phys}, there is another contribution to the large $K_\perp$ tail of the quark DTMD coming from the single hard scattering regime of the dipole S-matrix, i.e.~the domain where $\mathcal{D}(q_\perp,x_\po)$ falls like a $1/q_\perp^4$. This contribution is not suppressed when $x$ is small. To unravel this contribution, which exists for any $m$ values, let us first consider the simpler case $m=0$ and $x\ll 1$, such that \eq\eqref{eq:qDTMD-momentum-space} reduces to
\begin{align}
      \frac{\der x q_\po (x,\xpo,\Kb)}{\der^2\Kb}&\approx\frac{S_\perp \Nc x K_\perp^2}{8\pi^4}\left|\int\der^2\qb \ \mathcal{D}(\qb,x_\po)\left[\frac{(\Kb-\qb)^i}{(\Kb-\qb)^2}-\frac{\Kb^i}{\Kb^2}\right]\right|^2 \,,
\end{align}
where the $\approx$ symbol only refers to the $x\ll 1$ approximation here.
Performing the integral over the azimuthal angle between $\qb$ and $\Kb$, one finds
\begin{align}
         \frac{\der x q_\po (x,\xpo,\Kb)}{\der^2\Kb}&=\frac{S_\perp \Nc x K_\perp^2}{8\pi^4}\left|\frac{\Kb^i}{\Kb^2}\int\der^2\qb \ \mathcal{D}(\qb,x_\po)\left[\Theta(|\qb|\le |\Kb|)-1\right]\right|^2\\
         &=\frac{S_\perp \Nc x}{8\pi^4}\left|\int\der^2\qb \ \mathcal{D}(\qb,x_\po)\Theta(|\qb|\ge |\Kb|)\right|^2 \,,\label{eq:qDTMD-hard-contrib2}
\end{align}
which is the exact $x\to 0$ limit of the quark DTMD when $m=0$. It is clear that this contribution admits a different physical interpretation from that in Eq.\,\eqref{eq:qDTMD-hard-contrib1}. In the latter case, the $1/K_\perp^4$ behavior arises from the expansion of the splitting kernel, independently of the form of the dipole $S$-matrix at large $q_\perp$ (in particular, this $1/K_\perp^4$ behavior is also present in the GBW model, where the integral over $\qb$ in Eq.\,\eqref{eq:qDTMD-hard-contrib1} would actually be convergent). By contrast, the contribution in Eq.\,\eqref{eq:qDTMD-hard-contrib2} enters the hard tail of the quark DTMD only if the dipole $S$-matrix incorporates the single hard-scattering regime, i.e.~if $\mathcal{D}(\qb,x_\po)\sim Q_A^2/(2\pi q_\perp^4)$ at large $q_\perp$, as in the MV model. Indeed, in this case and for $K_\perp^2\gg Q_A^2$, one has
\begin{align}
    \int\der^2\qb \ \mathcal{D}(\qb,x_\po)\Theta(|\qb|\ge |\Kb|)&=\frac{Q_A^2}{2K_\perp^2}\,,
\end{align}
such that
\begin{align}
         \frac{\der x q_\po (x,\xpo,\Kb)}{\der^2\Kb}&=\frac{S_\perp \Nc x }{8\pi^4}\frac{Q_A^4}{4K_\perp^4} \,,\label{eq:hardkt-tail-x=0}
\end{align}
which is precisely the $x\to 0$ limit of Eq.\,\eqref{eq:DTMD-high-kt} in the main text. The momentum space expression of the quark DTMD enables a clear separation between these two contributions.

This calculation can easily be generalized to non-zero $x$ values. For $m=0$ (in this discussion, we do not care about $m$ since we are interested in the large $K_\perp\gg m$ limit), one has
\begin{align}
       \frac{\der x q_\po (x,\xpo,\Kb)}{\der^2\Kb}&=\frac{S_\perp \Nc x K_\perp^2}{8\pi^4}\left|\int\der^2\qb \ \mathcal{D}(\qb,x_\po)\left[\frac{(\Kb-\qb)^i}{(1-x)(\Kb-\qb)^2+xK_\perp^2}-\frac{\Kb^i}{\Kb^2}\right]\right|^2 \label{eq:qDTMD-square}\\
       &=\frac{S_\perp \Nc x}{8\pi^4}\left|2\pi \int_0^\infty\der q_\perp q_\perp \ \mathcal{D}(q_\perp,x_\po)\theta\left(\frac{\sqrt{1-x}q_\perp}{K_\perp},x\right)\right|^2 \,,\label{eq:qDTMD-hard-contrib2-xdep}
\end{align}
where the function $\theta(u,x)$ is defined as
\begin{align}
    \theta(u,x)&=\frac{u^2+(1-2x)\left[-1+\sqrt{1-2(1-2x)u^2+u^4}\right]}{2(1-x)\sqrt{1-2(1-2x)u^2+u^4}} \,.
\end{align}
The function $\theta(u,x)$ roughly behaves like the Heaviside distribution $\Theta(u-1)$: for $u\ge 1$, the function reaches rapidly its asymptotic value one at infinity, while for $u\ll 1$ it goes to $0$ like $2xu^2$. As a matter of fact, for $x=0$, one has exactly $\theta(u,0)=\Theta(u-1)$ and one thus recovers Eq.\,\eqref{eq:qDTMD-hard-contrib2}.
From Eq.\,\eqref{eq:qDTMD-hard-contrib2-xdep}, one can then easily obtain the two contributions to the hard $K_\perp$ tail of the quark DTMD. One is coming from the regime where $u\ge 1$ --- meaning $q_\perp^2\ge K_\perp^2/(1-x)$ --- where $\theta(u,x)\approx 1$. Therefore, 
\begin{align}
           \frac{\der x q_\po (x,\xpo,\Kb)}{\der^2\Kb}
       &=\frac{S_\perp N_c x}{8\pi^4}\left|\int_{K_\perp/\sqrt{1-x}}^\infty\der q_\perp q_\perp \ \frac{Q_A^2}{q_\perp^4}\right|^2\\
       &=\frac{S_\perp N_c x (1-x)^2}{8\pi^4}\frac{Q_A^4}{4K_\perp^4} \,.
\end{align}
This equation generalizes Eq.\,\eqref{eq:hardkt-tail-x=0} for non-zero $x$ values and gives the square of the first term in the square bracket of Eq.\,\eqref{eq:DTMD-high-kt}.
The other contribution, which vanishes for $x=0$, is coming from the region $u\ll 1$ --- meaning, $q_\perp^2\ll K_\perp^2/(1-x)$ --- where one can expand $\theta(u,x)\approx 2x u^2$:
\begin{align}
    \frac{\der x q_\po (x,\xpo,\Kb)}{\der^2\Kb}&= \frac{S_\perp N_c x}{8\pi^4}\frac{4x^2(1-x)^2}{K_\perp^4}\left|\int\der^2\qb \ \qb^2\mathcal{D}(\qb,x_\po)\right|^2 \,,
\end{align}
which is exactly Eq.\,\eqref{eq:qDTMD-hard-contrib1} for $m=0$ and gives the square of the second term in the square bracket of Eq.\,\eqref{eq:DTMD-high-kt}. Of course, since the quark DTMD for $m=0$ is a pure square, cf. Eq.\,\eqref{eq:qDTMD-square}, there is also an ``interference'' contribution coming from the product between these two well separated phase spaces, which then complete the asymptotic expression Eq.\,\eqref{eq:DTMD-high-kt} for the large $K_\perp$ tail of the quark DTMD.

\section{Gluon DTMD in momentum space}
\label{sec:gDTMD ms}

In this appendix, we derive the momentum space expression for the gluon DTMD, keeping track of its dependence on the total momentum transfer $\Deltab$. Our starting point is the tensor $G^{mn}$ defined in Eq.\,\eqref{eq: rel. soft: semi-hard tensor} that we write in a slightly different way to account for the impact parameter dependence of the gluon-gluon dipole:
\begin{align}\label{GBfull}
    G^{mn}(\Kb,\kbg,\Omega)&=\int\der^2\mathbf{k}\,\mathcal{C}(\kbg/2-\mathbf{k}-\Kb/2,\Kb+\kbg)\frac{\mathbf{k}^m\mathbf{k}^n-\frac{1}{2}k_\perp^2\delta^{mn}}{k_\perp^2+\Omega^2} \,, 
\end{align}
where $\mathcal{C}$ is related to the Fourier transform of the dipole correlator in the adjoint representation as
\begin{align}
    \mathcal{C}(\qb,\mathbf{\Delta})&=\int\frac{\der^2\rb}{(2\pi)^2}\frac{\der^2\bb}{(2\pi)^2}\ e^{-i\mathbf{\Delta}\cdot\bb-i\qb\cdot\rb} \left[ \frac{1}{N_c^2-1}\textrm{tr}(U_{\bb+\rb/2} U^\dagger_{\bb-\rb/2})-1\right]\,.
\end{align}
The generalization to non zero $\mathbf{\Delta}$ thus reads, after the change of variables $\qb=\kbg/2-\mathbf{k}-\Kb/2=\mathbf{\Delta}/2-\mathbf{k}-\Kb$ in the integral over $\mathbf{k}$ (with $\mathbf{\Delta}=\kbg+\Kb$):
\begin{align}
    &G^{nm}(\Kb,Y_\po,\mathbf{\Delta})
    =\frac{1}{2}\int\der^2\qb \ \mathcal{D}_A(\qb,x_\po,\mathbf{\Delta})\nonumber\\
    &\times\left\{\frac{2(\Kb+\qb-\mathbf{\Delta}/2)^n(\Kb+\qb-\mathbf{\Delta}/2)^m-(\Kb+\qb-\mathbf{\Delta}/2)^2\delta^{nm}}{(\Kb+\qb-\mathbf{\Delta}/2)^2+\Omega^2}-\frac{2K^nK^m-K^2\delta^{nm}}{K^2+\Omega^2}\right\} \,,
\end{align}
where the second term in the curly bracket comes from the  $-1$ term in the colour structure. The momentum space ajoint dipole $\mathcal{D}_A(q,x_\po,\Deltab)$ is defined as in Eq.\,\eqref{eq:DF-ms} with the Wilson lines and trace in the adjoint representation.
The gluon diffractive PDF is obtained by squaring this expression:
\begin{align}
    &G^{nm}(\Kb,Y_\po,\mathbf{\Delta})G^{rm*}(\Kb,Y_\po,\mathbf{\Delta})=\frac{1}{4}\int\der^2\qb\int\der^2\qb'\mathcal{D}_A(\qb,x_\po,\mathbf{\Delta})\mathcal{D}_A^*(\qb',x_\po,\mathbf{\Delta})\nonumber\\
    &\times \left\{\frac{2K^nK^m-K^2\delta^{nm}}{K^2+\Omega^2}-\frac{2(\Kb+\qb-\mathbf{\Delta}/2)^n(\Kb+\qb-\mathbf{\Delta}/2)^m-(\Kb+\qb-\mathbf{\Delta}/2)^2\delta^{nm}}{(\Kb+\qb-\mathbf{\Delta}/2)^2+\Omega^2}\right\}\nonumber\\
    &\times \bigg\{\frac{2K^{r}K^m-K^2\delta^{rm}}{K^2+\Omega^2}-\frac{2(\Kb+\qb'-\mathbf{\Delta}/2)^{r}(\Kb+\qb'-\mathbf{\Delta}/2)^m-(\Kb+\qb'-\mathbf{\Delta}/2)^2\delta^{rm}}{(\Kb+\qb'-\mathbf{\Delta}/2)^2+\Omega^2}\bigg\}.
\end{align}
Now, one should recall how $\Omega^2$ is expressed in terms of the target variables. We have
\begin{align}
    \Omega^2=\frac{\beta}{1-\beta}\kbg^2=\frac{\beta}{1-\beta}(\mathbf{\Delta}-\Kb)^2\,,
\end{align}
so that
\begin{align}
    &G^{nm}(\Kb,Y_\po,\mathbf{\Delta})G^{rm*}(\Kb,Y_\po,\mathbf{\Delta})=\frac{(1-\beta)^2}{4}\int\der^2\qb\int\der^2\qb'\mathcal{D}_A(\qb,x_\po,\mathbf{\Delta})\mathcal{D}_A^*(\qb',x_\po,\mathbf{\Delta})\nonumber\\
    &\times \bigg\{\frac{2K^nK^m-K^2\delta^{nm}}{(1-\beta)K^2+\beta(\Kb-\mathbf{\Delta})^2} \nonumber\\
    &\hskip 5ex-\frac{2(\Kb+\qb-\mathbf{\Delta}/2)^n(\Kb+\qb-\mathbf{\Delta}/2)^m-(\Kb+\qb-\mathbf{\Delta}/2)^2\delta^{nm}}{(1-\beta)(\Kb+\qb-\mathbf{\Delta}/2)^2+\beta(\Kb-\mathbf{\Delta})^2}\bigg\}\nonumber\\
    &\times \bigg\{\frac{2K^{r}K^m-K^2\delta^{rm}}{(1-\beta)K^2+\beta(\Kb-\mathbf{\Delta})^2} \nonumber \\
    &\hskip 5ex-\frac{2(\Kb+\qb'-\mathbf{\Delta}/2)^{r}(\Kb+\qb'-\mathbf{\Delta}/2)^m-(\Kb+\qb'-\mathbf{\Delta}/2)^2\delta^{rm}}{(1-\beta)(\Kb+\qb'-\mathbf{\Delta}/2)^2+\beta(\Kb-\mathbf{\Delta})^2}\bigg\} \,.
\end{align}
Including the prefactor from the jacobian between $z_3$ and $x$ and the colour factors, the gluon DTMD written differentially with respect to $\Deltab$ reads
\begin{align}
   &\frac{\der x\mathcal{G}^{nr}_\po (x,\xpo,\Kt)}{\der^2 \Kb\der^2\mathbf{\Delta}}=\frac{(\Nc^2-1)(1-x)}{4\pi^2}\int\der^2\qb\int\der^2\qb'\mathcal{D}_A(\qb,x_\po,\mathbf{\Delta})\mathcal{D}_A^*(\qb',x_\po,\mathbf{\Delta})\nonumber\\
    &\times \bigg\{\frac{2K^nK^m-K^2\delta^{nm}}{(1-x)K^2+x(\Kb-\mathbf{\Delta})^2}\nonumber \\
    & \hskip 5ex -\frac{2(\Kb+\qb-\mathbf{\Delta}/2)^n(\Kb+\qb-\mathbf{\Delta}/2)^m-(\Kb+\qb-\mathbf{\Delta}/2)^2\delta^{nm}}{(1-x)(\Kb+\qb-\mathbf{\Delta}/2)^2+x(\Kb-\mathbf{\Delta})^2}\bigg\} \nonumber\\
    &\times \bigg\{\frac{2K^{r}K^m-K^2\delta^{rm}}{(1-x)K^2+x(\Kb-\mathbf{\Delta})^2}\nonumber \\
    &\hskip 5ex -\frac{2(\Kb+\qb'-\mathbf{\Delta}/2)^{r}(\Kb+\qb'-\mathbf{\Delta}/2)^m-(\Kb+\qb'-\mathbf{\Delta}/2)^2\delta^{rm}}{(1-x)(\Kb+\qb'-\mathbf{\Delta}/2)^2+x(\Kb-\mathbf{\Delta})^2}\bigg\} \,.\label{eq:GTMD-ms-final}
\end{align}
Note that in general, for non zero $\Deltab$, the gluon DTMD may have a non-trivial tensorial structure (not only $\delta^{nr}$ contributes), which may lead to interesting correlations once contracted with the hard factor $\mathcal{H}^{nr}(z_1,z_2,P_\perp,Q,m)$. When $\Deltab=\boldsymbol{0}$, Eq.\,\eqref{eq:GTMD-ms-final} is equivalent to the expression given in Ref.\,\cite{Hatta:2022lzj} ; yet, our $\Deltab$ dependence is different as it non only appears in the adjoint dipole operator but also in the splitting kernel.

The relation between Eq.\,\eqref{eq:GTMD-ms-final} and the unintegrated gluon distribution of the Pomeron defined in coordinate space, cf.~Eq.\,\eqref{GDTMD} of the main text, is obtained by contracting Eq.\,\eqref{eq:GTMD-ms-final} with $\delta^{nr}$ and integrating over the total momentum transfer $\Deltab$:
\begin{align}
     \frac{\der x\mathcal{G}_\po (x,\xpo,\Kt)}{\der^2 \Kb}&=\int\der^2\Deltab  \frac{\der x\mathcal{G}^{nn}_\po (x,\xpo,\Kt)}{\der^2 \Kb\der^2\mathbf{\Delta}}\,.
\end{align}
As a cross-check of the overall prefactor in Eq.\,\eqref{eq:GTMD-ms-final}, let us compute the $K_\perp\to 0$ limit of the distribution from our momentum space expression, assuming a homogeneous target. In this case, $\mathcal{D}_A(\qb,x_\po,\Deltab)=\delta(\Deltab)\mathcal{D}_A(\qb,x_\po)$ and therefore, using again Eq.\,\eqref{eq:deltaofzero},
\begin{align}
     \lim_{K_\perp\to 0}\frac{\der x\mathcal{G}_\po (x,\xpo,\Kt)}{\der^2 \Kb}&=\frac{(N_c^2-1)(1-x)S_\perp}{4\pi^2(2\pi)^2}\int\der^2\qb\int\der^2\qb'\mathcal{D}_A(\qb,x_\po)\mathcal{D}_A^*(\qb',x_\po)\nonumber\\
    &\times \bigg\{\frac{2K^nK^m-K^2\delta^{nm}}{K_\perp^2} -\frac{2\qb^n\qb^m-\qb^2\delta^{nm}}{(1-x)\qb^2}\bigg\}\nonumber\\
    &\times\bigg\{\frac{2K^nK^m-K^2\delta^{nm}}{K_\perp^2} -\frac{2\qb'^n\qb'^m-\qb'^2\delta^{nm}}{(1-x)\qb'^2}\bigg\}\\
    &=\frac{(N_c^2-1)(1-x)S_\perp}{8\pi^4}
\end{align}
in agreement with the result of Ref.\,\cite{Iancu:2022lcw}.

\bibliographystyle{JHEP-2modlong}
\bibliography{refs}
\end{document}